\documentclass{article}

\usepackage{arxiv}

\usepackage[utf8]{inputenc} % allow utf-8 input
\usepackage[T1]{fontenc}    % use 8-bit T1 fonts
\usepackage{hyperref}       % hyperlinks
\usepackage{url}            % simple URL typesetting
\usepackage{booktabs}       % professional-quality tables
\usepackage{amsfonts}       % blackboard math symbols
\usepackage{amsmath}        % align environment and more
\usepackage{amssymb}        % for math symbols like \triangledown
\usepackage{nicefrac}       % compact symbols for 1/2, etc.
\usepackage{subcaption}     % for subfigure environment
\usepackage{microtype}      % microtypography
\usepackage{lipsum}
\usepackage{graphicx}
\usepackage{mathMacros}
\graphicspath{ {./images/} }

\newcommand{\beqn}{\begin{equation}}
\newcommand{\eeqn}{\end{equation}}

\newcommand\bx{\boldsymbol x}

\newcommand\by{\boldsymbol y}

\newcommand\bu{\boldsymbol u}

\newcommand\bv{\boldsymbol v}
\newcommand\bV{\boldsymbol V}

\newcommand\bW{\boldsymbol W}

\newcommand\bC{\boldsymbol C}

\newcommand\bS{\boldsymbol S}
\newcommand\bT{\boldsymbol T}

\newcommand\bPhi{\boldsymbol{\Phi}}
\newcommand\bSigma{\boldsymbol{\Sigma}}

\newcommand\bR{\boldsymbol{R}}

\title{The coherent structures of EVP Fluid Flow Past a Circular Cylinder}

\author{
 Adrián Corrochano \\
  School of Aerospace Engineering\\
  Universidad Politécnica de Madrid\\
  Madrid, E-28040 \\
  \texttt{adrian.corrochanoc@upm.es} \\
   \And
 Kazi Tassawar Iqbal \\
  SeRC and FLOW, Engineering mechanics\\
  KTH Royal Institute of Technology\\
  Stockholm, SE-10044 \\
  \texttt{ktiqbal@kth.se} \\
  \And
 Saeed Parvar \\
  SeRC and FLOW, Engineering mechanics\\
  KTH Royal Institute of Technology\\
  Stockholm, SE-10044 \\
  \texttt{s.parvar@hotmail.com} \\
   \And
   Soledad Le Clainche \\
  School of Aerospace Engineering\\
  Universidad Politécnica de Madrid\\
  Madrid, E-28040 \\
   \texttt{soledad.leclainche@upm.es} \\
   \And
   Outi Tammisola \\
  SeRC and FLOW, Engineering mechanics\\
  KTH Royal Institute of Technology\\
  Stockholm, SE-10044 \\
   \texttt{outi@mech.kth.se} \\
}

\begin{document}
\maketitle
\begin{abstract}
This study investigates the impact of elasticity and plasticity on two-dimensional flow past a circular cylinder at Reynolds number $Re = 100$. Ten direct numerical simulations were performed using the Saramito-Herschel–Bulkley model to represent viscoelastic and elastoviscoplastic (EVP) fluids. The flow evolves from a periodic von Kármán vortex street to chaotic-like regimes. Proper Orthogonal Decomposition (POD) and Higher Order Dynamic Mode Decomposition (HODMD) are applied to extract dominant flow structures and their temporal dynamics. For viscoelastic fluids, increasing the Weissenberg number $Wi$ elongates the recirculation bubble and shifts it downstream, resulting in more intricate but still periodic behavior. In EVP fluids, seven cases explore variations in Bingham number $Bn$, solvent viscosity ratio $\beta_s$, and power law index $n$, aiming to qualitatively assess their influence rather than determine critical thresholds. Results indicate that stronger plastic effects, especially with $n \ge 1$, lead to increased flow complexity. Three dynamic regimes are identified: (i) periodic; (ii) transitional, with elongated recirculation and disrupted periodicity; and (iii) fully complex, with breakdown of recirculation. Overall, the study highlights the interplay between inertia, elasticity, and yield stress in non-Newtonian flows past obstacles and identifies key parameters driving the transition from periodic to complex regimes.
\end{abstract}


\section{Introduction}\label{section:introduction}
Vortex shedding behind a cylinder plays a crucial role in fluid dynamics and has gained significant attention due to its relevance for engineering applications, such as the flow past cables, bridges, heat exchangers, piers, buildings, and automobiles. The flow behind a circular cylinder also serves as a benchmark for developing new methodologies \cite{corrochano2022structural,Abadia2025}. Extensive research has been conducted on the flow pattern of Newtonian fluids past a cylinder, and it has been thoroughly documented \cite{Posdziech2007,Rajani2009,Qu2013,Constant2017}. The Newtonian flow around a cylinder becomes unsteady and time-periodic at Reynolds number, which compares the inertial and viscous terms, $Re\approx47$, based on the cylinder diameter and inflow velocity. The origin of this unsteadiness is a supercritical Hopf bifurcation that has been well characterised in the literature \cite{Williamson1988,Williamson1996, StrykowskiSreenivasan90}. When the Reynolds number increases, the wake then becomes three-dimensional at around $Re=180$. Increasing the Reynolds number even more, the wake becomes turbulent. The effect of numerous parameters on this onset of the first bifurcation has been studied, including but not limited to confinement \cite{Sahin2004,Juniper2011}, rotation \cite{Pralits2010}, suction and blowing \cite{Delaunay2001,Kim2005,Tammisola2014}, and shape changes \cite{Bearman1998,Tammisola2017}.     

The present work studies the flow past a circular cylinder, but in a non-Newtonian fluid at moderate Reynolds number where inertial terms play an important role. More specifically, this study focuses on elastoviscoplastic (EVP) fluids, which are a class of materials that exhibit a combination of elastic, viscous, and plastic behaviors when subjected to stress. The presence of these combined behaviors is prevalent in both natural and industrial applications, including flows through porous materials, geological structures, lab-on-a-chip devices, and human biology \cite{Balmforth2014,Mossaz2010,Saramito2016}. In this context, the interaction between the shear layer, separation region, and wake in the cylinder flow gives rise to highly complex flow patterns. When combined with the rheological properties of the fluid, the complexity of the problem escalates, necessitating a meticulous and comprehensive investigation.

The use of machine learning algorithms for the analysis of complex flows, such as the ones related to non-Newtonian fluids, is widely spread in the community \cite{Brunton2020, LeClainche2023}. Regarding non-Newtonian fluid studies, Grilli et al. \cite{Grilli2013} conducted a study focused on simulating the flow of an Oldroyd-B fluid through a linear array of closely spaced cylinders in a channel at low Reynolds numbers. Their findings revealed a critical Weissenberg number at which the flow undergoes a transition to a purely elastic turbulent regime. To identify the key modes responsible for unsteady behavior, the researchers employed a stability analysis based on Dynamic Mode Decomposition (DMD) \cite{Schmid10}. Remarkably, they discovered that the most energetic modes corresponded to the stretching and relaxation of the polymer molecules by the flow. Ribau et al. \cite{Ribau2021} utilized Proper Orthogonal Decomposition (POD) \cite{Lumley1967} to investigate the von Kármán vortex street phenomenon in 2D flow around single and multiple cylinders with varying radii. The analysis focused on both Newtonian and power-law fluids exhibiting shear-thinning and shear-thickening behaviors. Shear-thinning and shear-thickening fluids are non-Newtonian fluids whose viscosity decreases or increases, respectively, with increasing shear rate. By leveraging the power of POD, a reduced-dimensional representation of the data was obtained, enabling a deeper comprehension of the flow characteristics. Hamid et al. \cite{Hamid22} numerically investigated the influence of fluid viscoelasticity on laminar vortex shedding past a circular cylinder over a range of Weissenberg numbers and polymer viscosity ratios. Using DMD, the analysis reveals that viscoelasticity significantly alters the dominant flow structures, suppresses vortex shedding, and reduces velocity fluctuation energy compared to the Newtonian case. More recently, Raffi et al. \cite{Raffi2024} studied the elastoviscoplastic flow past a circular cylinder, focusing on the change in the Weissenberg and Bingham numbers. The Bingham number quantifies the relative importance of a fluid's yield stress compared to the viscous stresses. The extensive research elucidates the effects of these two parameters on the streamlines, vorticity and the lift and drag forces. By means of DMD, the vortex-shedding frequency was calculated, as well as the coherent structure associated. This study also analyses the changes in the main flow patterns when varying the dimensionless parameters, discovering differences on the DMD modes.

Higher Order Dynamic Mode Decomposition (HODMD) \cite{LeClaincheVega17} is an extension of the DMD for the analysis of complex flows.  HODMD has proven its robustness to identify the main dynamics of the problem under study and their frequencies, even in challenging scenarios such as noisy experimental data and turbulent flows. In the field of the non-Newtonian fluids, Le Clainche et al. \cite{Clainche2020,LeClaincheRosti2022} conducted different studies using HODMD to investigate the coherent structures in turbulent channel flows of Newtonian and elastoviscoplastic (EVP) fluids. They analyzed non-equidistant temporal data to explore the role of near-wall streaks, their breakdown, and the interplay between turbulent dynamics and non-Newtonian effects. They identified six high-amplitude modes that characterized the yielded flow or the interaction between yielded and unyielded regions. They found that the interaction of streamwise velocity structures of different speeds was critical for streak breakdown, especially in Newtonian turbulence. Additionally, they observed that increasing elasticity and plasticity enhanced streamwise correlation, with long streaks intermittently disrupted by localized disturbances, resulting in drag reduction. Amor et al. \cite{Amor_2024} studied a planar jet when experiencing elastic turbulence. The findings provided the first characterization of the primary flow structures responsible for sustaining elastic turbulence in viscoelastic planar jets at low Reynolds numbers. Recently, Foggi et al. \cite{Foggi2025} proposed a unified view of fully developed elastic and elasto-inertial turbulence in channel flows. By means of HODMD, they analysed a viscoelastic channel flow at different conditions and correlated the structures appearing in the fluid, unifying the different regimes.

As in the previously discussed cases, HODMD is applied to a non-Newtonian fluid in the present study. The results are compared with those obtained from POD to provide a comprehensive characterization of the investigated cases. This study explores the impact of elastoviscoplasticity on the flow around a circular cylinder by employing the Saramito Herschel–Bulkley model \cite{Saramito2009} to describe the fluid's rheology. To overcome the challenges posed by high Weissenberg  numbers \cite{Fattal2004,Fattal2005,Vaithianathan2003,Vaithianathan2006}, the log-conformation approach is utilized. Comprehensive simulations are conducted to obtain velocity contour snapshots for various rheological characteristics. The obtained data is then subjected to analysis using POD and HODMD to elucidate the roles of yield stress, elasticity, shear-thinning, and shear-thickening on the wake dynamics, as well as the intricate interplay between vortex shedding and non-Newtonian effects. Three fluid types are examined: a Newtonian fluid, a purely viscoelastic fluid, and an elastoviscoplastic fluid. This approach allows the isolated assessment of the individual effects of plasticity and elasticity on the flow behavior.

The complexity of this configuration arises from the intricate interplay between various temporal and spatial scales, presenting significant challenges in its analysis. The present paper can be seen as an extension of the work made by Hamid et al. \cite{Hamid22}, where they investigated the change in Weissenberg numbers and polymer viscosity ratios; and Raffi et al. \cite{Raffi2024}, where they thoroughly examined the change in Weissenberg and Bingham numbers. While these studies only face periodic cases at Reynolds number $Re = 100$, the present study encounters, for the first time to the authors' knowledge, complex dynamics on the elastoviscoplastic flow past a circular cylinder. These complex dynamics also motivates the use of HODMD over the classical DMD method, as HODMD has proven its robustness to identify the main dynamics even in turbulent scenarios. 

This paper is structured as follows to achieve its objectives. First, the flow of interest is introduced and  an overview of the computational domain is provided in Sec. \ref{section:Numerical_Simulation_and_flow_description}. The extraction of the databases and the machine learning algorithms for the extraction of flow patterns are introduced in Sec. \ref{sec:num}. A brief characterization of the flow structures appearing in the numerical simulations is explained in Sec. \ref{section:Review_numerical_simulation_results}. Sections \ref{sec:POD_results} and \ref{sec:DMD} describe the main findings of the extracted flow patterns using POD and HODMD, respectively.
Finally, the conclusions of the study are summarized in Sec. \ref{sec:conclusions}.

\section{Numerical Simulation and flow description}\label{section:Numerical_Simulation_and_flow_description}

\begin{figure} %[h]1
\centering
\includegraphics[width=0.7\textwidth]{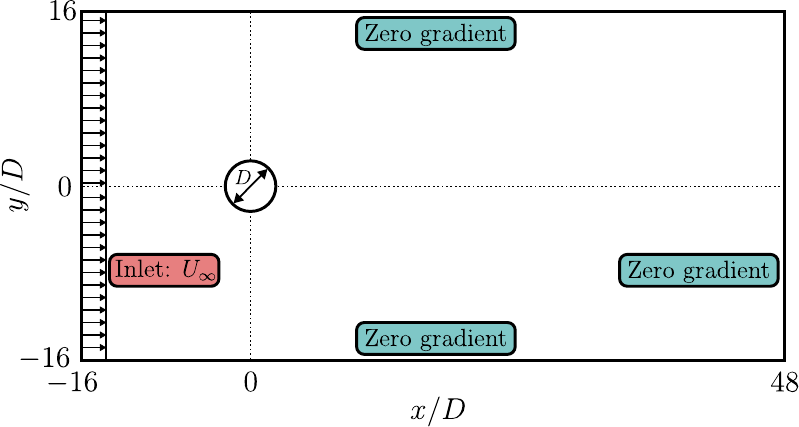}
\caption{\label{fig:image1}The schematic of flow past a cylinder}
\end{figure}

The flow setup is depicted in Fig. \ref{fig:image1}, where a circular cylinder with a diameter $D$ is positioned in an unbounded domain with a free stream velocity $U_{\infty}$ located far from the domain boundaries. The computational domain, centered at the origin, is scaled and normalized by $D$ and $U_{\infty}$, respectively. Time is normalized using the characteristic time scale $D/U_\infty$, while pressure differences and polymer stresses are normalized by the dynamic pressure of the free stream, denoted as $\rho U_\infty^2$, with $\rho$ representing the fluid density.

The dimensions of the computational domain span from [-16$D$, 48$D$] in the streamwise ($x$) direction and [-16$D$, 16$D$] in the vertical ($y$) direction relative to the origin. To enforce the no-slip/no-penetration condition on the cylinder surface, the immersed boundary method (IBM) is utilized, as explained in Sec. \ref{section:IBM0}. Although a grid size of $D/50$ is typically sufficient for numerical simulations of Newtonian fluid flow around a cylinder, as noted by Constant et al. \cite{Constant2017}, a finer mesh with a grid size of $D/64$ is employed in this study. The Eulerian grid is uniformly distributed in all directions, while the Lagrangian grid is equally distributed on the cylinder surface. The lateral boundaries are positioned far enough from the cylinder to have negligible confinement effects, and zero-gradient boundary conditions are applied. A uniform velocity profile is imposed as the inlet condition, and zero-gradient boundary conditions are enforced at the outlet. The simulations were performed on the HPE Cray EX supercomputer Dardel at the PDC Center for High-Performance Computing at KTH, utilizing 512 cores of the AMD EPYC\textsuperscript{TM} 7742 CPU for each simulation, with a wall-clock time of 192 hours per case.

\subsection{Governing Equations}\label{section:Governing_Equation}
This section provides a comprehensive overview of the governing equations, models, assumptions, and numerical methods employed in the study. To accurately capture the behavior of elastoviscoplastic (EVP) fluids with shear rate-dependent viscosity, the Saramito Herschel-Bulkley constitutive equation \cite{Saramito2009} is utilized.

The flow behavior is influenced by various dimensionless numbers, for which we have established reference scales. Specifically, we have previously defined the reference length and velocity scales. The reference viscosity is represented by the total viscosity, $\mu_0 = \mu_s + \mu_p$, where $\mu_s$ and $\mu_p$ correspond to the viscosities of the solvent and polymer, respectively. Moreover, the Reynolds number, $Re$, is defined as follows:

\begin{equation}
    Re = \frac{\rho U_{\infty}D}{\mu_{0}},
    \label{eq:3.5}
\end{equation}

\noindent where $\rho$ is the density of the fluid, $U_\infty$ is the free stream velocity, and $D$ is the diameter of the cylinder.
The Weissenberg number, denoted as $Wi$, represents the ratio between elastic and viscous forces and is expressed as:

\begin{equation}
    Wi = \frac{\lambda U_{\infty}}{D}.
    \label{eq:3.6}
\end{equation}

The parameter $\lambda$ represents the polymer relaxation time. The ratio of solvent viscosity, $\mu_s$, to total viscosity, $\mu_0$, is represented by $\beta_s$.

\begin{equation}
    \beta_{s} = \frac{\mu_{s}}{\mu_{0}} = \frac{\mu_{s}}{\mu_{p}+\mu_{s}}.
    \label{eq:3.7}
\end{equation}

The Bingham number, denoted as $Bn$, represents the ratio of the yield stress of the EVP fluid, $\tau_y$, to the viscous stress.

\begin{equation}
    Bn = \frac{\tau_{y}}{\mu_{0}(U_{\infty}/D)},
    \label{eq:3.8}
\end{equation}

The governing equations include the continuity equation, the momentum equation, and a constitutive equation for the polymer stress. Specifically, the continuity and momentum equations are: 
% continuity
\begin{align}
\label{eq:cont}
\divg{\tens{u}} = 0
\end{align}

\begin{align} % momentum
\label{eq:mom}
\pddt{\tens{u}} + \advec{\tens{u}} = 
-\grad{p} + 
\frac{1}{Re}\divg{\btau^s} + 
\divg{\btau^p} + 
\tens{f}.
\end{align}

The velocity vector, pressure, and elastoviscoplastic extra stress tensor are denoted by ${\tens{u}}={\tens{u}(\tens{x},t)}$, $p=p(\tens{x},t)$, and $\btau^p = \btau^p(\tens{x},t)$, respectively. The solvent stress contribution is calculated using $\btau^s=2\beta_s \tens{S}(\tens{u})$, where $\beta_s$ is the ratio of solvent viscosity to total viscosity and $\tens{S}(\tens{u})=(\grad{\tens{u}}+\trans{\grad{\tens{u}}})/2$ is the strain rate tensor. The momentum equation \eqref{eq:mom} includes a term $\tens{f}$, which is a body force used to apply no-slip/no-penetration boundary conditions on the cylinder's surface via the immersed boundary method. More information on this method can be found in Sec. \ref{section:IBM0}.

The constitutive equation of the EVP fluid explains how the EVP stress tensor $\btau^p$ evolves over time, as follows:
% SHB
\begin{equation}
Wi\, \ucd{\btau^p}+
F\btau^p -
\frac{2(1-\beta_s)}{Re}\tens{S}(\tens{u})=0,
\label{eq:SHB}
\end{equation}

\noindent where $F=\mathrm{max}\left(0,\dfrac{|\btau_d^p|-Bn/Re}{(2(1-\beta_s)/Re)^{1-n}|\btau_d^p|^n}\right)^{1/n}$. In \Eqref{eq:SHB}, the deviatoric stress tensor is $\btau_d^p = \btau^p - (\trace{\btau^p}/\trace{\tens{I}})\, \tens{I}$ (where $\tens{I}$ is the unit tensor and $\trace{:}$ is the trace of the tensor), and the second invariant of $\btau_d^p$ is $|\btau_d^p| \equiv \sqrt{\btau_d^p : \btau_d^p /2}$. The material behaves like a viscoelastic solid if the magnitude of $\btau_d^p$ is less than or equal to $\tau_{y}$. On the other hand, if the magnitude of $\btau_d^p$ exceeds $\tau_{y}$, the material flows as fluid. The Saramito model is a generalisation of the viscoplastic Herschel-Bulkley model, where the power-law index $n$ determines the degree of shear-thinning ($n < 1$), or in the case of $n > 1$, shear-thickening. While shear-thickening is not a common property for EVP fluids, in this work, we include such cases to qualitatively demonstrate how the instabilities change with the power law index. The Saramito Oldroyd-B model is a special case of the Herschel-Bulkley model when $n=1$ \cite{Saramito2007}.

In Eq. \ref{eq:SHB}, $\ucd{\btau^p}$ is the upper-convected derivative
\footnote{It is noted that the upper-convected derivative in \Eqref{eq:ucd} is written by defining the velocity gradient such that $(\grad{\tens{u}})_{ij}=\partial u_i/\partial x_j$.} is defined as follows,
\begin{align}
\label{eq:ucd}
\ucd{\btau^p} \equiv \pddt{\btau^p} + \advec{\btau^p} - (\grad{\tens{u}})\, \btau^p - \btau^p\, \trans{\grad{\tens{u}}}
\end{align}
which is used for the time derivative of $\btau^p$. To ensure the accuracy and reliability of our findings, we have employed a robust numerical approach. The numerical techniques utilized in this study are concisely described in the subsequent subsections.

\subsection{Immersed boundary method to model the flow past a circular cylinder}\label{section:IBM0}

There are various numerical methods for numerical simulation of the fluids and solids interaction. In this study, the immersed boundary method (IBM) was selected because it enables the use of highly efficient computational algorithms, such as the fast Fourier transform (FFT)-based pressure solver, which offers scalability and computational speed. The IBM was initially introduced by Peskin \cite{Peskin1972} and has since undergone several modifications, including the computation of body force ($\tens{f}$) on the right-hand side of Eq. \eqref{eq:mom} to act as a virtual boundary. A detailed review by Mittal and Iaccarino \cite{Mittal2005} provides further information on these developments.

In this study, the discrete forcing method was employed to represent the cylinder surface, incorporating improvements introduced by Breugem \etal \cite{Breugem2012}, resulting in second-order spatial accuracy. The simulation utilized a uniform Cartesian and Eulerian grid ($\Delta x = \Delta y$), while the Lagrangian grid was uniformly distributed to generate the solid cylinder surface. The initial velocity prediction was obtained using Eq. \ref{eq:cont} and Eq. \ref{eq:mom}, followed by velocity interpolation from the Eulerian to Lagrangian grid on the cylinder surface. It should be noted that the cylinder was considered stationary, with the application of no-slip and no-penetration conditions at the cylinder surface. For further details, interested readers are directed to consult the works of Breugem et al. \cite{Breugem2012}, Luo et al. \cite{Luo2007}, and Izbassarov et al. \cite{Izbassarov2018}.

A wide range of constitutive equations for viscoelastic and elastoviscoplastic materials involve the evolution of a conformation tensor, denoted as $\tens{A}$, following a general form given by Eq. \eqref{eq:AeqnGeneral}.

\begin{align}
\label{eq:AeqnGeneral}
\pddt{\tens{A}} + \advec{\tens{A}} - \grad{\tens{u}}\,\tens{A} - \tens{A}\trans{\grad{\tens{u}}} = \frac{1}{\lambda} \tilde{g}(\tens{A})
\end{align}

This equation encompasses terms related to advection, strain rate, and the function $\tilde{g}(\tens{A})$ specific to the employed model. To address the inherent numerical instability associated with the high Weissenberg number problem (HWNP), the log-conformation technique, proposed by Fattal and Kupferman \cite{Fattal2004,Fattal2005}, is adopted in this study. For the Saramito 2009 model, the relationship between the conformation tensor $\tens{A}$ and the polymer stress $\btau^p$ is described by Eq. \eqref{eq:Kramers}, where

\begin{equation}
\btau^p =\frac{1-\beta_s}{Wi}(\tens{A}-\tens{I}).
\label{eq:Kramers}
\end{equation}

To solve the governing equations, namely Eqs. \eqref{eq:cont}, \eqref{eq:mom}, and \eqref{eq:SHB}, a second-order central finite difference scheme is implemented on a uniform staggered grid. However, for the advection term in Eq. \eqref{eq:SHB}, a fifth-order weighted essentially non-oscillatory (WENO) scheme \cite{Shu2009,Sugiyama2011} is employed for improved accuracy. Time integration is achieved through a fractional-step, third-order explicit Runge-Kutta scheme \cite{Kim1985}. Further details can be found in the works of Izbassarov et al. \cite{Izbassarov2018,Izbassarov2021}.

\begin{table}[!h]
%  \begin{center}
\def~{\hphantom{0}}
  \begin{tabular}{cccccc}
   \hline 
   ($C_D$, $St$) &  &   &  &  & \\           
   \hline
    $Wi$    & $Re$ & Oliveira~\cite{Oliveira2001}  & Peng et al.~\cite{Peng2021} &~Minaeian et al.~\cite{Minaeian2022}~  &  ~Present simulations  \\[3pt]
    \hline
   
       0 & 100 & (1.370, 0.167) & (1.361, 0.165) & --- & (1.359, 0.164) \\
       1  & 100 & & ---  & (1.2006, -) & (1.21933, -) \\
  \hline
  \end{tabular}
  \caption{Drag coefficient and Strouhal number of a cylinder in an infinite unidirectional flow.}
  \label{tab:Cd}
%  \end{center}
\end{table}

\begin{figure} %DMD_1
    \centering
    \includegraphics[width=1\linewidth]{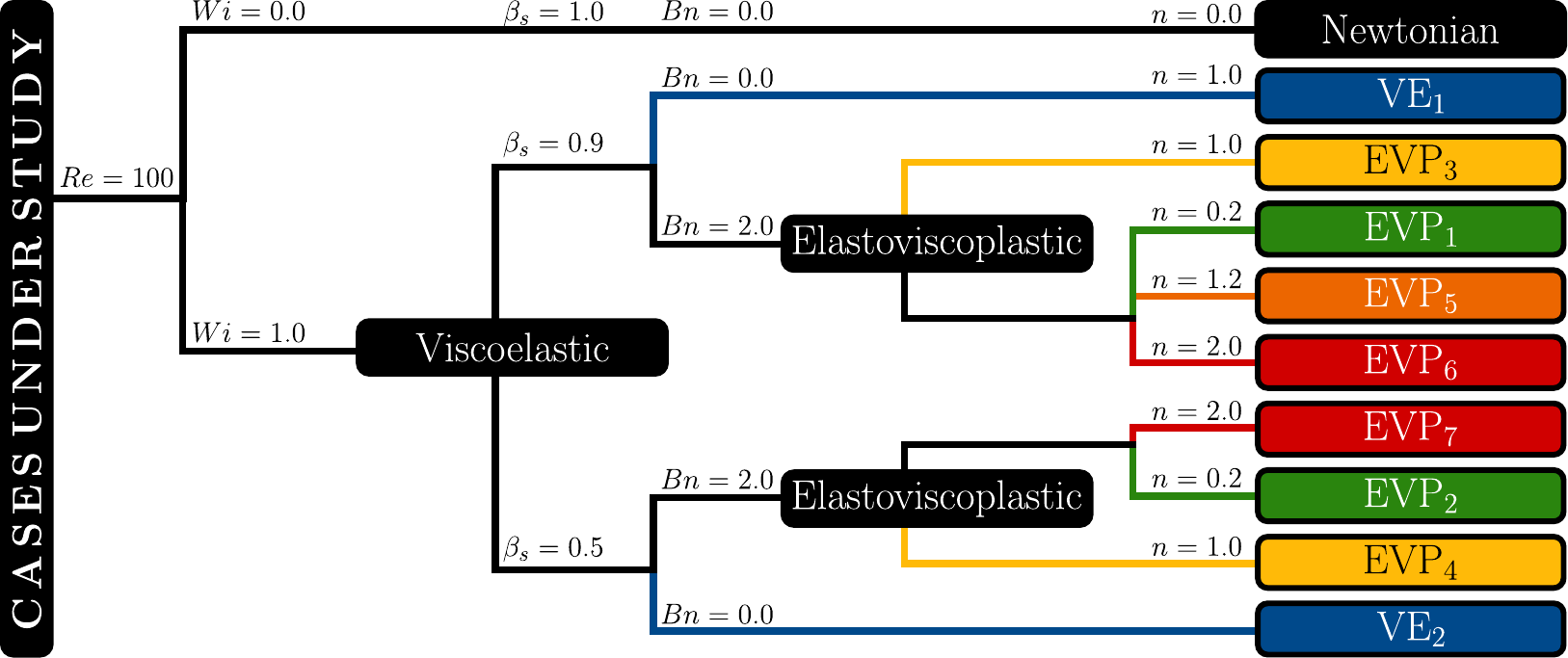}
    \caption{Schematic representation of the simulated cases under study. Each case is categorized as Newtonian, viscoelastic (VE), or elastoviscoplastic (EVP), with corresponding labels indicating the specific configuration.The Reynolds number is fixed at \( Re = 100 \) for all cases. The Weissenberg number is set to \( Wi = 1 \) for the non-Newtonian simulations. The Bingham number is set to \( Bn = 2 \) for the EVP cases and \( Bn = 0 \) for the VE cases. Different values of the power-law index \( n \) are considered to analyze the effect of shear-thinning and shear-thickening behavior. }
    \label{fig:cases}
\end{figure}

Numerical validations of the setup were performed for Newtonian and viscoelastic vortex shedding around a cylinder at $Re=100$ (summarised in Table \ref{tab:Cd}).  

In the present study, a total of ten numerical simulations at $Re=100$ was performed, systematically varying key dimensionless parameters. Three fluid classes are considered: Newtonian, purely viscoelastic (VE), and elastoviscoplastic (EVP). The details of each simulation setup are summarized in Fig.~\ref{fig:cases}, which outlines the classification of cases based on the Weissenberg number ($Wi$), solvent viscosity ratio ($\beta_s$), Bingham number ($Bn$), and power-law index ($n$). Each simulation is labeled according to its fluid type and parameter configuration. The notation for all of the cases will be followed on the next sections.

\section{Modal decomposition techniques \label{sec:num}}
\subsection{Formulation of the snapshot matrix}

In order to examine the unsteady characteristics of a field variable, such as velocity, a vector field $\bv(x,y,t_{k})$ is analyzed at different time steps $t_{k}$ (for convenience, expressed as $\bv_k$). Machine learning algorithms involve collecting and grouping a set of $K$ time-equidistant snapshots $\bv_k$ into the so-called \textit{snapshot matrix} $\bV_1^K$ of dimensions $J\times K$, where $J$ is the total number of grid points defining the spatial domain. For two-dimensional computational domains with a uniform and structured mesh, $J$ is calculated as $N_x \times N_y$, where $N_x$ and $N_y$ are the numbers of points along the streamwise and normal directions as follows:

\beqn
    \bV_1^K = [\bv_{1},\bv_{2},\ldots,\bv_{k},\bv_{k+1},\ldots,\bv_{K-1},\bv_{K}] \in \mathbb{R}^{J\times K}.
\label{ab0}
\eeqn

If the analysis is performed over the fluctuations of the velocity, the temporal mean $\overline{\bv}(x,y)$ is subtracted from the instantaneous field to study the flow fluctuations as

\begin{equation}
\tilde{\bv}(x,y,t_k)=\bv_k-\overline{\bv}(x,y), \; \text{for} \;k = 1,...,K,
\label{pod_1}
\end{equation}

where $\tilde{\bv}(x,y,t_k)$ (for convenience $\tilde{\bv}_k$) denotes the fluctuating component of the vector data at a time instant $t_k$, with its temporal mean subtracted. By stacking the $K$ snapshots into matrix form, the flow fluctuations snapshot matrix $\tilde{\bV}_1^K$ is obtained:

\begin{equation}
 \tilde{\bV}_1^K=[\tilde{\bv}_1,\tilde{\bv}_2,...,\tilde{\bv}_k,...,\tilde{\bv}_{K-1},\tilde{\bv}_K ] \in \mathbb{R}^{J\times K}.
\label{pod_2}
\end{equation}

For the POD computation, the snapshot matrix Eq. \eqref{pod_2} is composed by 400 snapshots equidistant in time at time interval $\Delta t = 0.15$, employing a grid size of $1024 \times 171$ points on the streamwise and normal directions, respectively. The spatial domain considered in this analysis is $-16D \le x \le 48D$ and $-5D \le y \le 5D$. Both the fluctuations of the streamwise and normal velocity components are considered for the analysis. As a result, the matrix has dimensions $J \times K = 350208 \times 400$, where $J = 1024 \times 171 \times 2$ represents the spatial degrees of freedom and $K = 400$ the number of time snapshots.

\begin{table} % # 1
   \centering
   \caption{Summary of the database sizes used in the analysis of the different simulations. Here, \( n_x \) and \( n_y \) denote the number of spatial points extracted in the streamwise and normal directions, respectively, and \( K \) is the number of time snapshots collected at uniform time intervals \( \Delta t \). The notation of the different cases is the same as in Fig. \ref{fig:cases}.}   \label{tab:DMD}
   \begin{tabular}{lcccc}
   $ Case $         & $n_x$  & $n_y$  & $K$  & $\Delta t$ \\[2pt]\hline
   $N$              & 1024   &  171   & 400  & 0.15 \\ \hline
   $VE_1-VE_2$      & 1024   &  171   & 400  & 0.15 \\
   \hline
   $EVP_1 - EVP_3$  & 1024   &  171   & 400  & 0.15 \\
   $EVP_4$          & 1536   &  171   & 400  & 0.9  \\
   $EVP_5$          & 1536   &  171   & 400  & 0.9  \\
   $EVP_6$          & 1024   &  171   & 400  & 0.45 \\
   $EVP_7$          & 1024   &  171   & 400  & 0.45 \\ \hline
   \end{tabular}
 \end{table}

For the application of Higher Order Dynamic Mode Decomposition (HODMD), the databases are organised in tensor form (as explained in Sec. \ref{sec:hodmd_meth}) of size $2 \times n_x \times n_y \times K$ being $n_x$ and $n_y$ the number of grid points in the streamwise and normal spatial directions, respectively, and $K$ the number of snapshots taken. The domain in the normal direction is reduced to $-5D \le y \le 5D$ and in the streamwise direction, the full domain of the simulation is taken. The snapshots are extracted at regular intervals of $\Delta t$. The values of $n_x$, $n_y$, $K$ and $dt$ are specified in Tab. \ref{tab:DMD}. The different values in some of the cases is related to the complexity of the problem. As further discussed in Sec. \ref{sec:DMD}, some of the most complex cases exhibit low-frequency dynamics, requiring longer simulation time spans to ensure accurate identification.

\subsection{Proper orthogonal decomposition (POD)}

POD is a valuable technique for extracting coherent patterns in fluid mechanics, proposed by Lumley in 1967 \cite{Lumley1967}. The objective of the POD algorithm is to decompose a dataset of a given field variable into a minimal set of modes (basis functions) that capture the maximum possible energy. This ensures that the POD modes are optimal in minimizing the mean-square error between the original signal and its reconstructed representation. To find the most suitable representation for the given dataset $\tilde{\bV}_1^K$, the eigenvectors $\bPhi$ and their corresponding eigenvalues $\lambda_{j}$ are computed using the equation shown below:

\begin{equation}
 \bC \boldsymbol{\Phi}_{j} 
 = \lambda_{j} \bPhi_{j},  \; \boldsymbol{\Phi}_{j} \in \mathbb{R}^{J} , \; \lambda_{1}\geq...\geq \lambda_{n} \geq  0,
\label{pod_3}
\end{equation}

The eigenvectors of the matrix are used to generate the POD modes, which organize the modes in a hierarchical manner based on the amount of captured energy. The eigenvalues corresponding to each eigenvector $\bPhi_j$ (\Eqref{pod_3}) indicate how well the original data is represented in a least-squares optimal sense. Finally, in \Eqref{pod_3}, $\bC$ states the covariance matrix of the input data, defined as (\Eqref{pod_4}).

\begin{equation}
 \bC=\sum_{i=1}^K \hat{\bv}(t_{i}) \hat{\bv}^{T}(t_{i})= \tilde{\bV}\tilde{\bV}^{T} \in \mathbb{R}^{J \times J}.
\label{pod_4}
\end{equation}

An alternative approach to computing the POD algorithm is through the singular value decomposition (SVD) method, introduced by Sirovich in 1987 \cite{Sirovich1987}. By directly applying the SVD method to the snapshot matrix $\tilde{\bV}_1^K$, the left singular vectors $\bPhi$ and the right singular vectors $\boldsymbol{\Psi}$ are obtained as shown below:

\begin{equation}
 \tilde{\bV}_1^K=\bPhi\bSigma\boldsymbol{\Psi}^{T}.
\label{pod_5}
\end{equation}

The matrix $\bSigma\in \mathbb{R}^{J \times K}$ contains singular values ($\sigma_{1}, \sigma_{2},...,\sigma_{N}$) along its diagonal, which are related to the eigenvalues as $\sigma_{j}^{2}=\lambda_{j}$. The left singular vectors $\boldsymbol{\Phi} \in \mathbb{R}^{J \times J}$ are the eigenvectors of the matrix $\tilde{\bV}\tilde{\bV}^{T}$, while the right singular vectors $\boldsymbol{\Psi}\in\mathbb{R}^{K\times K}$ are the eigenvectors of the matrix $\tilde{\bV}^{T}\tilde{\bV}$. 

\subsection{Higher order dynamic mode decomposition (HODMD) \label{sec:hodmd_meth}}

Higher Order Dynamic Mode Decomposition (HODMD) \cite{LeClaincheVega17} is an advanced technique that builds upon the well-known Dynamic Mode Decomposition (DMD)  \cite{Schmid10} approach. HODMD is a widely used algorithm within the fluid mechanics community, with its robustness and accuracy validated across a broad range of applications. HODMD analysis primarily focuses on complex fluid flows, encompassing noisy experiments turbulent flows \cite{CHEN2024105719,LeClaincheVegaSoria17, LeClaincheetalJAircraft18,HODMDMalaga}, turbulent flows \cite{Corrochano22JAE, Lazpita} and reacting flows \cite{CORROCHANO2023108219,corrochano2024hierarchical}.

To perform HODMD,the spatio-temporal data is decomposed into $M$ modes $\bu_m$, where each mode is weighted by an amplitude $a_m$ as 

\begin{equation}
\bv(x,y,t_{k})\simeq  
\sum_{m=1}^M a_{m}\bu_m(x,y)e^{(\delta_m+i \omega_m)t_k}.
\label{ab00}
\end{equation}

The modes obtained through HODMD oscillate in time with frequency $\omega_m$ and may exhibit growth, decay, or remain neutral in time, as determined by the growth rate $\delta_m$. 

The robustness of the HODMD algorithm and its suitability for the analysis of complex flows comes from the sliding window process. Following the notation from Eq. \ref{ab0}, a snapshot is related with its previous d snapshots using higher-order Koopman assumption defined as 

\beqn
\bV_{d+1}^K \simeq \bR_1 \bV_1^{K-d}+ \bR_2 \bV_2^{K-d+1} + \ldots + \bR_d \bV_d^{K-1}.
\label{hoKoopman}
\eeqn
 
The HODMD algorithm can be simplified into two major steps explained below: the dimensionality reduction and the sliding window process. A thorough understanding of the algorithm is shown in Ref. \cite{LeClaincheVega17,LeClaincheVegaSoria17} and its implementation in Matlab can be found in \cite{VegaLeClaincheBook20}. HODMD has been recently implemented in Python in \cite{Hetherington2024ModelFLOWs, ModelFLOWs2023}.

\begin{itemize}
    \item \textbf{Step 1: Dimensionality reduction via SVD:}\\ SVD is applied to the snapshot matrix (\ref{ab0}) to reduce its dimensionality, eliminate spatial redundancies, and filter out noise, as follows:

    \beqn
        \bV_1^{K}\simeq\bW\,\bSigma\,\bT^T. \label{ab20}
    \eeqn
    
    As in the previous section, $\bSigma$ contains the singular values, $\bW$ the spatial POD modes, and the associated temporal coefficients are collected in $\bT$. A tolerance value $\varepsilon_1$ is selected to determine the number of linearly independent vectors to retain. This user-defined parameter $\varepsilon_1$ sets the number $N$ of SVD modes to be preserved, according to the following criterion:

    \beqn
        \sigma_{N+1}/\sigma_{1}\leq \varepsilon_1. \label{ab21}
    \eeqn

    The original snapshot data set has a spatial dimension of $J$, but after applying SVD, the result is a set of linearly independent vectors of dimension $N$, where $N<J$ represents the spatial complexity. The {\it reduced  snapshot matrix}, of dimension $N\times K$, is then defined from Eq. (\ref{ab20}) as

    \begin{equation}
        \widehat{\bV}_1^K=\bSigma\,\bT^T. \label{ab22}
    \end{equation}
    
    \item \textbf{Step 2: DMD-d algorithm:}
    The high-order Koopman assumption from Eq. \eqref{hoKoopman}, is applied to the reduced snapshot matrix, as

    \beqn
    \widehat{\bV}_{d+1}^K\simeq \widehat{\bR}_1 \widehat{\bV}_1^{K-d}+ \widehat{\bR}_2 \widehat{\bV}_2^{K-d+1} + \ldots + \widehat{\bR}_d \widehat{\bV}_d^{K-1},
    \label{ab26}
    \eeqn

    in which $\widehat{\bR}_k=\bW^T\bR_k\bW$ for $k=1,\ldots,d$. Equation \ref{ab26} divides the snapshot matrix into $d$ blocks. Each block contains $K-d$ snapshots but time-delayed, as seen in Fig. \ref{fig:DMD_1}. It is possible to combine the Koopman matrices $\widehat{\bR}_k$ into the so-called modified Koopman matrix $\tilde{\bR}$ and express the previous equation as 

    \beqn
    \tilde{\bV}_2^{K-d+1} = \tilde{\bR} \, \tilde{\bV}_1^{K-d}.
    \label{ab30}
    \eeqn

    \begin{figure} %DMD_1
    \centering
    \includegraphics[width=0.3\linewidth]{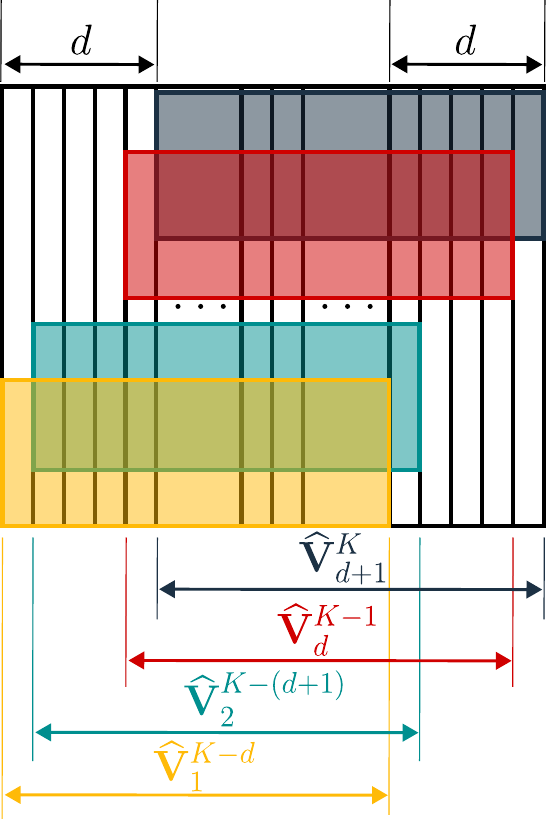}
    \caption{Sketch representing the DMD-d sliding window process over the reduced snapshot matrix, defined in Eq. (\ref{ab26}).
    \label{fig:DMD_1}}
    \end{figure}

    An eigenvalue problem is solved over Eq. \eqref{ab30} to obtain the DMD modes, frequencies and growth rates in Eq. \eqref{ab00}. The amplitudes are calculated through a least square fitting of the original expansion and are sorted in decreasing order. The DMD expansion is then reduced with a second tunable tolerance $\varepsilon_{2}$ as
    
    \beqn
    a_{M+1}/a_1\leq \varepsilon_2,
    \label{b66}
    \eeqn

    where $M$ is the number of DMD modes selected for the DMD mode expansion in Eq. \eqref{ab00}, and $a_1$ denotes the amplitude of the dominant mode. The accuracy of the HODMD reconstruction (\ref{ab00}) is evaluated using the relative root mean square (RRMS) error, which is computed as:

    \beqn
    RRMSE=\sqrt{\frac{\sum_{k=1}^K||\textbf{v}_k-\textbf{v}^{DMD}_k||^2}{\sum_{k=1}^K||\textbf{v}_k||^2}},
    \label{RRMS}
    \eeqn

    where $||\cdot||$ is the usual Euclidean norm and $\textbf{v}^{DMD}_k$ is the reconstruction of the snapshot $k$ using the DMD mode expansion in Eq. \ref{ab00}.

\end{itemize}

In the present study, the multi-dimensional HODMD is used to analyse the data, as this version is a more efficient version of HODMD, suitable for analyzing complex multi-dimensional data. This algorithm, described in \cite{LeClaincheVegaSoria17}, uses high order singular value decomposition (HOSVD) \cite{Tucker66} instead of the classic SVD used in Step 1 of the method. Specifically, instead of organizing the data in the snapshot matrix (\ref{ab0}), it is organized in tensor form $V(c_i,x_j, y_l, t_k)=V_{ijlk}$. In tensor form, $i=1,I$ represents the number of variables analysed, which typically are the 2 or 3 components of the velocity, $j=1,J$ and $l=1,L$ represent the number of grid points related to the streamwise and normal spatial components $\bx$, $\by$, and finally $k=1,K$ is the number of snapshots. The method applies standard SVD to the four matrices, each of which has columns formed by one of the three data variables, similar to the fibers of a tensor, resulting in the following decomposition.

 \begin{equation}
 V_{ijlk}\simeq\sum_{p_1=1}^{P_1}\sum_{p_2=1}^{P_2}\sum_{p_3=1}^{P_3}\sum_{n=1}^{N} \bS_{p_1p_2p_3n}
   \bW^{(c)}_{ip_1}\bW^{(x)}_{jp_2} \bW^{(y)}_{lp_3} \bT_{kn},    
   \label{c10}
 \end{equation}

In the given equation, $\bS_{p_1p_2p_3n}$ refers to a fourth-order tensor known as the {\it core tensor}, and the columns of matrices $\bW^{(c)}$, $\bW^{(x)}$, $\bW^{(y)}$, and $\bT$ are referred to as the {\it modes} of the decomposition, which includes the components analysed, the two spatial dimensions and the temporal one. The reduction mentioned in Eq. \eqref{ab21} is implemented on the spatial and temporal matrices to achieve better cleaning. Finally, "Step 2" is performed only on the temporal modes $\bT$. The implementation in Python of the multi-dimensional HODMD can be found in \cite{Hetherington2024ModelFLOWs, ModelFLOWs2023}.

\begin{figure} %[h] 2
\begin{subfigure}{1.0\textwidth}
\centering
\includegraphics[width=0.49\linewidth,height=2.5cm]{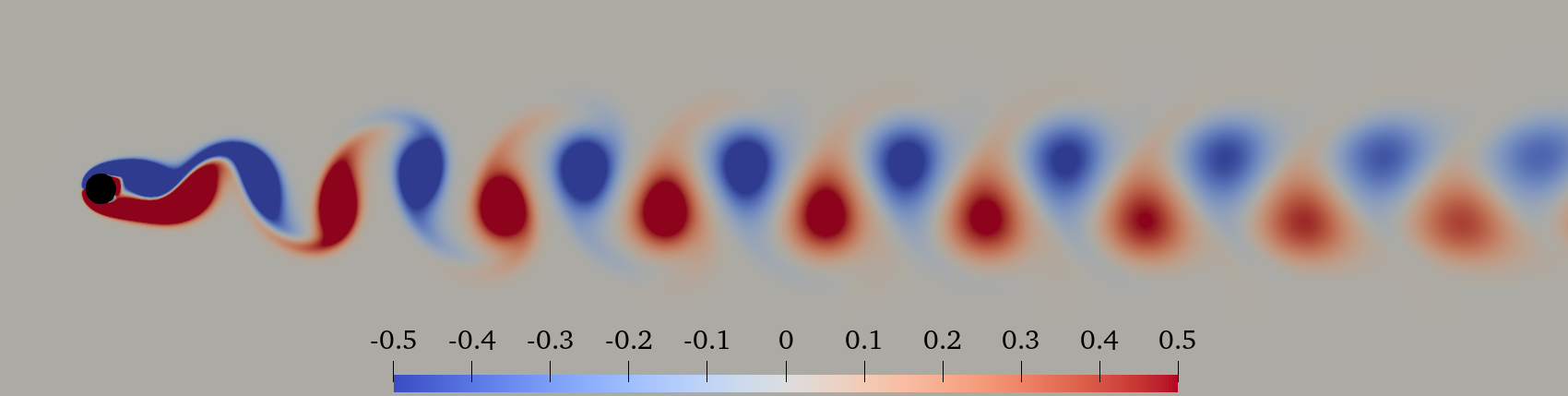}
\caption{Newtonian Fluid}
\label{fig:subim2_1}
\end{subfigure}
\begin{subfigure}{0.49\textwidth}
\centering
\includegraphics[width=1.0\linewidth, height=2.5cm]{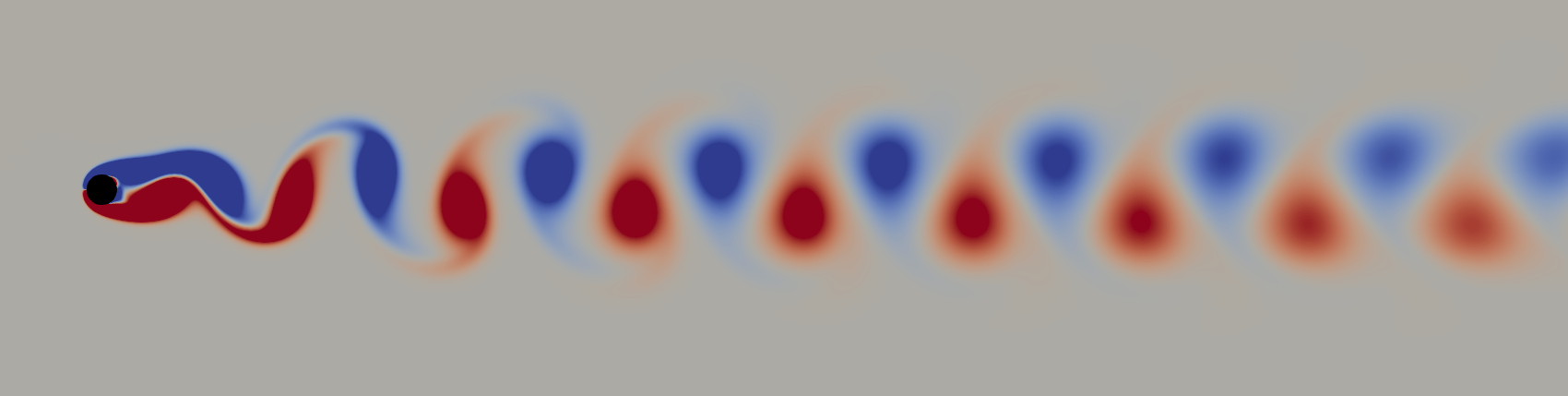}
\caption{$VE_1$ : $\beta_{s}=0.9,~n=1.0,~Bn=0.0$ }
\label{fig:subim2_2}
\end{subfigure}
\begin{subfigure}{0.49\textwidth}
\centering
\includegraphics[width=1.0\linewidth, height=2.5cm]{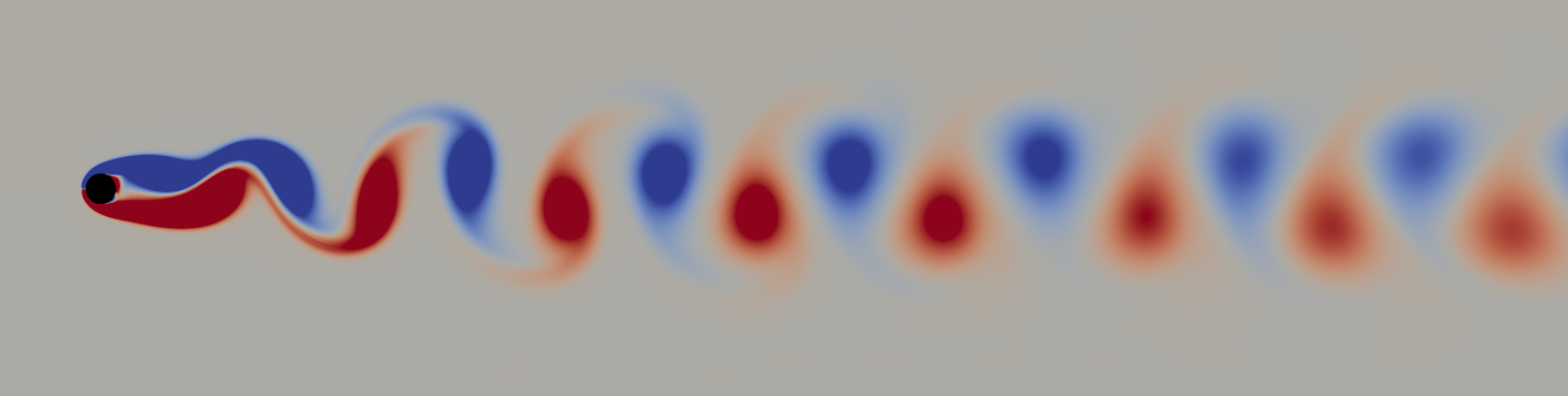}
\caption{$VE_2$ : $\beta_{s}=0.5,~n=1.0,~Bn=0.0$}
\label{fig:subim2_3}
\end{subfigure}
\caption{The influence of elasticity on vorticity $\boldsymbol{\zeta}_{z}$ contour flow past a cylinder, $Wi=1.0$, $\beta_{s}=0.9,~0.5$, $Bn=0.0$. Blue and red colors denote regions of cyclonic and anticyclonic vorticity, respectively, highlighting areas of counter-rotating flow structures, while gray represent the neutral regions. The color scale is the same for all contours. The case notation corresponds to that shown in Fig.~\ref{fig:cases}.}
\label{fig:image2}
\end{figure}

\section{Coherent structures in the Newtonian and elastoviscoplastic fluids of cylinder flows} \label{section:Review_numerical_simulation_results}

This study focuses on the analysis and characterization of flow structures using data obtained from direct numerical simulations of an elastoviscoplastic (EVP) fluid flow past a circular cylinder. Three distinct fluid types are investigated: a Newtonian fluid, a purely viscoelastic fluid, and an elastoviscoplastic fluid, following the approach of Parvar \etal \cite{PARVAR20231}. By examining these different fluid types, it is possible to isolate and examine the individual influences of plasticity and elasticity on the flow behavior.

In this study, the dynamics of unsteady two-dimensional flow behind a circular cylinder are studied at Reynolds number $Re=100$. This particular Reynolds number is commonly used in the literature for comparison and validation purposes \cite{Posdziech2007,Williamson1996,PARVAR20231}. At $Re=100$, the flow is unstable, with an unsteady wake and periodic fluctuations. To examine the flow structures, the temporal evolution of vorticity contours ($\boldsymbol{\zeta}_{z}$) is analysed for the Newtonian case in Fig. \ref{fig:subim2_1}. The well-known von Kármán vortex street phenomenon emerges, characterized by alternating rows of cyclonic (negative) and anticyclonic (positive) vortices shedding behind the bluff body in parallel with the cylinder axis. These vortices create a recirculation region attached to the rear of the cylinder. As they progress downstream, the vortices transform into tear-shaped configurations. The blue and red colors represent regions of cyclonic and anticyclonic vorticity, respectively, indicating counter-rotating areas of vorticity.

In addition, the influence of elasticity on the von Kármán vortex street is investigated in Fig. \ref{fig:subim2_2} and Fig. \ref{fig:subim2_3}. The findings reveal that at $Wi=1$, the impact of elasticity on the vorticity pattern is subtle so the pattern is resembling that of a Newtonian fluid flow, forming a "Haladie" shape close to the cylinder. However, even at this elasticity level, the two vortices near the cylinder extend in the streamwise direction, resulting in an increased bubble length. These observations align with previous numerical and experimental studies such as Oliveira \cite{Oliveira2001,Oliveira2005}, Coelho \cite{Coelho2003I,Coelho2003II}, Richter \cite{Richter2010} and Peng \cite{Peng2021}.

\begin{figure} %[h] 3
\begin{subfigure}{0.49\textwidth}
\centering
\includegraphics[width=1.0\linewidth, height=2.5cm]{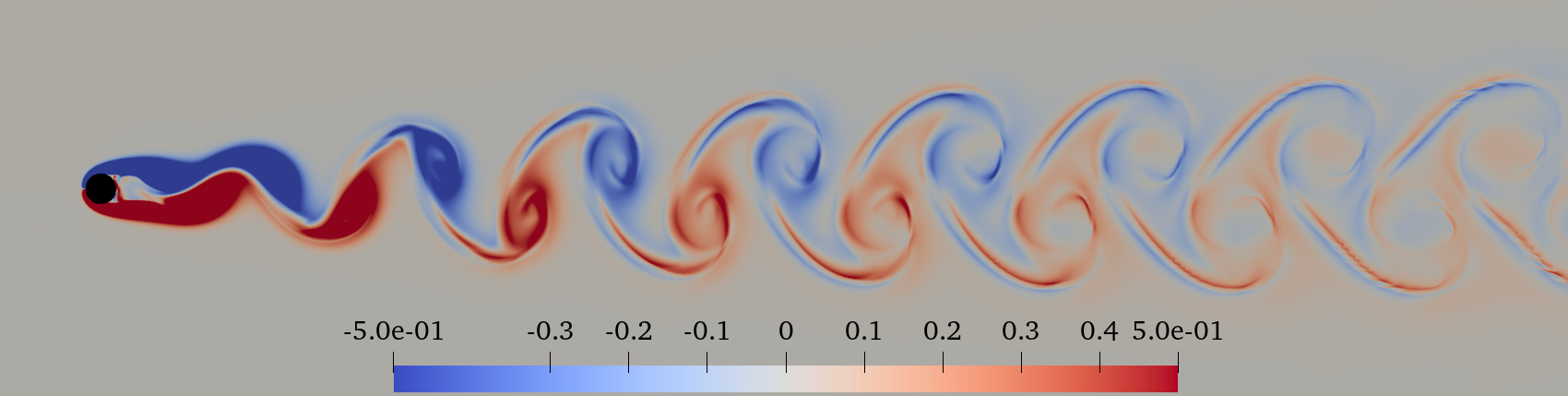}
\caption{$EVP_1$: $\beta_{s}=0.9,~n=0.2,~Bn=2.0$ }
\label{fig:subim3_1}
\end{subfigure}
\begin{subfigure}{0.49\textwidth}
\centering
\includegraphics[width=1.0\linewidth, height=2.5cm]{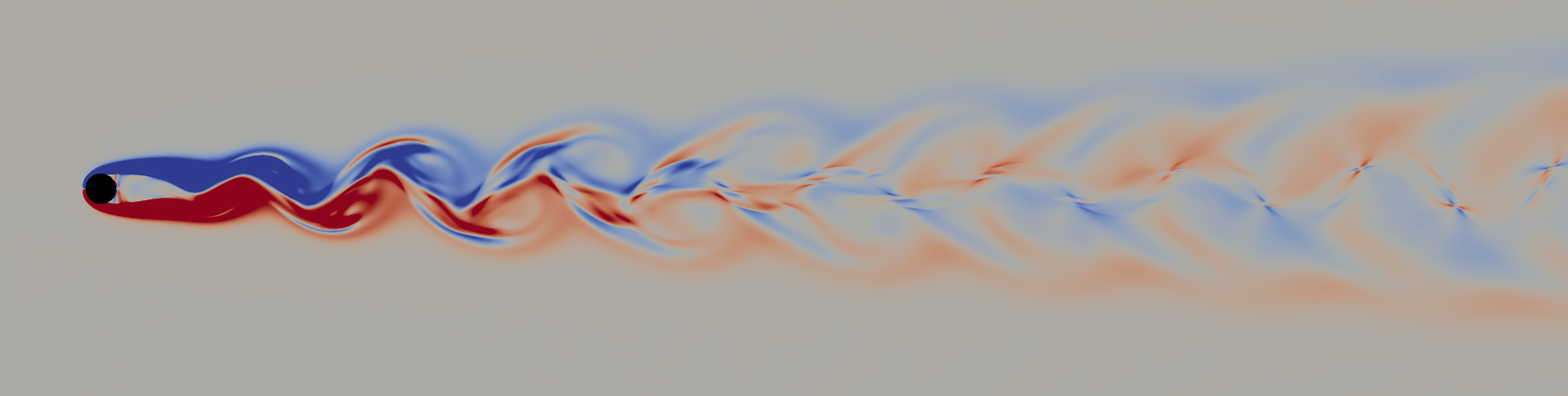}
\caption{$EVP_2$: $\beta_{s}=0.5,~n=0.2,~Bn=2.0$ }
\label{fig:subim3_2}
\end{subfigure}
\begin{subfigure}{0.49\textwidth}
\centering
\includegraphics[width=1.0\linewidth, height=2.5cm]{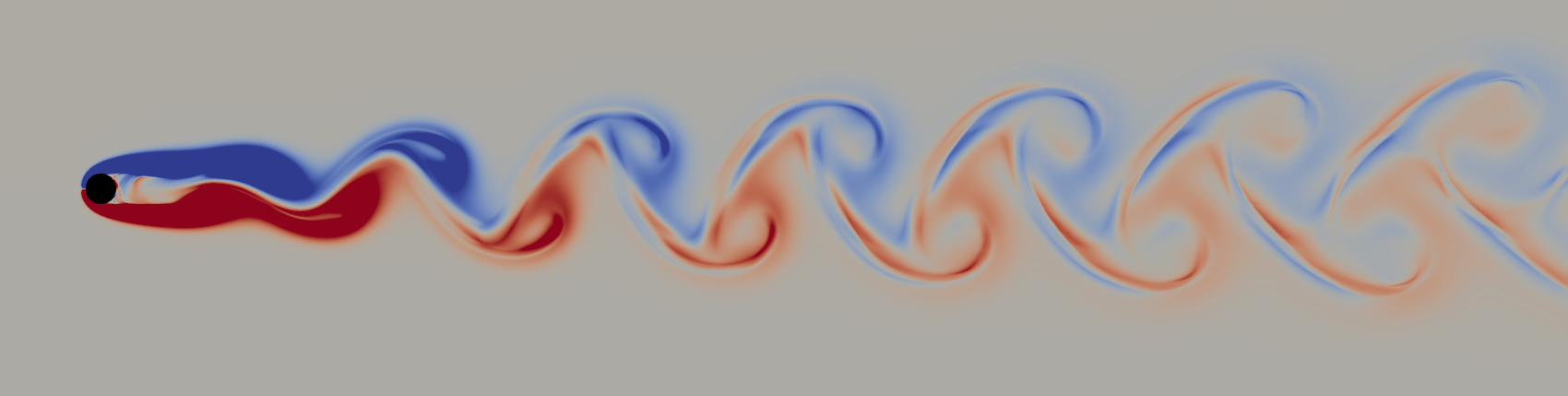}
\caption{$EVP_3$: $\beta_{s}=0.9,~n=1.0,~Bn=2.0$ }
\label{fig:subim3_3}
\end{subfigure}
\begin{subfigure}{0.49\textwidth}
\centering
\includegraphics[width=1.0\linewidth, height=2.5cm]{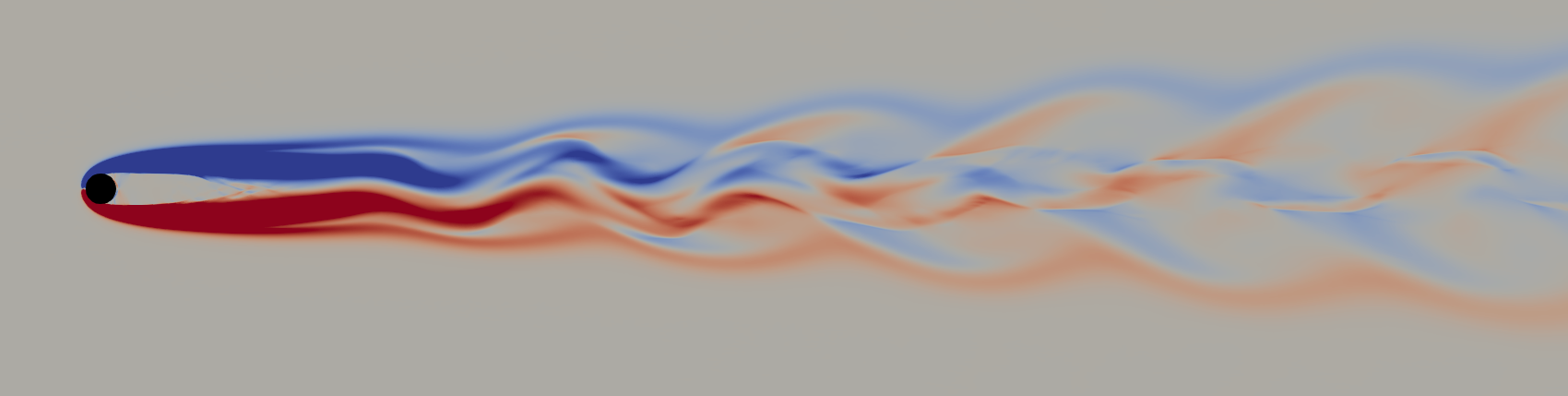}
\caption{$EVP_4$: $\beta_{s}=0.5,~n=1.0,~Bn=2.0$ }
\label{fig:subim3_4}
\end{subfigure}
\begin{subfigure}{0.49\textwidth}
\centering
\includegraphics[width=1.0\linewidth, height=2.5cm]{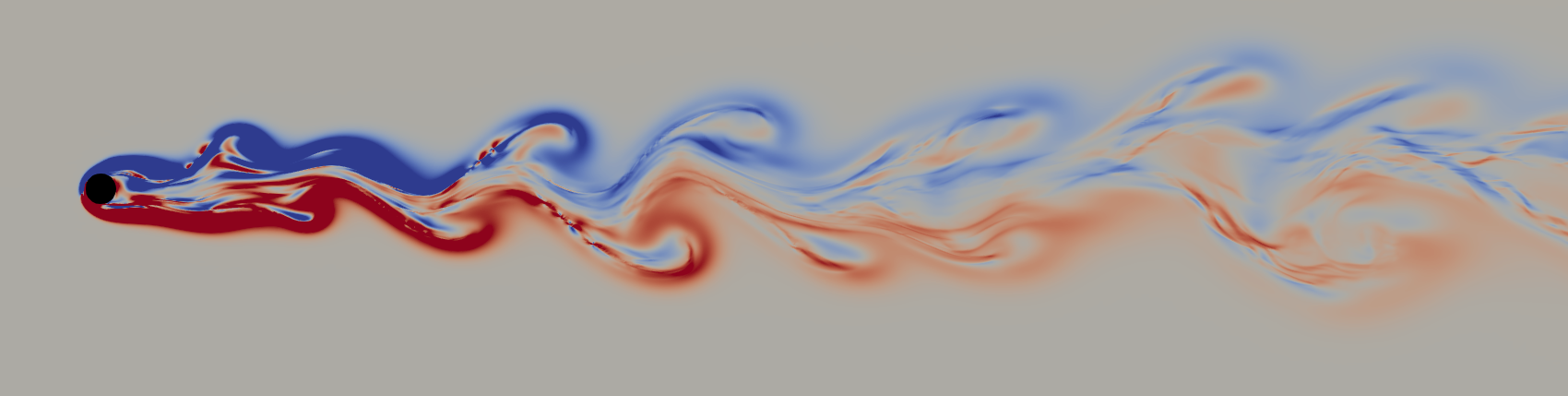}
\caption{$EVP_6$: $\beta_{s}=0.9,~n=2.0,~Bn=2.0$ }
\label{fig:subim3_5}
\end{subfigure}
\begin{subfigure}{0.49\textwidth}
\centering
\includegraphics[width=1.0\linewidth, height=2.5cm]{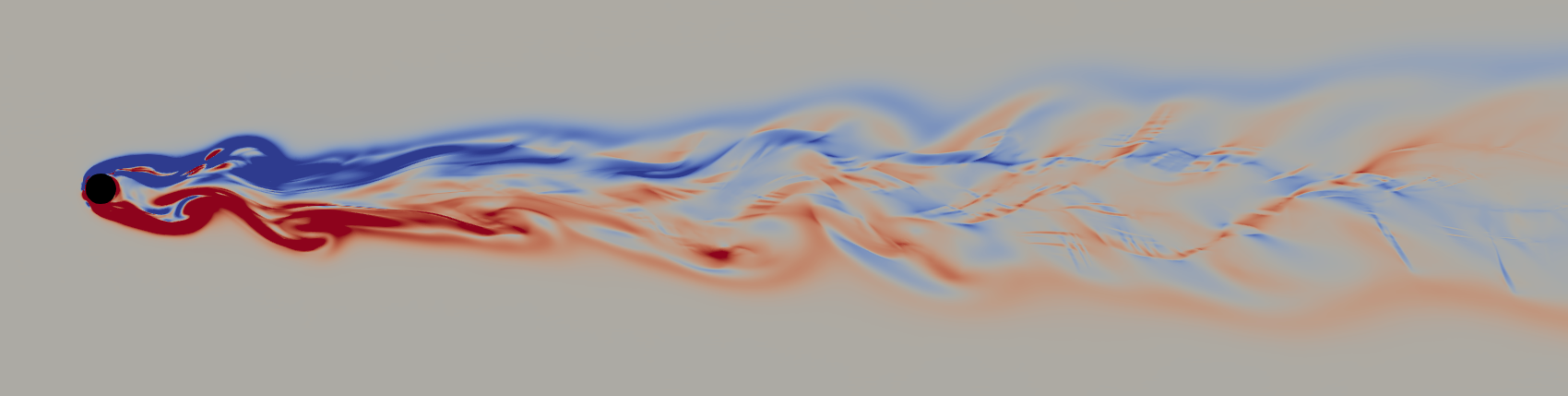}
\caption{$EVP_7$: $\beta_{s}=0.5,~n=2.0,~Bn=2.0$ }
\label{fig:subim3_6}
\end{subfigure}
\caption{The influence of elastoviscoplasticity on vorticity $\boldsymbol{\zeta}_{z}$ contour flow past a cylinder, $Wi=1.0$, $\beta_{s}=1.0~0.5$, $Bn=2.0$. The color scale is the same for all contours. The case notation corresponds to that shown in Fig.~\ref{fig:cases}.}
\label{fig:image3}
\end{figure}

In the presence of yield stress, however, the flow pattern exhibits distinct characteristics, even at $Wi=1$, particularly further downstream from the cylinder. As shown in Fig. \ref{fig:subim3_1} and Fig. \ref{fig:subim3_3}, for both shear-independent ($n=1$) and shear-thinning ($n=0.2$) material with $Wi=1$, $Bn=2$ and $\beta_{s}=0.9$, Haladie-like vortices appear near the cylinder, similar to the Newtonian and viscoelastic cases. However, these vortex shapes undergo transformation downstream. In contrast, when considering a fluid with a lower solvent viscosity contribution ($\beta_s=0.5$), more representative for real life EVP fluids, in both shear-thinning ($n=0.2$) and shear-independent ($n=1$) cases, the typical cyclonic and anticyclonic vortex structures are no longer present. As a result, the formation of a von Kármán vortex street is inhibited. Instead, the vortices tend to extend in the streamwise direction, leading to the formation of a cellular structure downstream of the cylinder. Due to numerical constraints, we were not able to lower the value of $\beta_s$ further, but based on the above we can assume that this results in further deviation from the periodic von Kármán vortex mode present in Newtonian flow.

In contrast, when increasing the power-law index of the fluid to $n=2$ (which corresponds to shear-thickening, although less common in EVP fluids), the instabilities in the flow grow significantly. This results in the breakdown and absence of a vortex street formation and the transition to a chaotic flow pattern occurs. Lower $\beta_s=0.5$ combined with the yield stress lead to a more complex flow pattern, as indicated in Fig. \ref{fig:subim3_5} and Fig. \ref{fig:subim3_6}. The features of these flow patterns indicate the onset of new instabilities that affect not only the region near the cylinder but also modify the downstream flow structure, ultimately suppressing the formation of the characteristic unsteady von Kármán vortex street.

\begin{figure} % [h!] 
\begin{subfigure}{1.0\textwidth}
\centering
\includegraphics[width=0.5\linewidth, height=2.5cm]{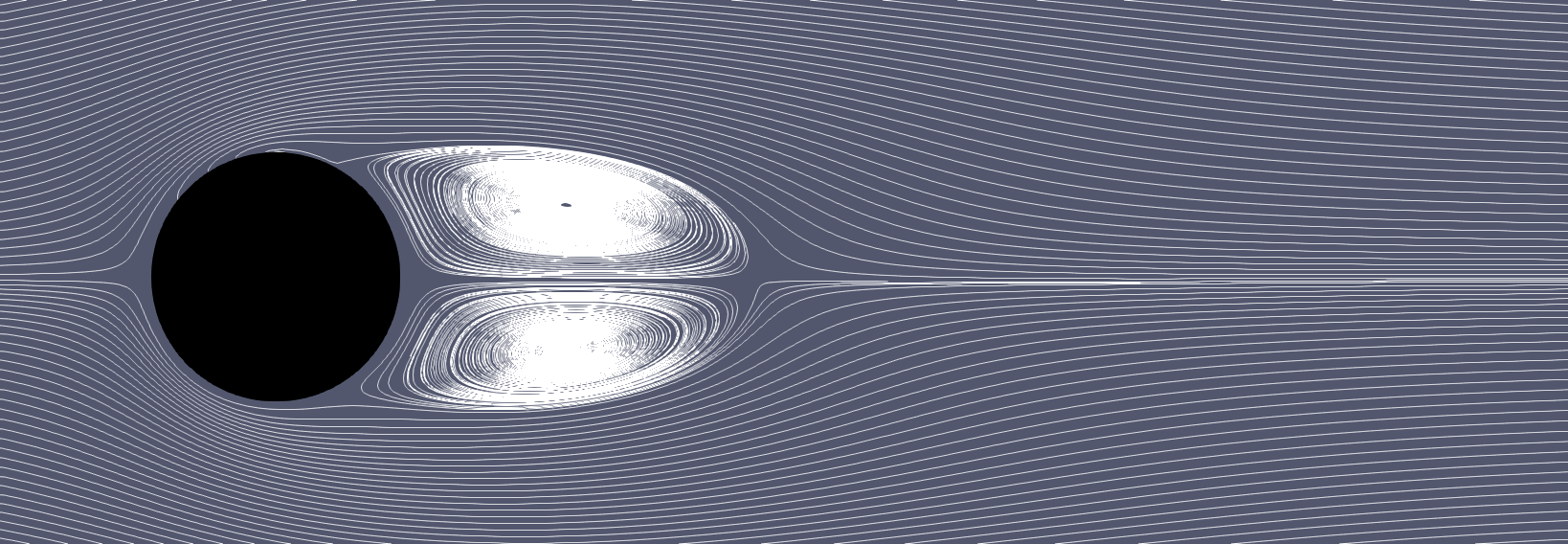}
\caption{\label{StreamLine_Re100_HB_Wi1.0_n0.2_Bs0.9_Bn2.0} Newtonian Fluid }
\label{fig:subim200_1}
\end{subfigure}
\begin{subfigure}{0.5\textwidth}
\centering
\includegraphics[width=1.0\linewidth, height=2.5cm]{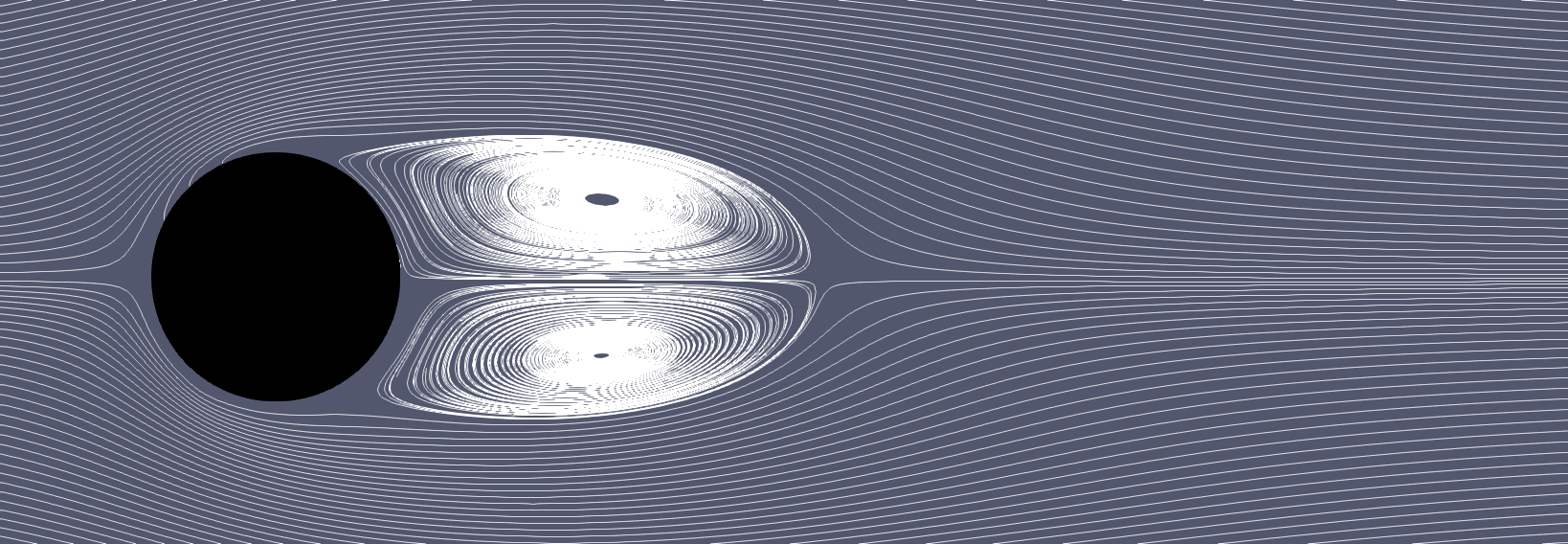}
\caption{\label{StreamLine_Re100_HB_Wi1.0_n1.0_Bs0.9_Bn2.0} $VE_1$ : $\beta_{s}=0.9,~n=1.0,~Bn=0.0$}
\label{fig:subim200_2}
\end{subfigure}
\begin{subfigure}{0.5\textwidth}
\centering
\includegraphics[width=1.0\linewidth, height=2.5cm]{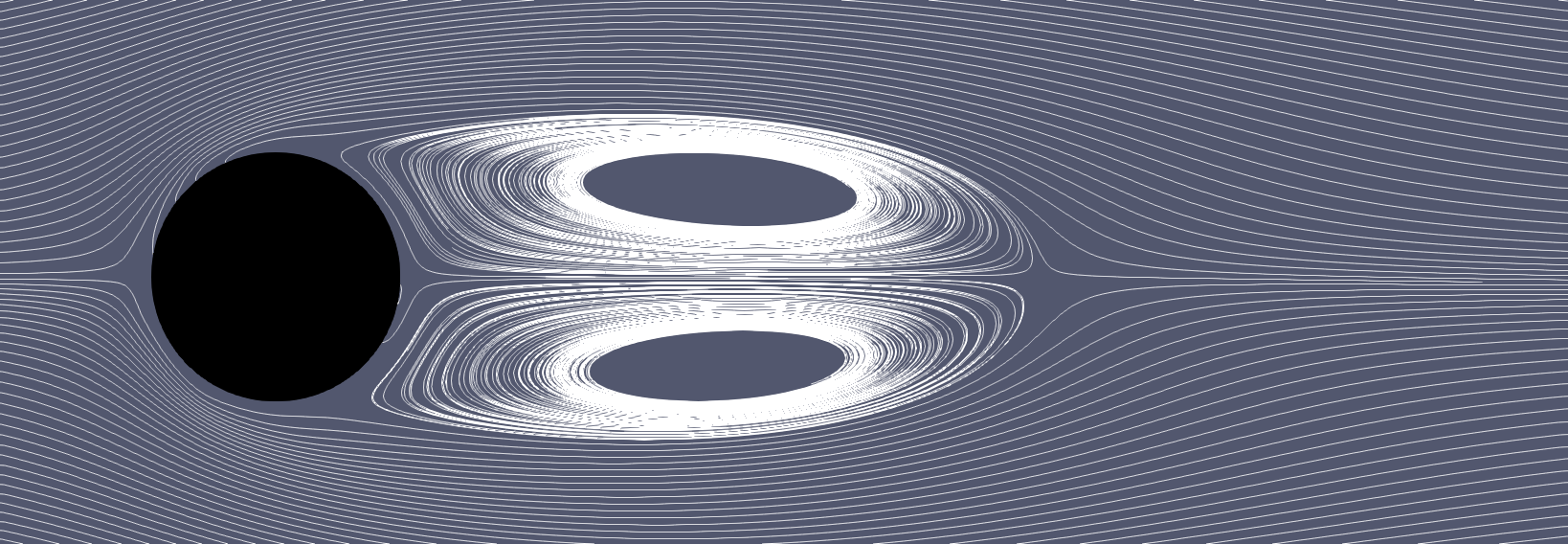}
\caption{\label{StreamLine_Re100_HB_Wi1.0_n0.2_Bs0.9_Bn2.0} $VE_2$ : $\beta_{s}=0.5,~n=1.0,~Bn=0.0$}
\label{fig:subim200_3}
\end{subfigure}
\caption{Streamlines of the mean flow for the Newtonian and viscoelastic fluids $VE_1$ and $VE_2$. The case notation corresponds to that shown in Fig.~\ref{fig:cases}. }
\label{fig:image200}
\end{figure}

Figure \ref{fig:image200} illustrates the time-averaged streamlines for both the Newtonian and viscoelastic fluids. When compared to the Newtonian case depicted in \Figref{fig:subim200_1}, the length of the recirculation bubble increases as elasticity is introduced, measured with $Wi$ (\Figref{fig:subim200_2} and \Figref{fig:subim200_3}). Furthermore, both recirculation centers shift downstream in the streamwise direction. These findings align with previous numerical and experimental studies \cite{Oliveira2001,Peng2021}. The change in the recirculation bubble is more noticeable than the change in the vorticity contours shown in \Figref{fig:image2}. The recirculation bubble is larger in the more concentrated case ($VE_2$) than in the diluted case ($VE_1$), as a lower value of $\beta_s$ increases the influence of the polymer stress tensor $\btau^p$ in Eq.~\eqref{eq:Kramers}. Conversely, a more diluted polymer solution, corresponding to a $\beta_s$ value closer to one, reduces the magnitude of $\btau^p$, making the flow behavior more similar to the Newtonian case.

\begin{figure} %[h] 300
\begin{subfigure}{0.5\textwidth}
\centering
\includegraphics[width=1.0\linewidth, height=3.0cm]{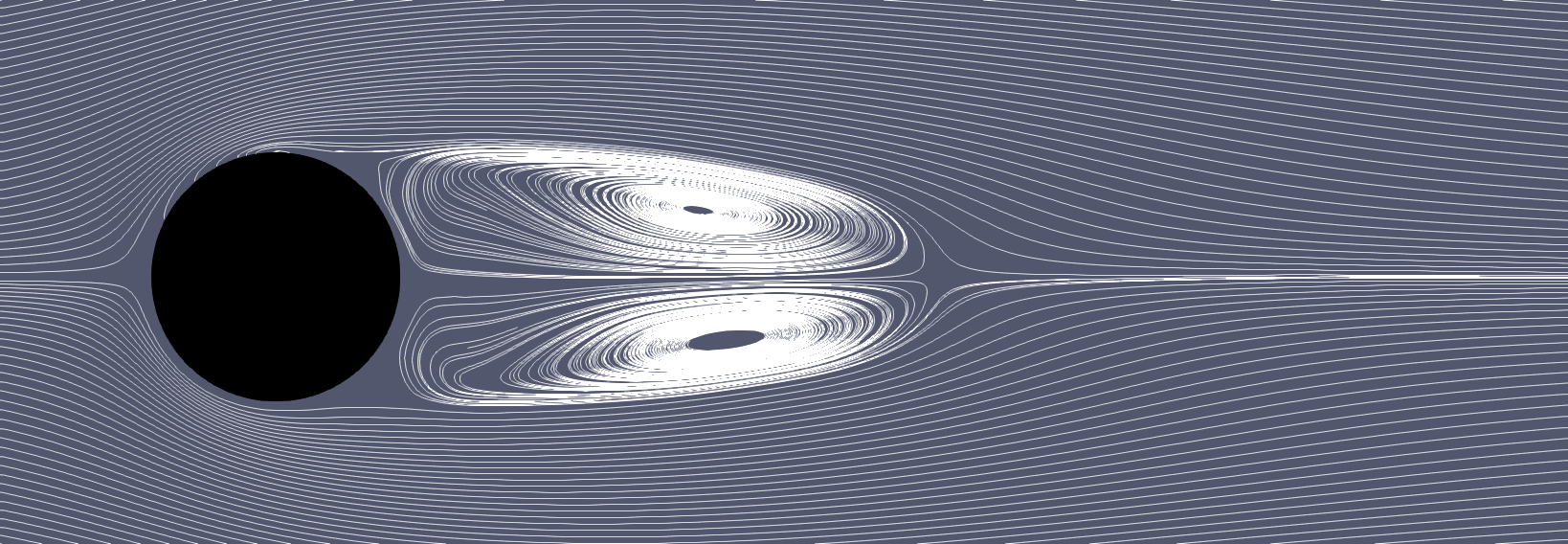}
\caption{\label{StreamLine_Re100_HB_Wi1.0_n0.2_Bs0.9_Bn2.0} $EVP_1$: $\beta_{s}=0.9,~n=0.2,~Bn=2.0$ }
\label{fig:subim300_1}
\end{subfigure}
\begin{subfigure}{0.5\textwidth}
\centering
\includegraphics[width=1.0\linewidth, height=3.0cm]{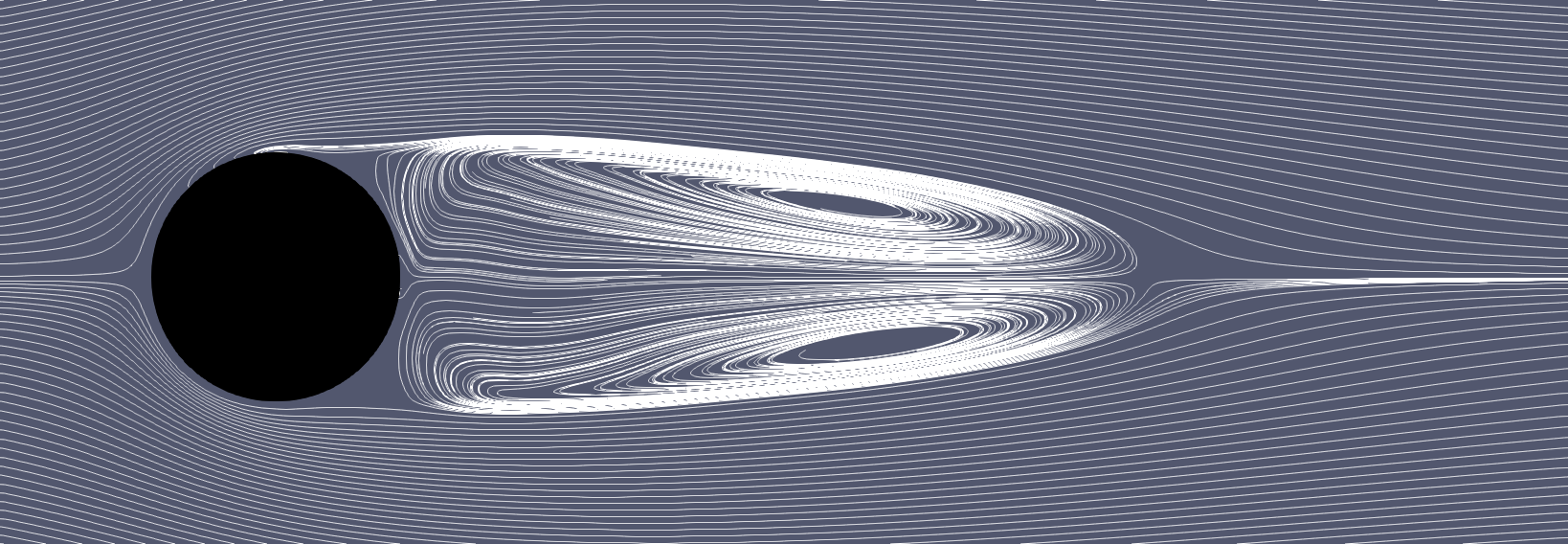}
\caption{\label{StreamLine_Re100_HB_Wi1.0_n0.2_Bs0.9_Bn2.0} $EVP_2$: $\beta_{s}=0.5,~n=0.2,~Bn=2.0$ }
\label{fig:subim300_2}
\end{subfigure}
\begin{subfigure}{0.5\textwidth}
\centering
\includegraphics[width=1.0\linewidth, height=3.0cm]{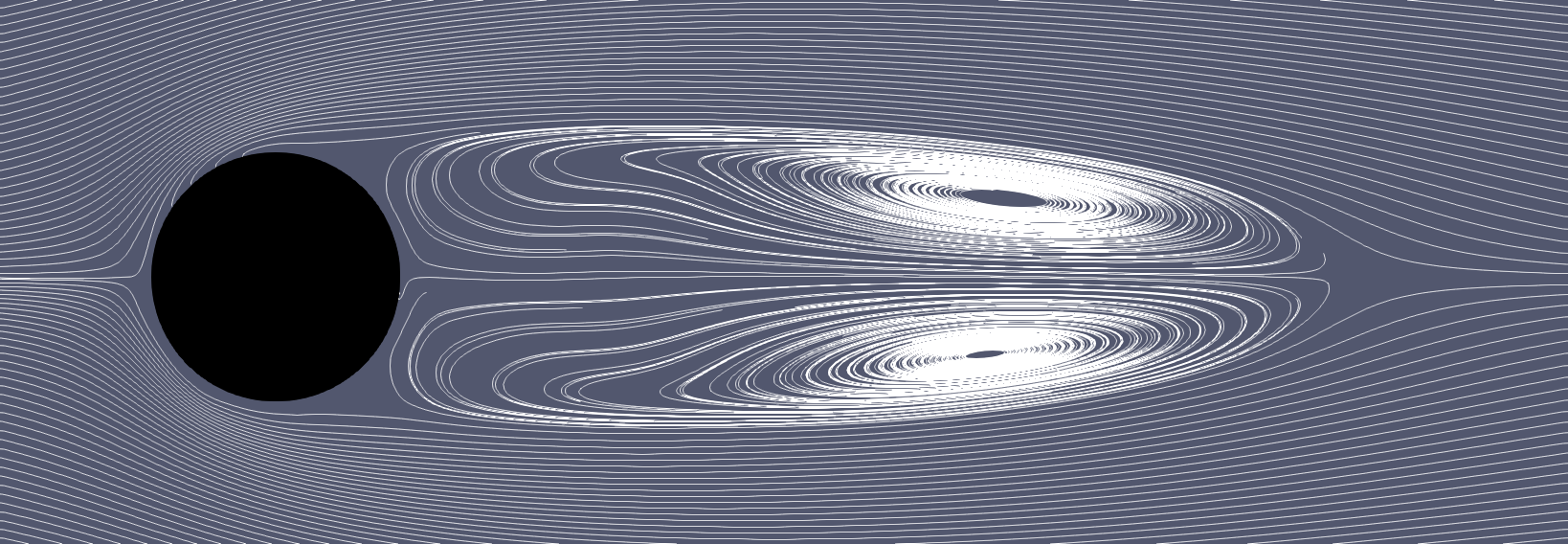}
\caption{\label{StreamLine_Re100_HB_Wi1.0_n1.0_Bs0.9_Bn2.0} $EVP_3$: $\beta_{s}=0.9,~n=1.0,~Bn=2.0$ }
\label{fig:subim300_3}
\end{subfigure}
\begin{subfigure}{0.5\textwidth}
\centering
\includegraphics[width=1.0\linewidth, height=3.0cm]{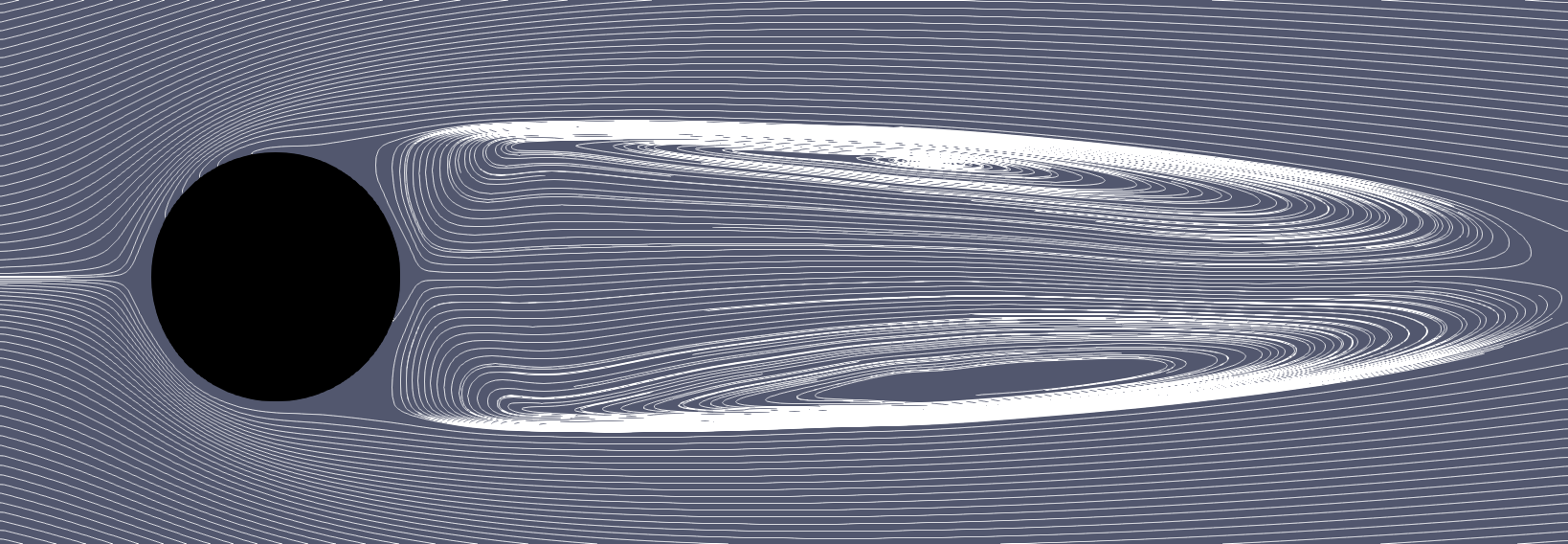}
\caption{\label{StreamLine_Re100_HB_Wi1.0_n1.0_Bs0.5_Bn2.0} $EVP_4$: $\beta_{s}=0.5,~n=1.0,~Bn=2.0$ }
\label{fig:subim300_4}
\end{subfigure}
\begin{subfigure}{0.5\textwidth}
\centering
\includegraphics[width=1.0\linewidth, height=3.0cm]{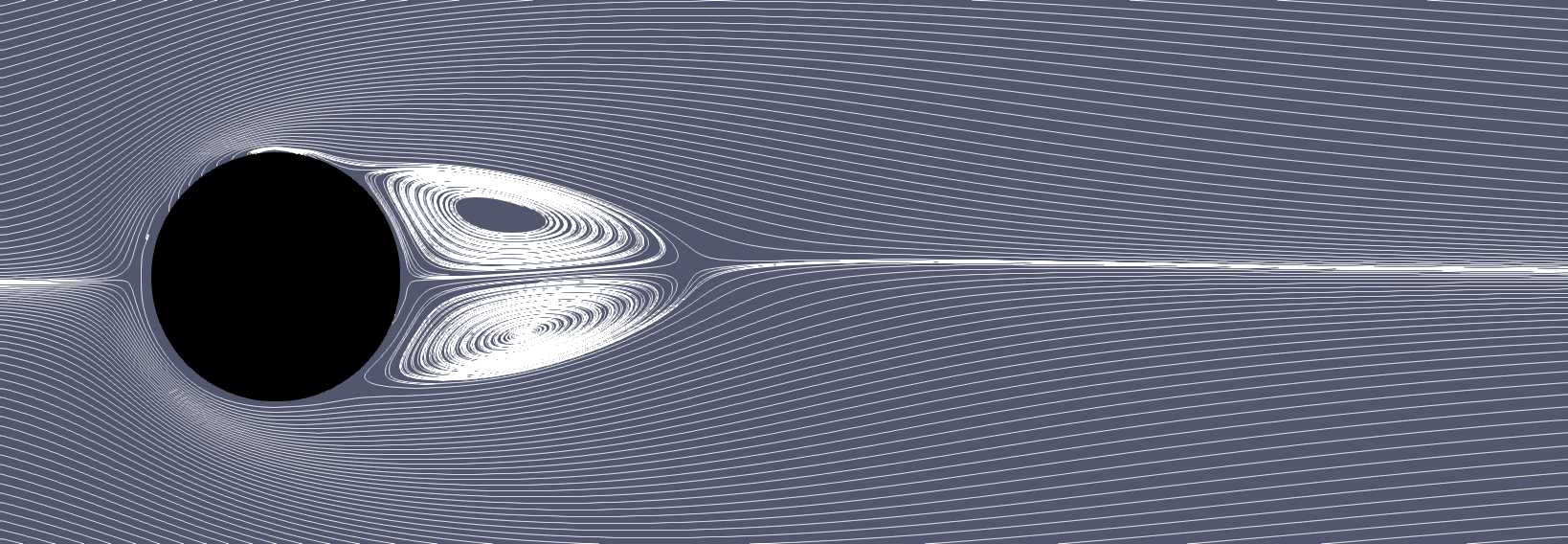}
\caption{\label{StreamLine_Re100_HB_Wi1.0_n2.0_Bs0.9_Bn2.0} $EVP_6$: $\beta_{s}=0.9,~n=2.0,~Bn=2.0$ }
\label{fig:subim300_5}
\end{subfigure}
\begin{subfigure}{0.5\textwidth}
\centering
\includegraphics[width=1.0\linewidth, height=3.0cm]{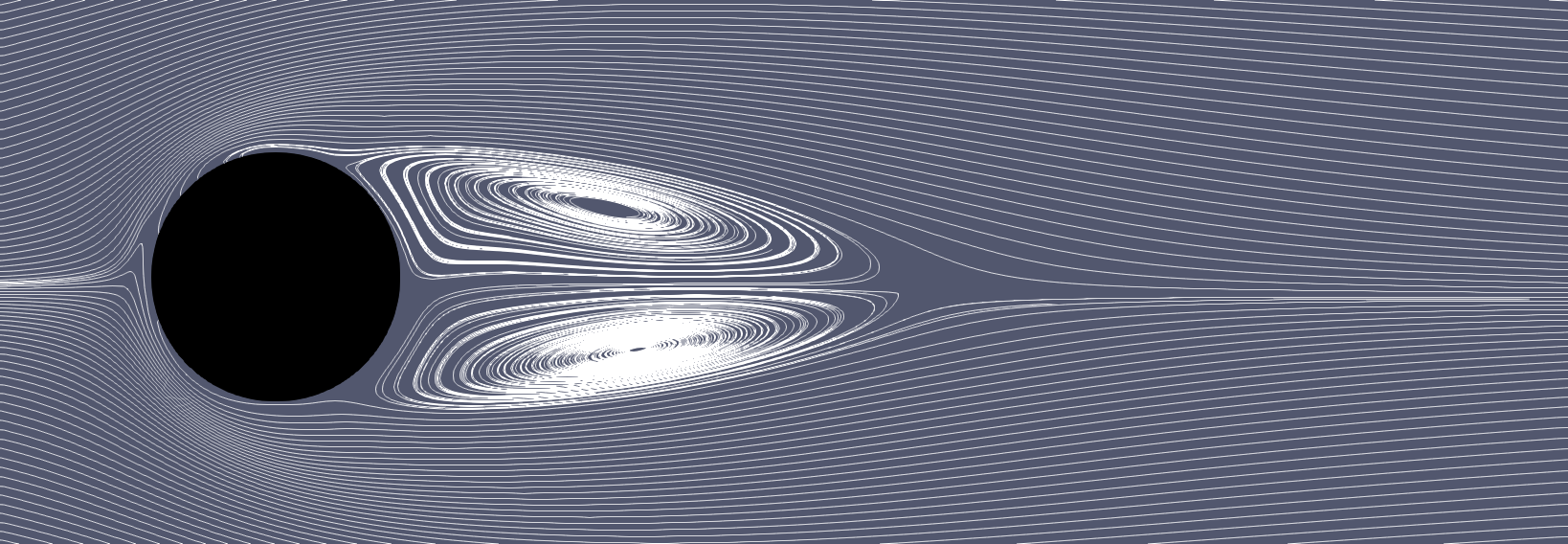}
\caption{\label{StreamLine_Re100_HB_Wi1.0_n2.0_Bs0.5_Bn2.0} $EVP_7$: $\beta_{s}=0.5,~n=2.0,~Bn=2.0$ }
\label{fig:subim300_6}
\end{subfigure}
\caption{Streamlines of the mean flow for EVP fluids under shear-thinning ($n = 0.2$, $EVP_{1,2}$), shear-independent ($n = 1.0$, $EVP_{3,4}$) and shear-thickening ($n = 2.0$, $EVP_{6,7}$) effects . The case notation corresponds to that shown in Fig.~\ref{fig:cases}.}
\label{fig:image300}
\end{figure}

The images presented in Fig. \ref{fig:image300} visualize the combined effects of elasticity, yield stress, and power law index on the mean flow streamlines. These images reveal significant changes compared to the Newtonian and Oldroyd-B fluids flow. Specifically, for EVP flow with $n=1$, there is a noticeable elongation of the recirculation bubble, and its center shifts further downstream. On the other hand, increasing shear-thinning ($n=0.2$) shortens the recirculation bubble length and the recirculation center to move closer to the cylinder in dilute EVP solution. Conversely, the influence of shear-thickening on the recirculation bubble length and its central location is more intricate. In the case of shear-thickening EVP fluid with $n=2$, the recirculation bubble length decreases, and the center shifts towards the cylinder. Additionally, with $\beta_{s}=0.5$, the recirculation extends in the streamwise direction in both shear-thinning and shear-thickening flow contributing to a downstream shift of the recirculation center. However, in the shear-independent flow of a concentration EVP fluid the recirculation bubble length reduces compared to dilute EVP solution with $n=1$, indicating the complex interplay between these factors.

\section{Proper orthogonal decomposition to identify coherent flow structures\label{sec:POD_results}}

The following section presents the results obtained from applying POD to the circular cylinder flow databases, as described in Sec. \ref{section:Review_numerical_simulation_results}. 

\begin{figure} %[H] %POD_1
\begin{subfigure}{0.49\textwidth}
\centering
\includegraphics[width=1.0\linewidth]{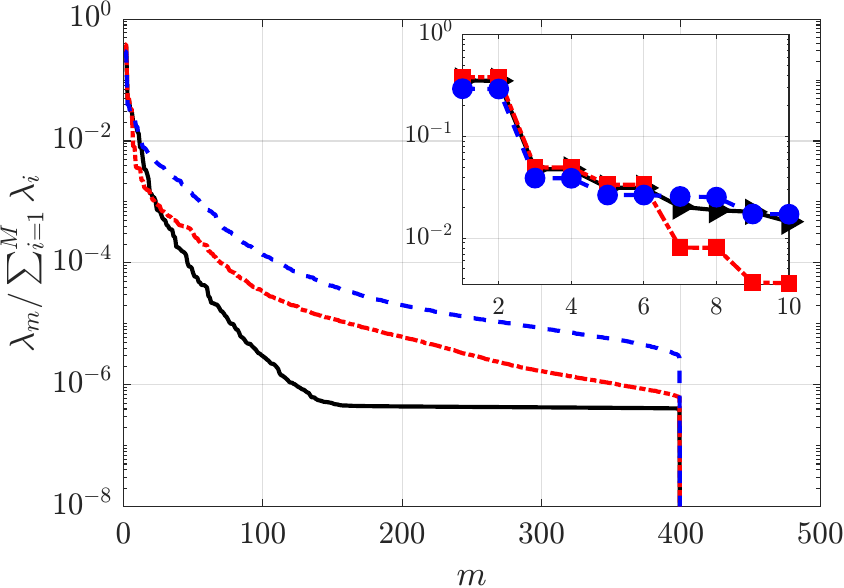}
\label{fig:POD_1_1}
\end{subfigure}
\begin{subfigure}{0.49\textwidth}
\centering
\includegraphics[width=1.0\linewidth]{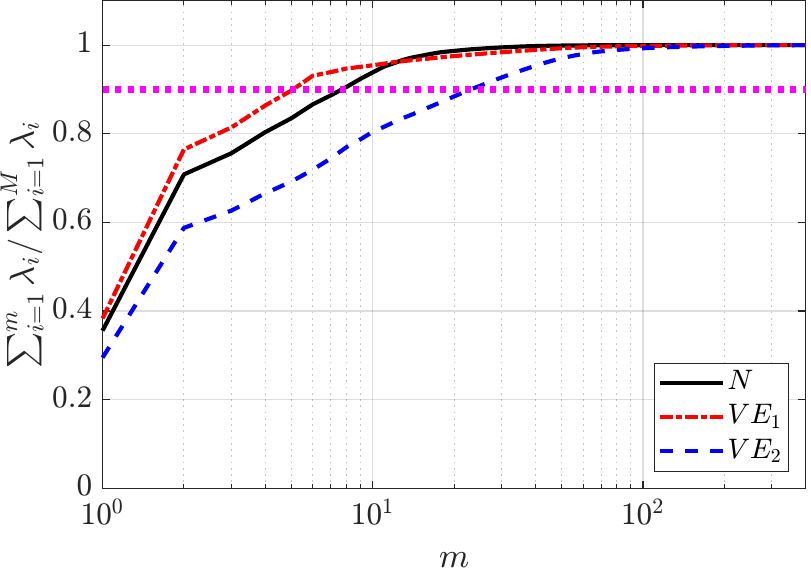}
\label{fig:POD_1_2}
\end{subfigure}
\caption{Singular-value distribution of the POD modes (left) and cumulative energy distribution (right) corresponding to the complete set of velocity components (streamwise and normal components) of the (black) Newtonian, (red) $VE_1$ and (blue) $VE_2$ cases. The number of the POD mode is represented with $m$ and $\lambda_{m} = \sigma_{m}^{2}$ as in Eq. \eqref{pod_5}. The pink dashed line illustrates the number of modes required to capture $90\%$ of the cumulative energy. }
\label{fig:POD_1}
\end{figure}

\begin{figure} %[h] POD_2
\begin{subfigure}{0.99\textwidth}
\centering
\includegraphics[width=1.0\linewidth]{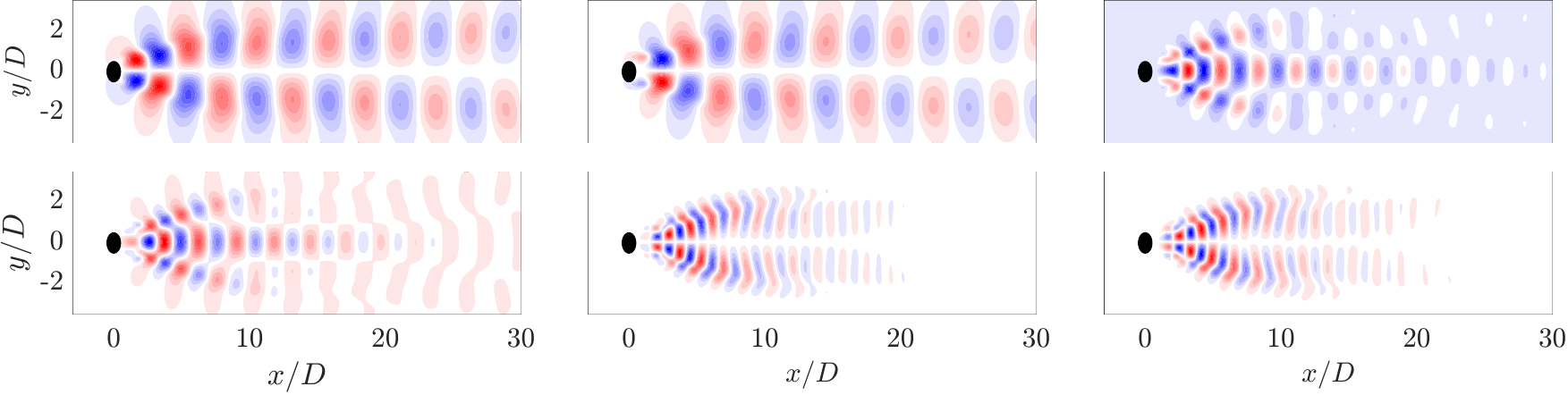}
\caption{Newtonian Fluid}
\label{fig:POD_2_1}
\end{subfigure}
\begin{subfigure}{0.99\textwidth}
\centering
\includegraphics[width=1.0\linewidth]{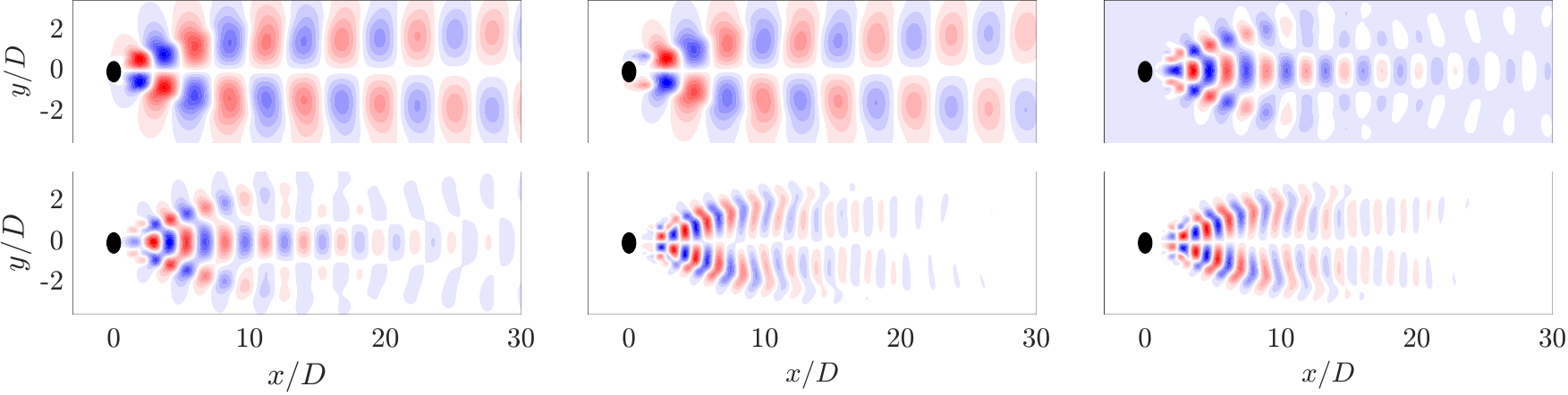}
\caption{$VE_1$ : $\beta_{s}=0.9,~n=1.0,~Bn=0.0$}
\label{fig:POD_2_2}
\end{subfigure}
\begin{subfigure}{0.99\textwidth}
\centering
\includegraphics[width=1.0\linewidth]{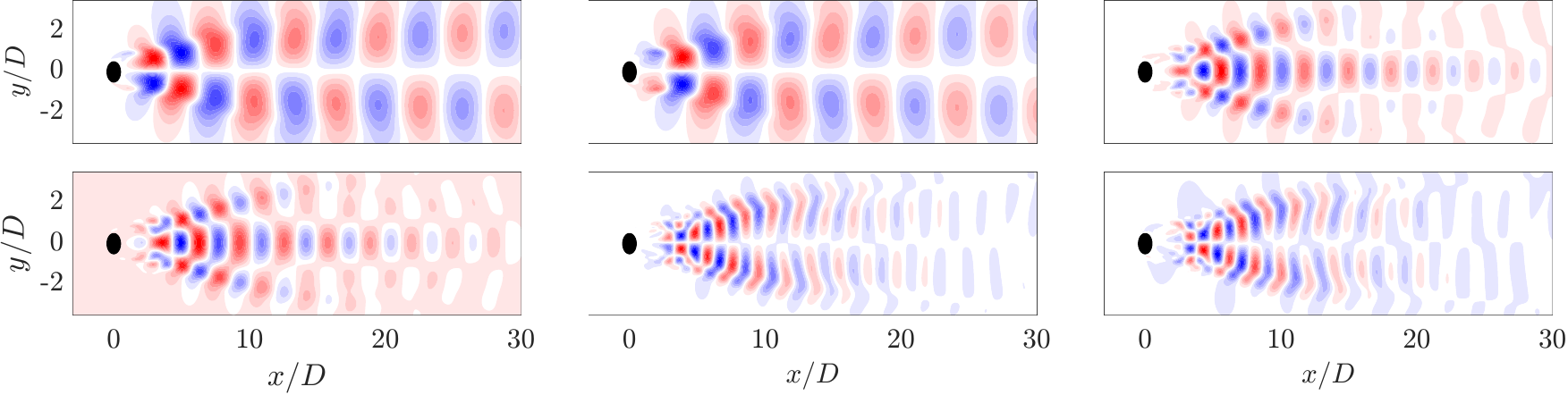}
\caption{$VE_2$ : $\beta_{s}=0.5,~n=1.0,~Bn=0.0$}
\label{fig:POD_2_3}
\end{subfigure}
\caption{POD orthogonal basis of the streamwise $v_x$ velocity field. Newtonian fluid (\Figref{fig:POD_2_1}), $VE_1$ (\Figref{fig:POD_2_2}), and $VE_2$ (\Figref{fig:POD_2_3}).}
\label{fig:POD_2}
\end{figure}

\begin{figure} %[h] POD_3 
\begin{subfigure}{0.99\textwidth}
\centering
\includegraphics[width=0.26\linewidth]{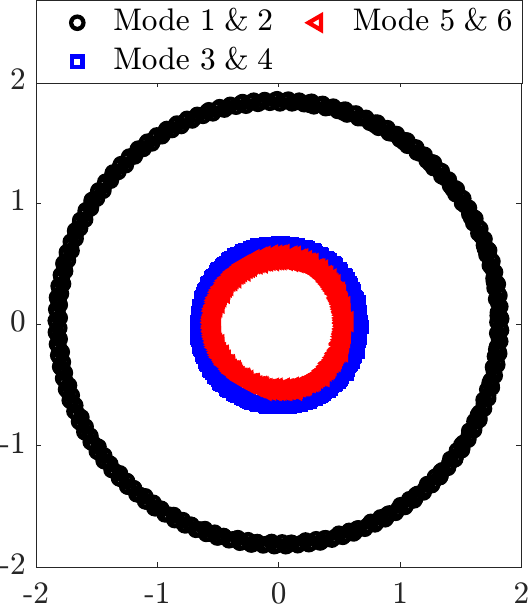}
\includegraphics[width=0.38\linewidth]{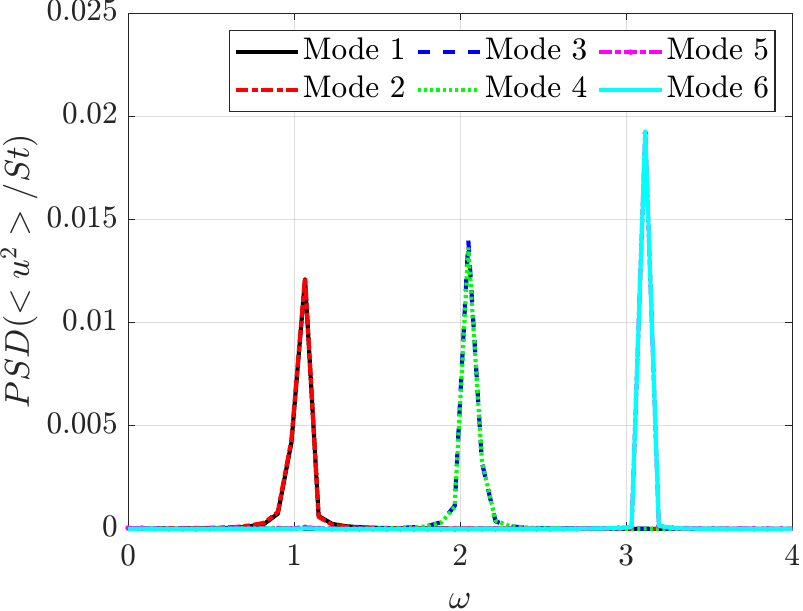}
\caption{Newtonian}
\label{fig:POD_3_1}
\end{subfigure}
\begin{subfigure}{0.99\textwidth}
\centering
\includegraphics[width=0.26\linewidth]{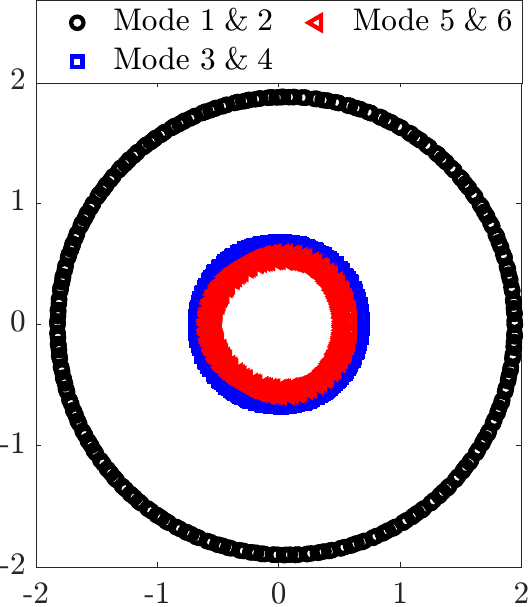}
\includegraphics[width=0.38\linewidth]{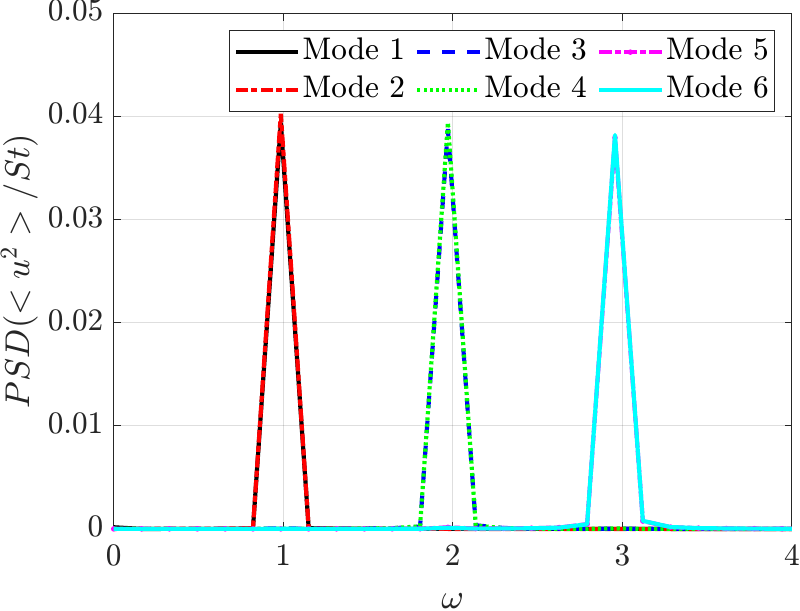}
\caption{$VE_1$ : $\beta_{s}=0.9,~n=1.0,~Bn=0.0$}
\label{fig:POD_3_2}
\end{subfigure}
\begin{subfigure}{0.99\textwidth}
\centering
\includegraphics[width=0.26\linewidth]{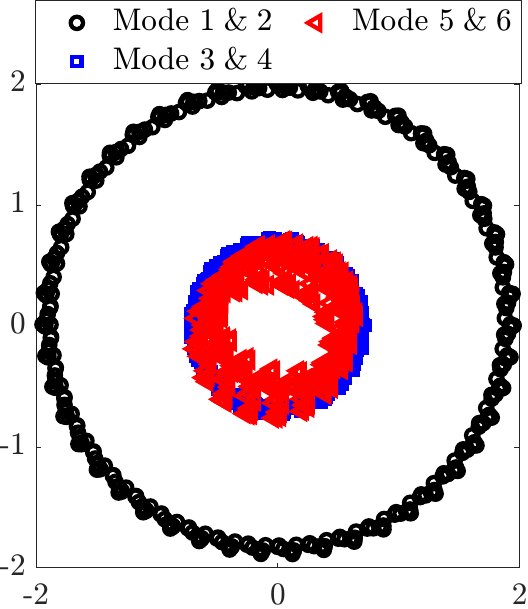}
\includegraphics[width=0.38\linewidth]{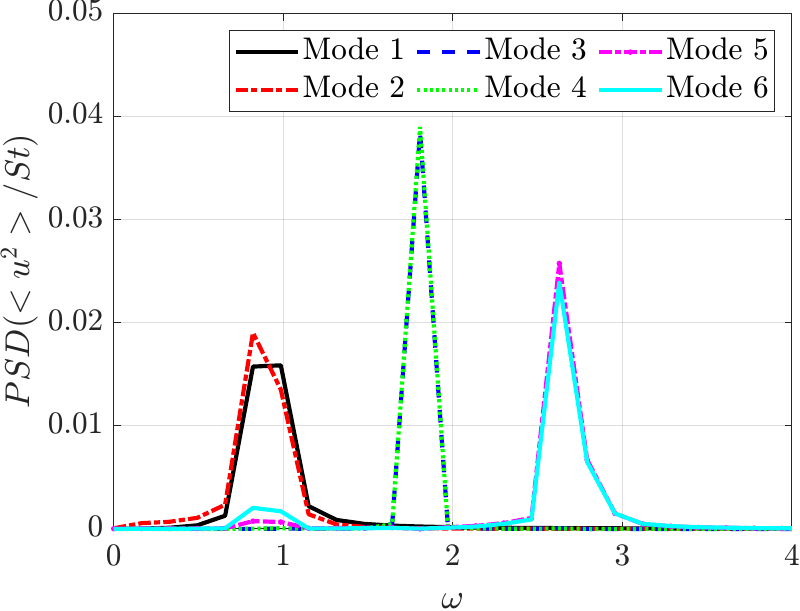}
\caption{$VE_2$ : $\beta_{s}=0.5,~n=1.0,~Bn=0.0$}
\label{fig:POD_3_3}
\end{subfigure}

\caption{The left column is the pairs of time coefficients of consecutive modes, and the right column is the FFT of the time coefficients of the same POD modes. Newtonian fluid (\Figref{fig:POD_3_1}), $VE_1$ (\Figref{fig:POD_3_2}), $VE_2$ (\Figref{fig:POD_3_3}).}
\label{fig:POD_3}
\end{figure}

\subsection{Coherent structures in Newtonian and viscoelastic fluids \label{sec:POD_N_VE}}

This section presents a comprehensive analysis of the flow past a cylinder, focusing on three distinct fluid types: Newtonian ($N$), viscoelastic with $\beta_s=0.9$ ($VE_1$) and viscoelastic with $\beta_s=0.5$ ($VE_2$). In purely viscoelastic fluids, a lower value of beta can represent more concentrated polymer solutions. The left column of Fig. \ref{fig:POD_1} illustrates the distribution of the singular values ($\lambda$), normalized by the sum of all the singular values obtained for each analysis. The right column showcases the cumulative energy of the POD modes. The cumulative energy represents the cumulative contribution of all the modes up to a specific mode, providing insights into the energy distribution across the modes. As seen, in the three cases, the modes are organized in pairs of modes with similar energy level. This fact suggests a periodic nature of the flow.

As depicted in the left column of Fig. \ref{fig:POD_1}, the first and second modes emerge as the most influential modes, capturing approximately $35.4\%$, $38.2\%$, and $29.3\%$ of the kinetic energy for the Newtonian, dilute viscoelastic, and concentrated viscoelastic solutions, respectively. These modes collectively contribute to $71\%$, $76\%$, and $59\%$ of the total cumulative energy, highlighting their substantial role in characterizing the flow behavior. Notably, the first two modes encapsulate the most crucial information about the flow across all cases. Moving to modes three and four, their contribution to the total energy sharply diminishes to approximately $4.7\%$, $4.9\%$, and $3.8\%$ for the Newtonian, dilute viscoelastic, and concentrated viscoelastic solutions, respectively. 

The pink dashed line in Fig. \ref{fig:POD_1} illustrates the number of modes required to capture $90\%$ of the total kinetic energy in the Newtonian and viscoelastic cases. The difference in the number of required modes among the three cases stems from the fact that the leading POD modes capture coherent structures that govern the dominant flow dynamics, while the higher-order modes are associated with lower-energy content and tend to represent uncorrelated fluctuations or noise. In the dilute viscoelastic case ($VE_1$), the introduction of mild elasticity simplifies the flow dynamics, enabling a more efficient modal representation in which even the first two modes capture more energy than in the Newtonian case ($N$).
In contrast, concentrated viscoelastic solutions increase the complexity of the flow dynamics, causing each mode to contain less energy than in the previous cases ($N$ and $VE_1$) and thus requiring a greater number of modes to accurately capture the system's intricate behavior.

Figure \ref{fig:POD_2} presents the first six spatial POD modes for the Newtonian, high-$\beta_s$ viscoelastic ($VE_1$), and low-$\beta_s$ viscoelastic ($VE_2$) cases, while \Figref{fig:POD_3} shows the corresponding temporal coefficients. A frequency analysis of these coefficients is conducted using the Fast Fourier Transform (FFT), providing insight into the periodic nature of each case.

In all three configurations, the first two POD modes exhibit anti-symmetric structures that reflect the large-scale wake dynamics behind the cylinder. Their temporal coefficients form closed annular loops in phase space, consistent with periodic behavior. The dominant frequency extracted from the FFT is $\omega = 1.047$ for the Newtonian case, what matches the literature \cite{Williamson1988,Williamson1996}), $\omega = 0.99$ for $VE_1$, and $\omega = 0.83$ for $VE_2$, confirming the persistence of periodic vortex shedding, although with slightly shifted frequencies due to viscoelastic effects.

Modes three and four reveal symmetric structures about the centerline and are associated with the first harmonic of the dominant vortex shedding frequency. These modes also exhibit periodic behavior, with frequencies approximately twice that of the primary mode: $\omega = 2.094$ for $N$, $\omega = 1.98$ for $VE_1$, and $\omega = 1.8$ for $VE_2$. The phase-space trajectories remain closed, although slightly more dispersed in $VE_2$, indicating a more complex but still periodic regime. The fifth and sixth modes capture finer-scale anti-symmetric structures and correspond to the second harmonic in the three cases, with frequencies near three times the fundamental one. 

\begin{figure} %[h] POD_d1
\begin{subfigure}{0.49\textwidth}
\centering
\includegraphics[width=1.0\linewidth]{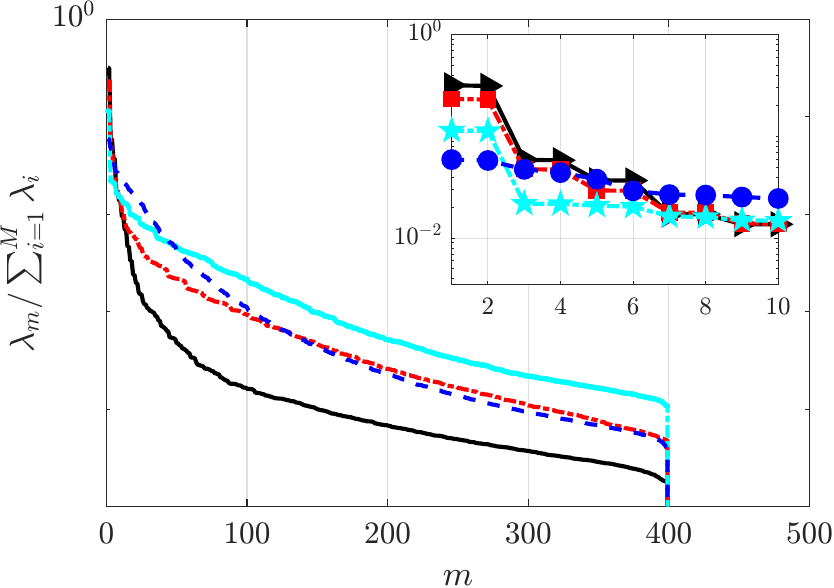}
\label{fig:POD_d1_1}
\end{subfigure}
\begin{subfigure}{0.49\textwidth}
\centering
\includegraphics[width=1.0\linewidth]{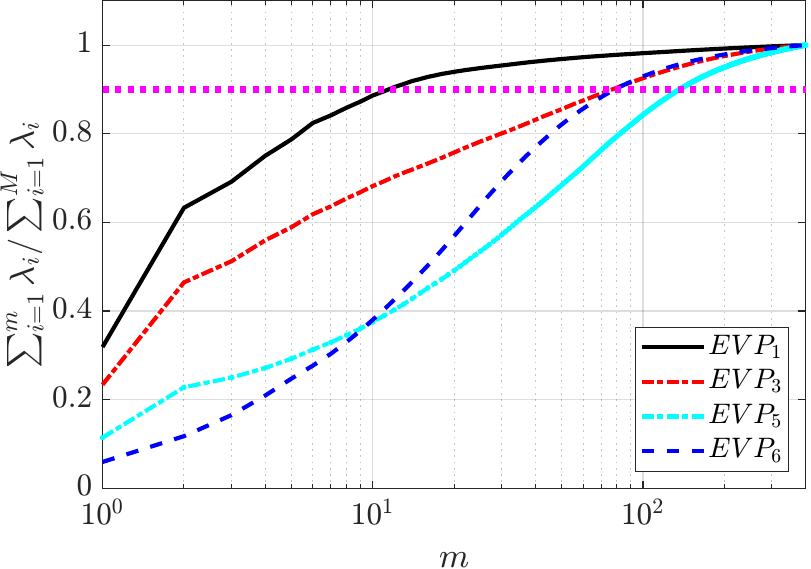}
\label{fig:POD_d1_2}
\end{subfigure}
\caption{Singular-value distribution of the POD modes (left) and cumulative energy distribution (right) corresponding to the complete set of velocity components (streamwise and normal components) of the shear-thinning $n=0.2$ ($EVP_1$, black), shear independent $n = 1$ ($EVP_3$, red) and shear-thickening $n = 1.2$ ($EVP_5$, cyan) and $n = 2$ ($EVP_6$, blue) diluted EVP cases. The number of the POD mode is represented with $m$ and $\lambda_{m} = \sigma_{m}^{2}$ as in Eq. \eqref{pod_5}. The pink dashed line illustrates the number of modes required to capture $90\%$ of the cumulative energy.}
\label{fig:POD_EVPd1}
\end{figure}

%%%%%%%%%%%%%%%%%%%%%%%%%%%%%%%%%%%%%%%%%%%%%%%%%%%%%%%%%%%%%%%%%%%%%%%%%%%%%%%
%%%%%%%%%%%%%%%%%%%%%%%%%%%%%%%%%%%%%%%%%%%%%%%%%%%%%%%%%%%%%%%%%%%%%%%%%%%%%%%

\begin{figure} %[h] POD_d2
\begin{subfigure}{0.99\textwidth}
\centering
\includegraphics[width=1.0\linewidth]{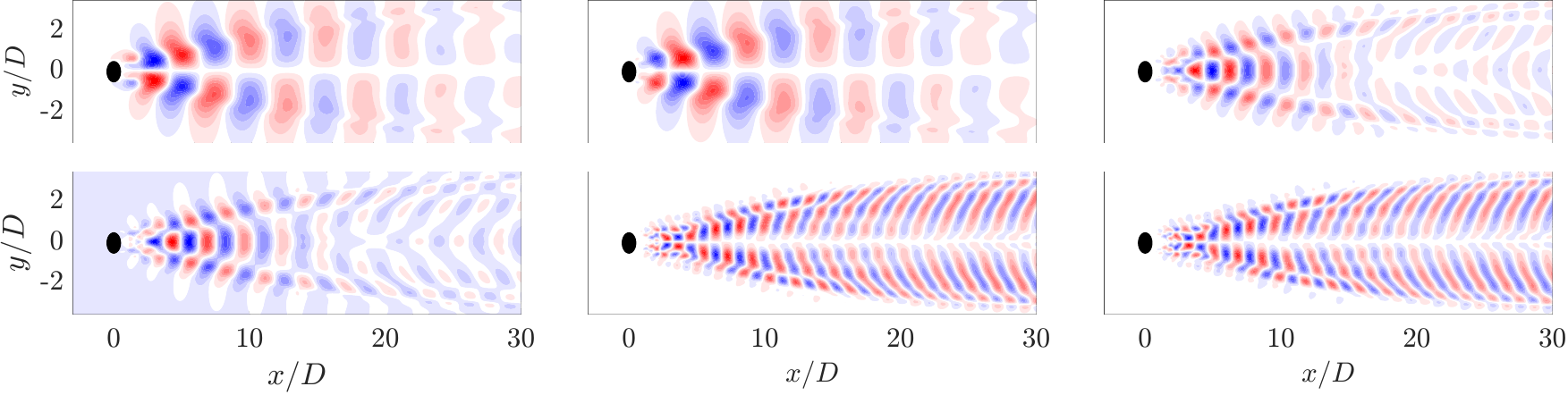}
\caption{$EVP_1$: $\beta_{s}=0.9,~n=0.2,~Bn=2.0$}
\label{fig:POD_d2_1}
\end{subfigure}
\begin{subfigure}{0.99\textwidth}
\centering
\includegraphics[width=1.0\linewidth]{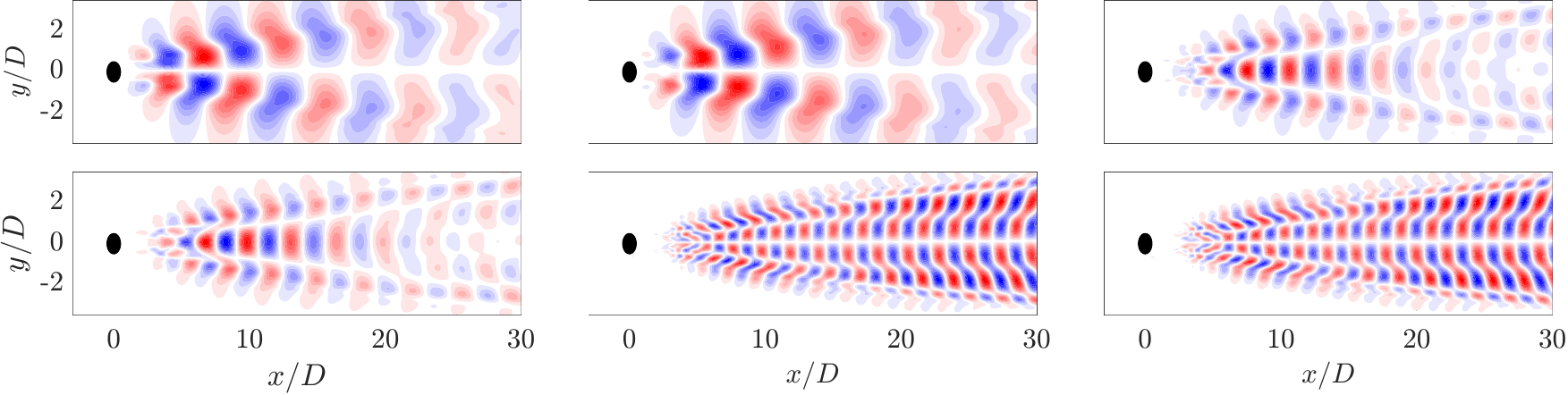}
\caption{$EVP_3$: $\beta_{s}=0.9,~n=1.0,~Bn=2.0$}
\label{fig:POD_d2_2}
\end{subfigure}
\begin{subfigure}{0.99\textwidth}
\centering
\includegraphics[width=1.0\linewidth]{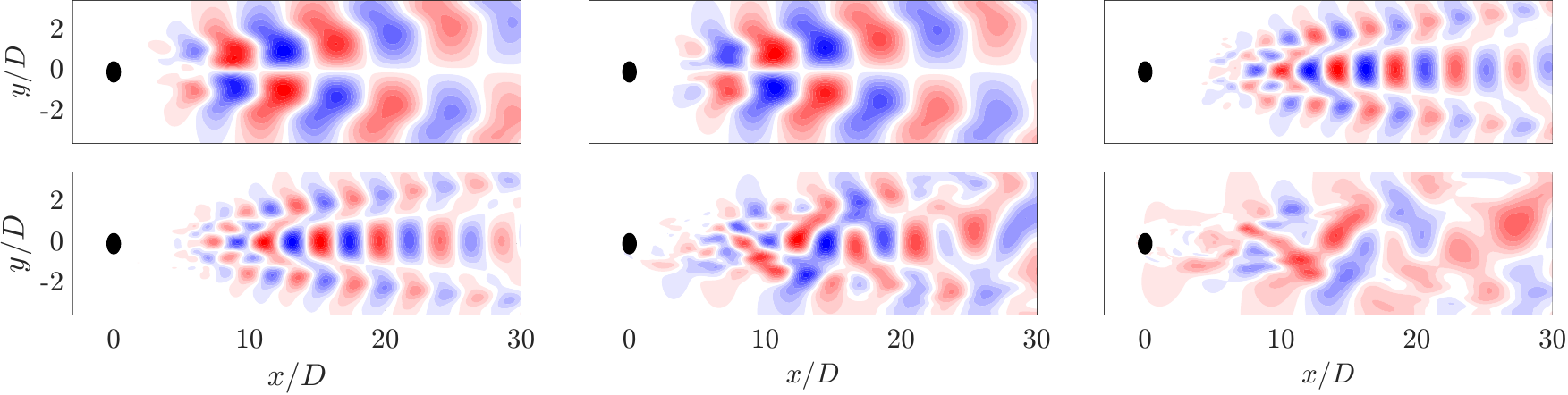}
\caption{$EVP_5$: $\beta_{s}=0.9,~n=1.2,~Bn=2.0$}
\label{fig:POD_d2_3}
\end{subfigure}
\begin{subfigure}{0.99\textwidth}
\centering
\includegraphics[width=1.0\linewidth]{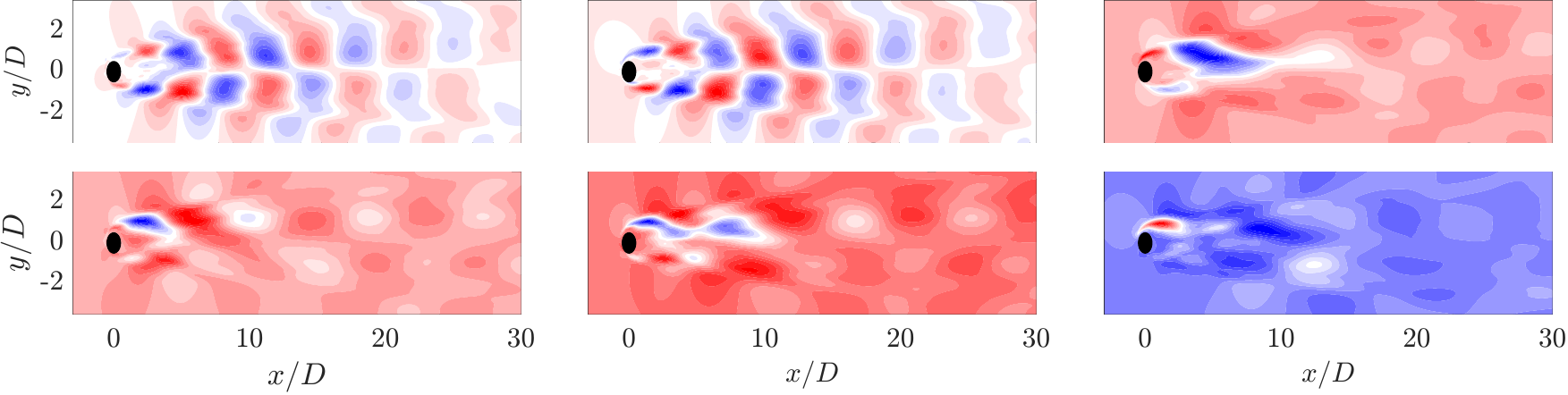}
\caption{$EVP_7$: $\beta_{s}=0.9,~n=2.0,~Bn=2.0$}
\label{fig:POD_d2_4}
\end{subfigure}
\caption{POD orthogonal basis of the streamwise $v_x$ velocity field for EVP fluids with $\beta_s = 0.9$ ($Bn = 2.0$, $Wi = 1.0$), at different values of power law index: $n = 0.2$ ($EVP_1$, \Figref{fig:POD_d2_1}), $n = 1.0$ ($EVP_3$, \Figref{fig:POD_d2_2}), $n = 1.2$ ($EVP_5$, \Figref{fig:POD_d2_3}) and $n = 2$ ($EVP_7$, \Figref{fig:POD_d2_4}).}
\label{fig:POD_d2}
\end{figure} 

\begin{figure} %[h] POD_d3 
\begin{subfigure}{0.99\textwidth}
\centering
\includegraphics[width=0.26\linewidth]{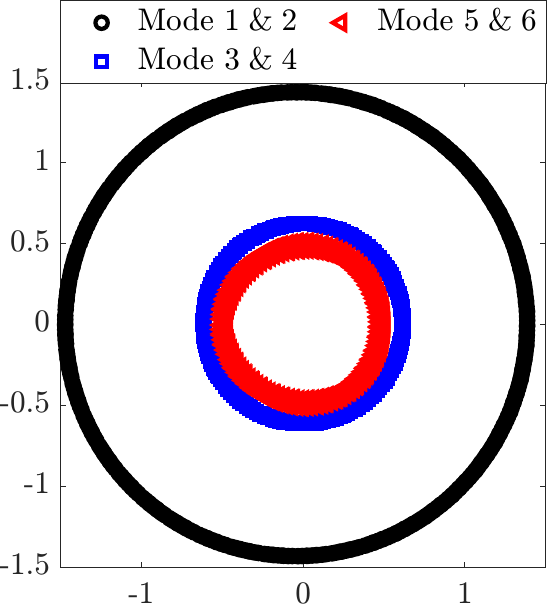}
\includegraphics[width=0.38\linewidth]{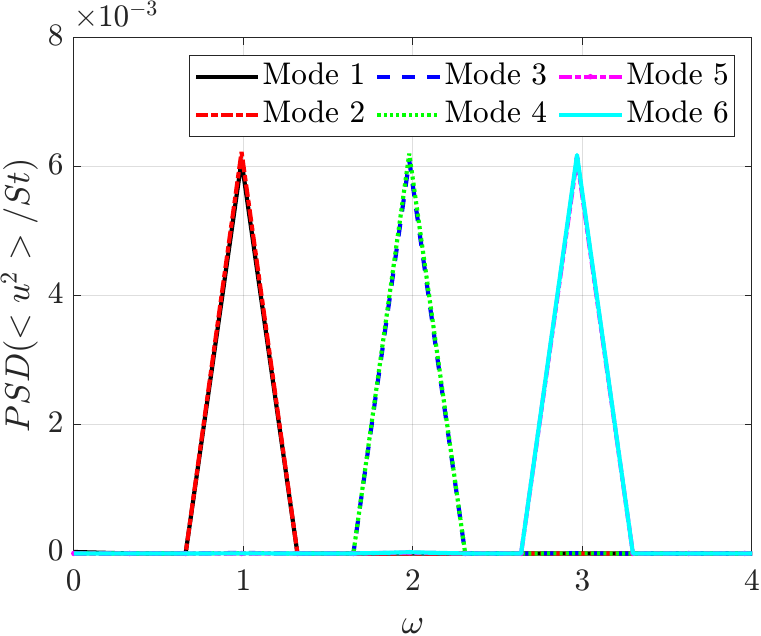}
\caption{$EVP_1$: $\beta_{s}=0.9,~n=0.2,~Bn=2.0$}
\label{fig:POD_d3_1}
\end{subfigure}
\begin{subfigure}{0.99\textwidth}
\centering
\includegraphics[width=0.26\linewidth]{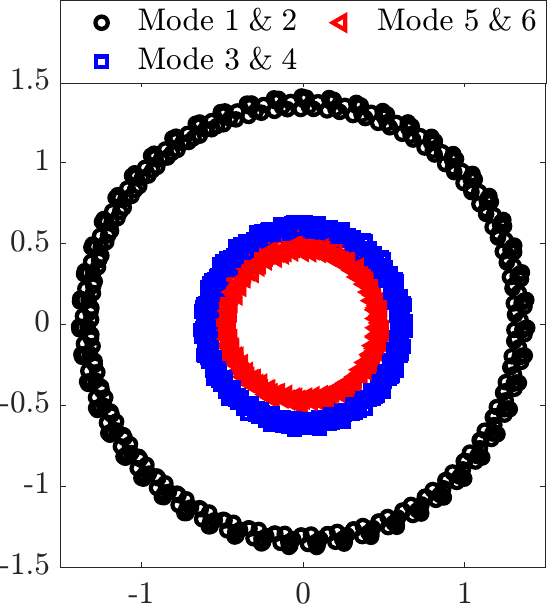}
\includegraphics[width=0.38\linewidth]{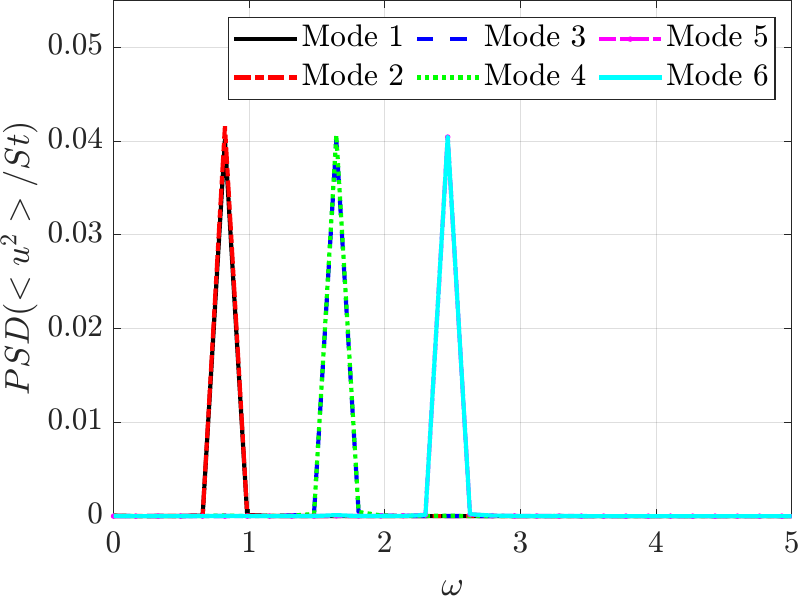}
\caption{$EVP_3$: $\beta_{s}=0.9,~n=1.0,~Bn=2.0$}
\label{fig:POD_d3_2}
\end{subfigure}
\begin{subfigure}{0.99\textwidth}
\centering
\includegraphics[width=0.26\linewidth]{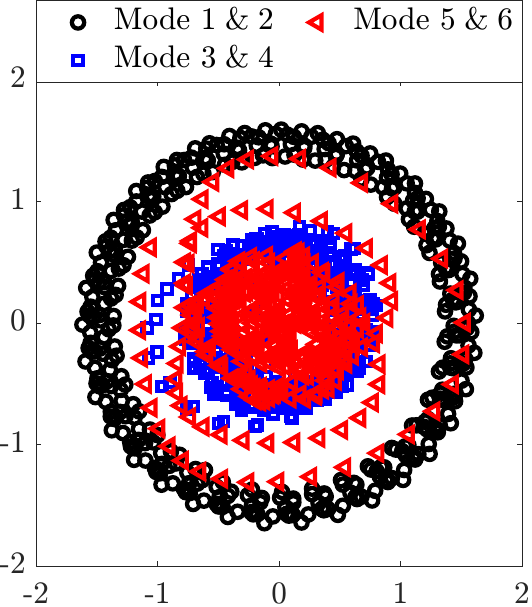}
\includegraphics[width=0.38\linewidth]{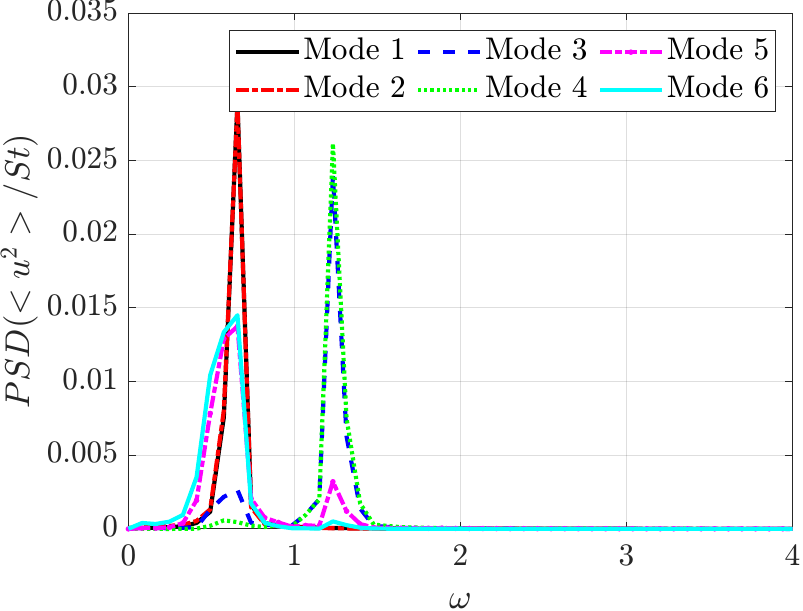}
\caption{$EVP_5$: $\beta_{s}=0.9,~n=1.2,~Bn=2.0$}
\label{fig:POD_d3_3}
\end{subfigure}
\begin{subfigure}{0.99\textwidth}
\centering
\includegraphics[width=0.26\linewidth]{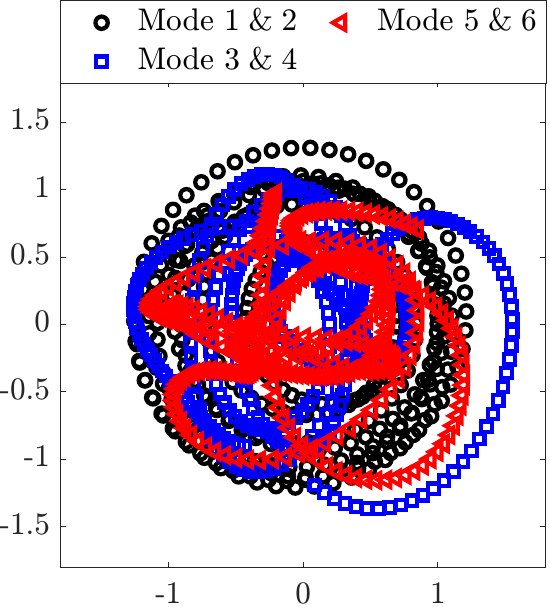}
\includegraphics[width=0.38\linewidth]{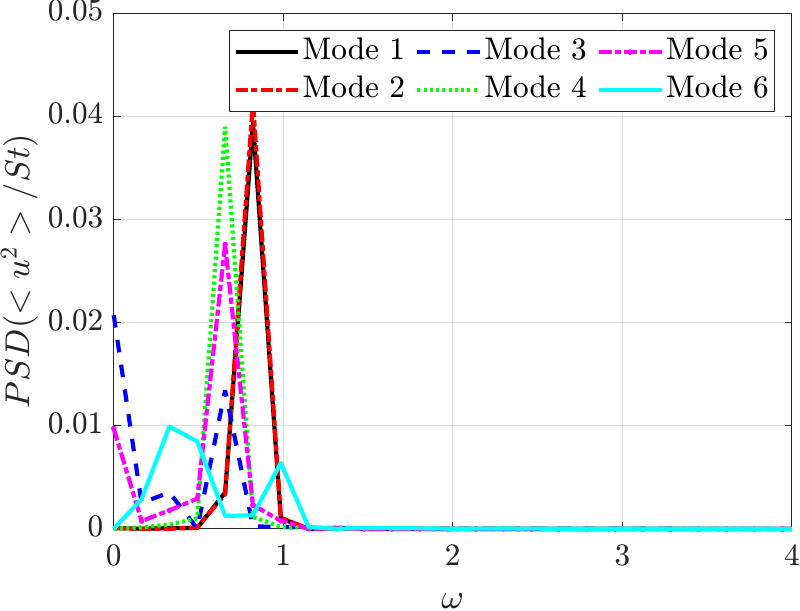}
\caption{$EVP_6$: $\beta_{s}=0.9,~n=2.0,~Bn=2.0$}
\label{fig:POD_d3_4}
\end{subfigure}
\caption{The left column presents the pairs of time coefficients of consecutive modes, and the right column is the FFT of the time coefficients of the same POD modes at different values of power law index: $n = 0.2$ ($EVP_1$, \Figref{fig:POD_d3_1}), $n = 1.0$ ($EVP_3$, \Figref{fig:POD_d3_2}), $n = 1.2$ ($EVP_5$, \Figref{fig:POD_d3_3}) and $n=2$ ($EVP_7$,\Figref{fig:POD_d3_4}).}
\label{fig:POD_d3}
\end{figure}

\subsection{Effect of yield stress on coherent structures at $\beta_s=0.9$ \label{sec:Di_Solution}}

In this section, we investigate how a finite Bingham number (making the fluid EVP) affects the wake with constant $\beta_s = 0.9$. Hence, we compare the viscoelastic wake ($Bn=0$) to an EVP wake with $Bn=2.0$. Following the analysis of the previous simulations, the decay of the singular values and the cumulative energy is presented in Fig. \ref{fig:POD_EVPd1}. First, the fluid with power law index $n = 1$ ($EVP_3$) is analysed. As in the previous case, the modes are grouped in pairs with similar energy level, suggesting the periodic nature of the flow. The first two modes capture approximately $23.2\%$ of the energy. Both modes contribute $46.5\%$ of the cumulative energy, value significantly lower than the viscoelastic high-$\beta_s$ case $VE_1$. The addition of plasticity increases the complexity of the dynamics, however there is still a sharp decrease between the singular value of these two modes and the rest. Modes 3 and 4 each capture approximately $4.7\%$ of the energy, what is similar to the previous cases.

Moving on to the strongly shear-thinning EVP fluid with $n=0.2$ ($EVP_{1}$), the first two modes account for $31.6\%$ of the kinetic energy or a combined $63.2\%$ of the cumulative energy, notably more than the shear-independent $EVP_3$ case. There is a large energy gap between the second and third mode, as modes 3 and 4 accounts for only $5.8\%$ of the energy, still greater than the shear-independent case. Also in this case, the modes are grouped in pairs with similar energy level, suggesting that the flow is periodic.

Let us now consider how the instability dynamics changes when n is increased, so that the flow becomes less shear-thinning and finally shear-thickening (cases $EVP_5$ and $EVP_6$). On the $EVP_5$ case, there is a huge decrease in the contribution to the total energy of the first two modes, in comparison to the shear independent case $EVP_3$. These modes contribute now to approximately $11.4\%$ of the kinetic energy, therefore a $22.8\%$ of the cumulative energy. Despite this, the first mode pair still defines a dominant coherent structure, as it has significantly higher energy than the third and fourth mode (approximately $2.2\%$ of energy). Increasing the value of $n$ further ($EVP_6$), decreases even more the importance of the first two modes, now accounting to just $5.9\%$ of the energy and a sum of $11.8\%$ of the cumulative energy. In this case, the energy gap mostly disappears, with an increase of the energy accounting to the third and fourth mode with respect to the other shear thickening case. Now these two modes each contributes to $4.7\%$ of the total energy.

The pink dashed line in the right column of Fig. \ref{fig:POD_EVPd1}  illustrates the number of modes required to capture 90\% of the total kinetic energy. The number of modes required in the shear-independent case $n = 1$ ($EVP_3$) is higher than the Newtonian and diluted viscoelastic cases, further illustrating that the addition of plasticity enhance the dynamic complexity of the problem. The decrease in the power law index $n$ from $1.0$ to $0.2$ leads to a decrease in the number of modes necessary to reach the $90\%$ of the cumulative energy. The shear-thinning effects rise the energy contain in the dominant modes, therefore the solution becomes more periodic. On the other hand, shear thickening increases the dynamic complexity of the solution. Although in the case with $n = 1.2$ ($EVP_5$) the first two modes accumulate more energy than the first two modes in the case with $n = 2$ ($EVP_6$), the number of modes necessary to reach the $90\%$ of the cumulative energy is surprisingly largest for $n=1.2$. 

Figure \ref{fig:POD_d2} present the contour of the first spatial POD modes, and Fig. \ref{fig:POD_d3}, their temporal coefficients. The cases of strongly shear-thinning fluid with $n=0.2$ ($EVP_{1}$) and shear-independent EVP fluid with $n=1$ ($EVP_{3}$) behave in a similar way. The cases are periodic and the singular values decay in pairs. The first two modes are well-organized and are similar to the Newtonian solution of the wake past a circular cylinder. The coefficients show a closed ring and the FFT exhibit just one frequency for these two modes, being $\omega = 0.99$ for the shear-thinning case ($EVP_1$) and $\omega = 0.82$ for the shear-independent case ($EVP_3$). In addition, the mode frequency decreases strongly with $n$, and the size of spatial structures increases.

The third and fourth modes are symmetric and behaves parallel to the cases presented in the previous section. The frequency presented is the double of the main one, namely $\omega = 1.98$ and $\omega = 1.84$ for the $EVP_1$ and $EVP_3$ cases, what is associated with the first harmonic of the dominant mode. The fifth and sixth modes together can be considered as the second harmonic, with a frequency being three times the primary mode frequency in each case. These results confirm that the dynamics is still periodic with dominant modes and their harmonics.

Increasing the value of the power law index $n$, the modes are more complex. The weak shear-thickening case $n = 1.2$ ($EVP_5$) presents the first four modes similar to the previous cases, with a well-organised antisymmetric pattern on the first two modes and symmetric on the following two modes. These four modes are located further downstream and present larger structures than the previous cases. The temporal coefficients (Fig. \ref{fig:POD_d3_3}) show that the annular closed ring is conserved, while scattered data is presented alongside the rings. Analysing these four temporal coefficients with an FFT, the first two modes present a clear frequency $\omega = 0.63$, much lower than previous cases, and the third and fourth $\omega = 1.23$, approximately the double of the first one, therefore it is a harmonic. In the fifth and sixth mode, the periodic pattern, presented in all previous cases, breaks and the shape of the modes is more difficult to interpret, as well as the pair of coefficients. Looking to the FFT associated, it appears that the modes have more than one frequency associated, therefore highlighting the importance of using a more sophisticated technique to analyze the frequency spectrum in detail. HODMD is used in the following section for this purpose.

Lastly, increasing even more the power law index $n$, the complexity of the strong shear-thickening case $n = 2.0$ ($EVP_6$) also increases. In this case, just the first two modes are clear and maintain the antisymmetric pattern, associated with the wake past the cylinder. The frequency associated with this two modes is $\omega = 0.82$. Next two modes are localised in the near field, just after the cylinder. Even though there is no annular ring in this pair of coefficients as they present more than one frequency, both have in the FFT $\omega = 0.66$ as the main one for these modes. Advancing in the number of the mode (5 and 6), the patterns become progressively more complex, as well as the number of frequencies associated increases.

The shear-thickening cases ($EVP_5$ and $EVP_6$) present a significantly lower main frequency. Modes do not present an unique main frequency and its harmonics, but instead lower secondary peaks in $\omega \in [0.4 - 0.8]$. This suggest that both cases present high dynamic complexity, which is typical of turbulent-like flows with increased temporal and spatial complexity.

\begin{figure} %[h] POD_c1
\begin{subfigure}{0.49\textwidth}
\centering
\includegraphics[width=1.0\linewidth]{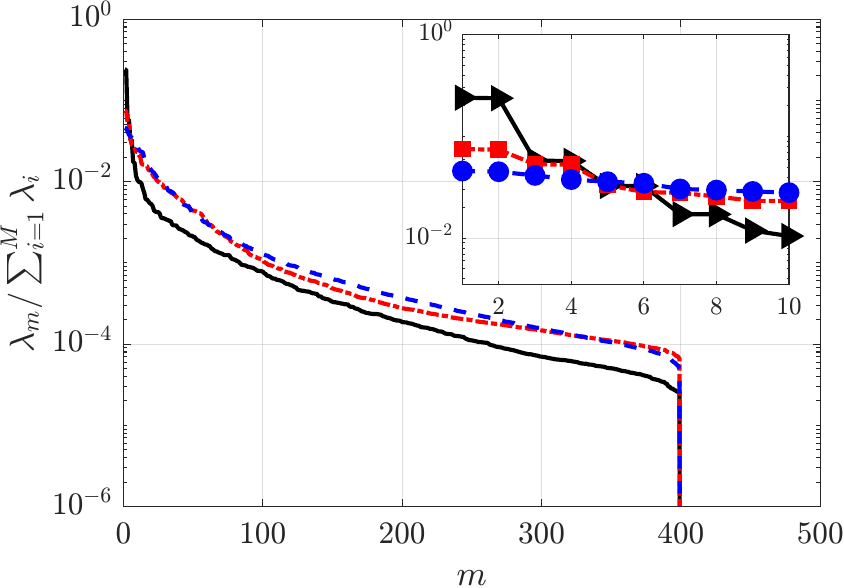}
\label{fig:POD_c1_1}
\end{subfigure}
\begin{subfigure}{0.49\textwidth}
\centering
\includegraphics[width=1.0\linewidth]{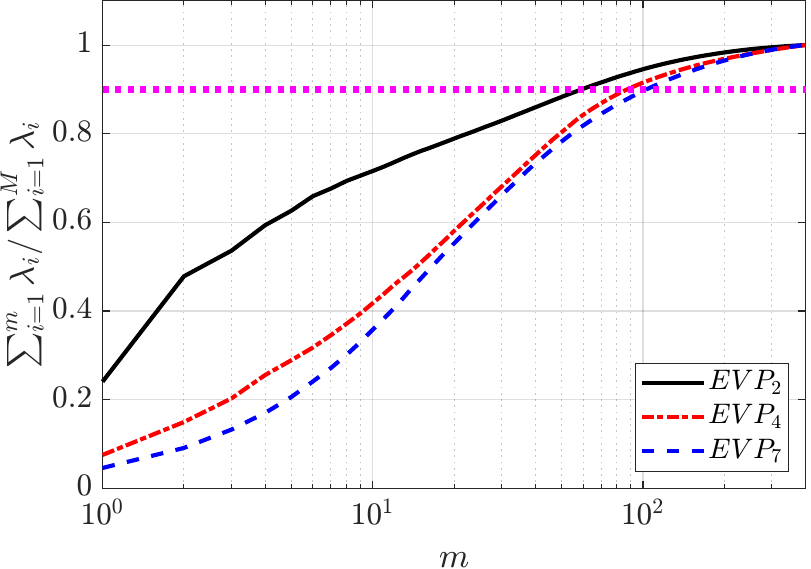}
\label{fig:POD_c1_2}
\end{subfigure}
\caption{Singular-value distribution of the POD modes (left) and cumulative energy distribution (right) corresponding to the complete set of velocity components (streamwise and normal components) of the shear-thinning $n=0.2$ ($EVP_2$, black), shear independent $n = 1$ ($EVP_4$, red) and shear-thicking $n = 2$ ($EVP_7$, blue) concentrated  EVP cases. The number of the POD mode is represented with $m$ and $\lambda_{m} = \sigma_{m}^{2}$ as in Eq. \eqref{pod_5}. The pink dashed line illustrates the number of modes required to capture $90\%$ of the cumulative energy.}
\label{fig:POD_EVPc1}
\end{figure}

\begin{figure} %[h] POD_c2
\begin{subfigure}{0.99\textwidth}
\centering
\includegraphics[width=1.0\linewidth]{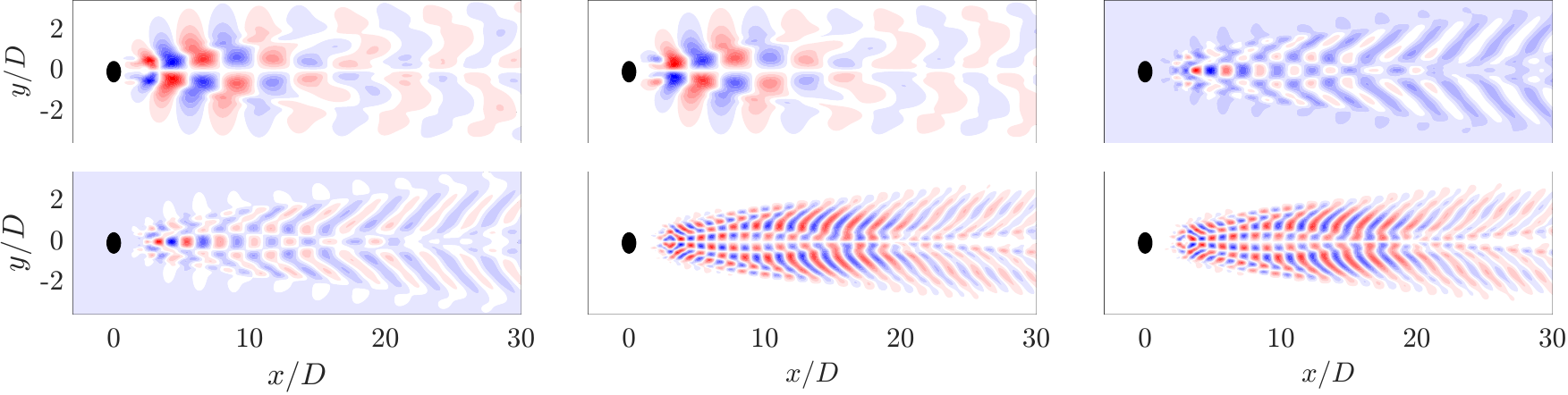}
\caption{$EVP_2$: $\beta_{s}=0.5,~n=0.2,~Bn=2.0$}
\label{fig:POD_c2_1}
\end{subfigure}
\begin{subfigure}{0.99\textwidth}
\centering
\includegraphics[width=1.0\linewidth]{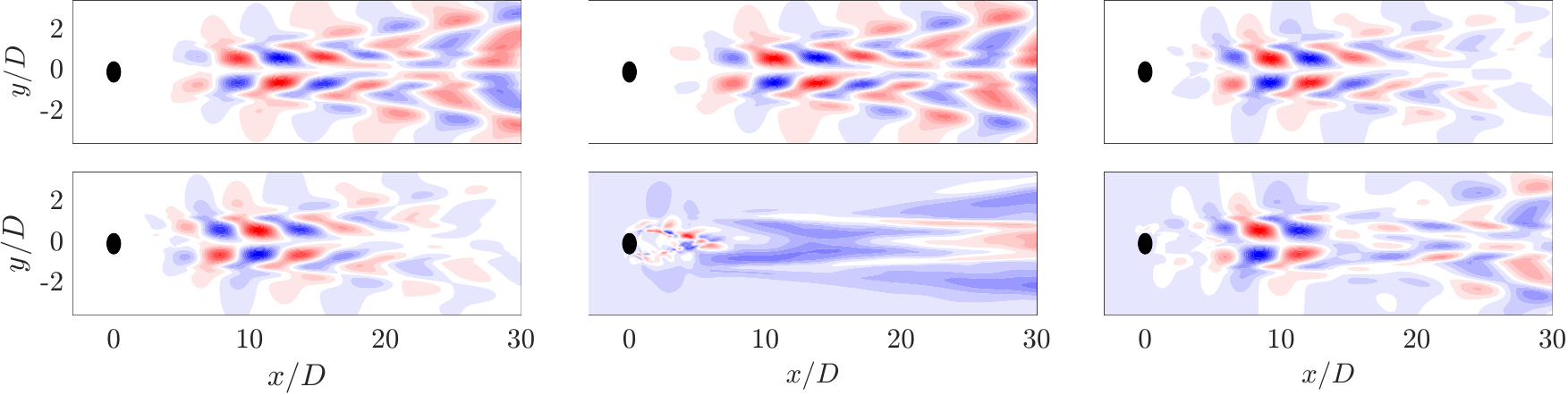}
\caption{$EVP_4$: $\beta_{s}=0.5,~n=1.0,~Bn=2.0$}
\label{fig:POD_c2_2}
\end{subfigure}
\begin{subfigure}{0.99\textwidth}
\centering
\includegraphics[width=1.0\linewidth]{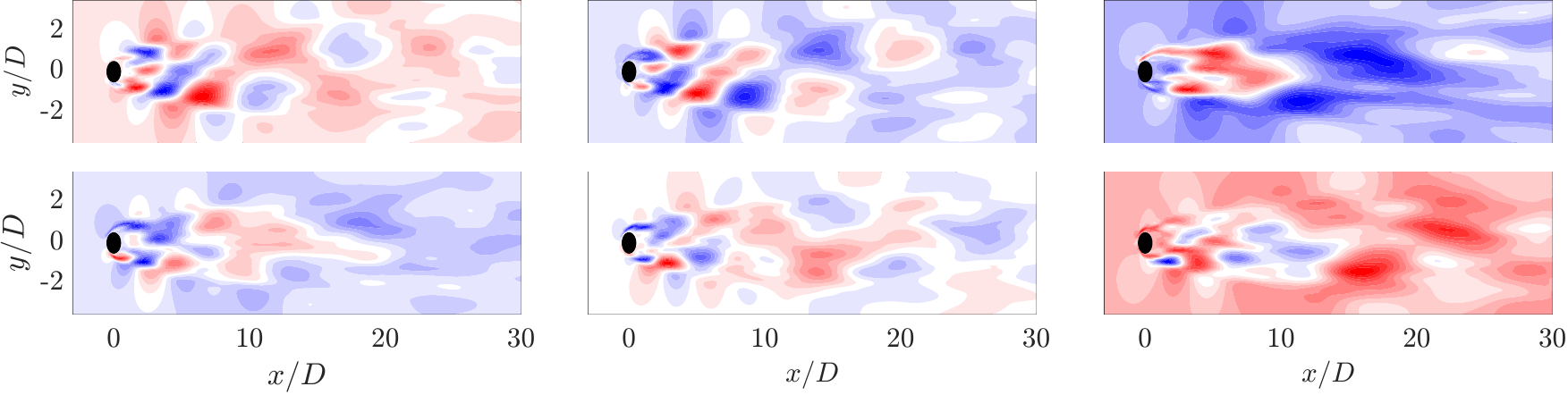}
\caption{$EVP_7$: $\beta_{s}=0.5,~n=2.0,~Bn=2.0$}
\label{fig:POD_c2_3}
\end{subfigure}
\caption{POD orthogonal basis of the streamwise $v_x$ velocity field for concentrated EVP fluids ($\beta_s = 0.5$, $Bn = 2.0$, $Wi = 1.0$). Shear-thinning $n = 0.2$ ($EVP_2$, \Figref{fig:POD_c2_1}), shear-independent $n = 1.0$ ($EVP_4$, \Figref{fig:POD_c2_2}) and shear-thickening $n = 2.0$ ($EVP_7$, \Figref{fig:POD_c2_3}).}
\label{fig:POD_c2}
\end{figure} 

\begin{figure}%[h!] POD_c3 
\begin{subfigure}{0.98\textwidth}
\centering
\includegraphics[width=0.28\linewidth]{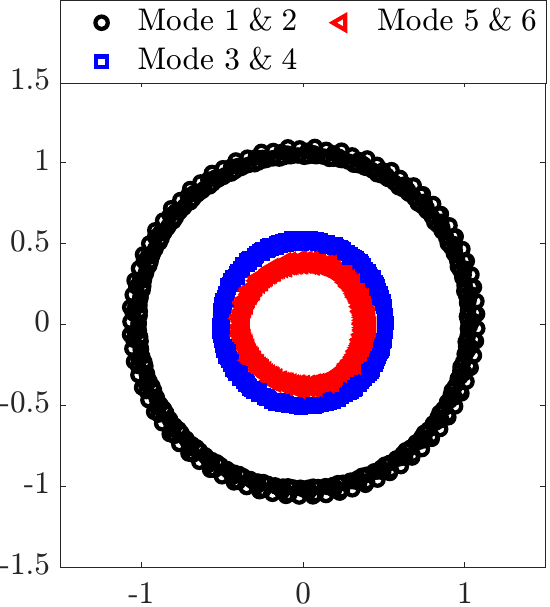}
\includegraphics[width=0.4\linewidth]{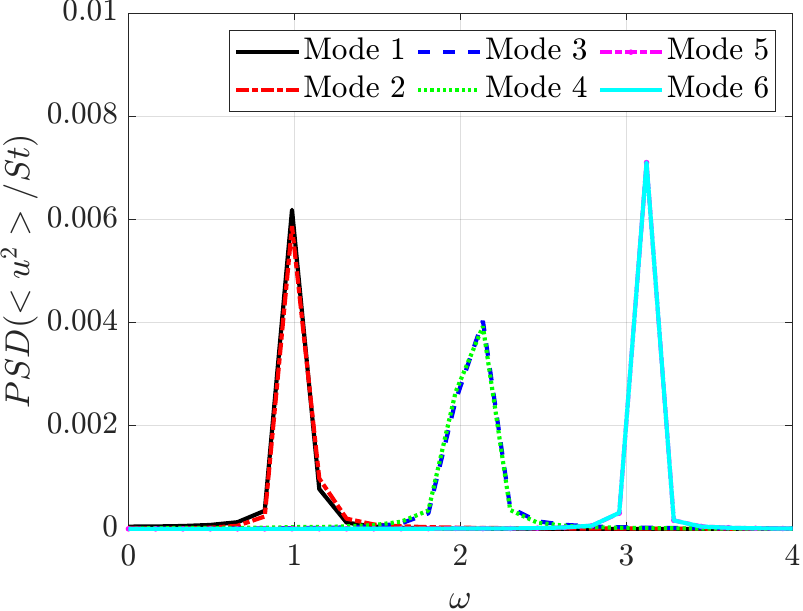}
\caption{$EVP_2$: $\beta_{s}=0.5,~n=0.2,~Bn=2.0$}
\label{fig:POD_c3_1}
\end{subfigure}
\begin{subfigure}{0.98\textwidth}
\centering
\includegraphics[width=0.28\linewidth]{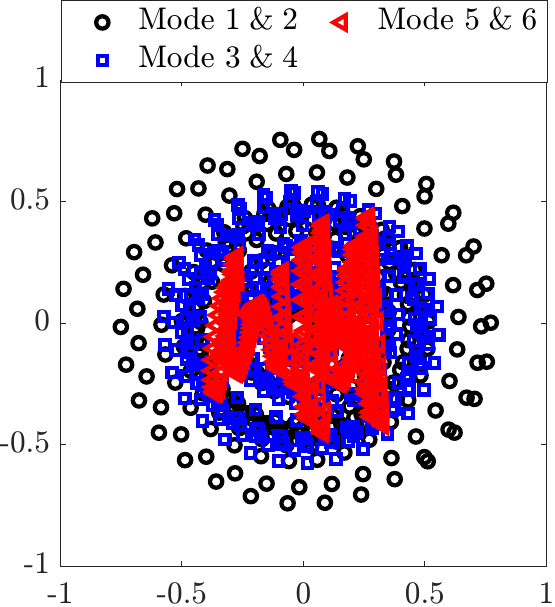}
\includegraphics[width=0.4\linewidth]{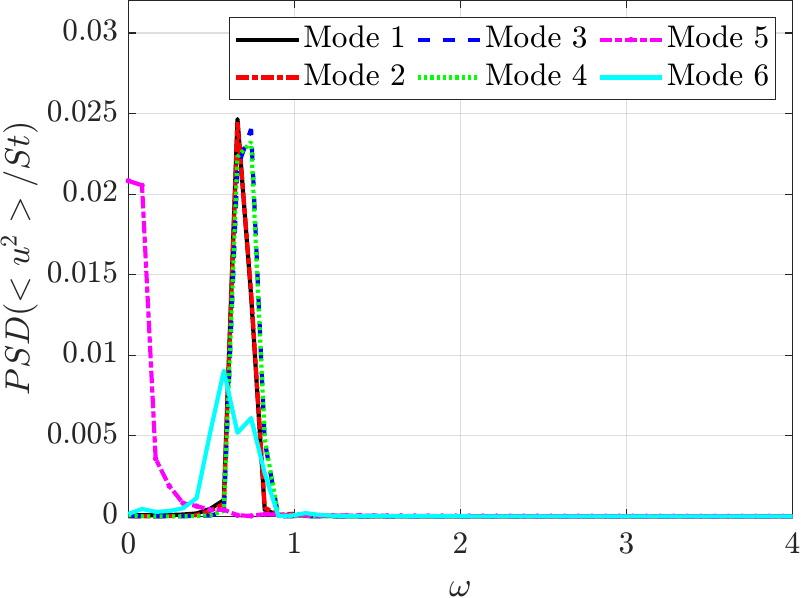}
\caption{$EVP_2$: $\beta_{s}=0.5,~n=1.0,~Bn=2.0$}
\label{fig:POD_c3_2}
\end{subfigure}
\begin{subfigure}{0.98\textwidth}
\centering
\includegraphics[width=0.28\linewidth]{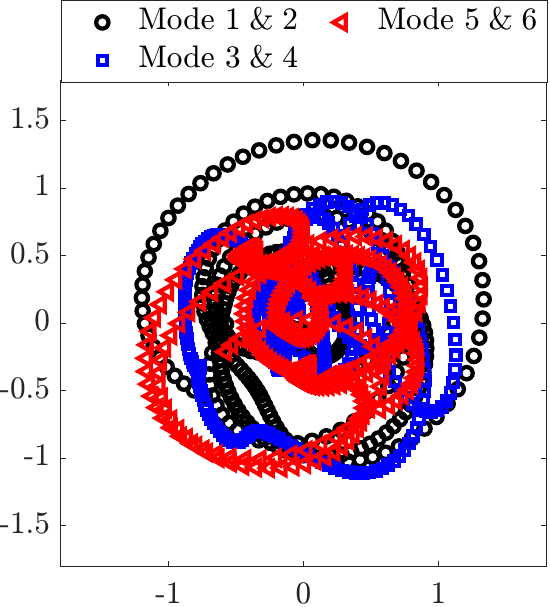}
\includegraphics[width=0.4\linewidth]{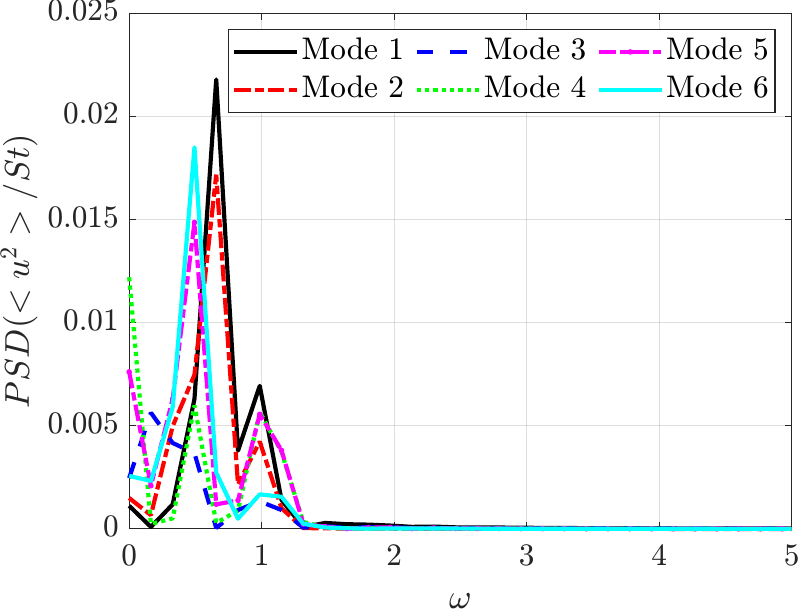}
\caption{$EVP_7$: $\beta_{s}=0.5,~n=2.0,~Bn=2.0$}
\label{fig:POD_c3_3}
\end{subfigure}
\caption{The left column presents the pairs of time coefficients of consecutive modes, and the right column is the FFT of the time coefficients of the same POD modes. Shear-thinning $n = 0.2$ ($EVP_2$, \Figref{fig:POD_c3_1}), shear-independent $n = 1.0$ ($EVP_4$, \Figref{fig:POD_c3_2}) and shear-thickening $n = 2.0$ ($EVP_7$, \Figref{fig:POD_c3_3}).}
\label{fig:POD_c3}
\end{figure}

\subsection{Effect of solvent viscosity ratio on coherent structures in EVP wakes \label{sec:Co_Solution}}

Next, we investigate qualitatively how solvent viscosity ratio affects the EVP wake dynamics, by lowering it to $\beta_s=0.5$. The decay of the singular values and the cumulative energy is presented in Fig. \ref{fig:POD_EVPc1}. Figure \ref{fig:POD_c2} presents the contour of the first six POD modes in each one of the 3 cases studied, while Fig. \ref{fig:POD_c3} presents the analysis of the temporal coefficients associated.

Looking to the number of modes required to reach the $90\%$ of the cumulative energy, it is clear that increasing the value of $n$ means an increase of the complexity and therefore, the number of modes needed to represent the main dynamics of the flow. Moreover, in general, we observe that this lower value of $\beta_s$ (more consistent with real life EVP fluids) makes the dynamics more complex, as a higher number of modes are required to represent the flow.

For the shear-thinning case $n = 0.2$ ($EVP_2$) in Fig. \ref{fig:POD_EVPc1}, the decay of the singular values is organised in pairs and the first two modes captures approximately $24\%$ of the energy, for a combined of $47.8\%$ of the cumulative energy. This value is significantly lower than the $EVP_1$ case. Despite the increased complexity of the dynamics, the solution still exhibits periodic behavior. The time coefficients of these two modes form a closed ring and both have the same frequency associated $\omega = 0.99$, as seen in Fig. \ref{fig:POD_c3_1}. The shape of the first two modes (\Figref{fig:POD_c2_1}) presents an organised anti-symmetric structure. These modes are similar to the Newtonian solution of the wake past the cylinder and, in comparison with the other shear-thinning case ($EVP_1$), the high intensity structures of the mode are smaller and more constrained to the centerline. 

Between the first two modes and the following pair, there remains an energy gap, as the contribution to the total energy significantly decreases to $5.7\%$. As well as in other cases, third and fourth mode present symmetric patterns and the frequency associated is $\omega = 2.14$, associated to the first harmonic of the leading mode. Finally, the fifth and sixth mode are again antisymmetric and have a frequency, $\omega = 3.12$, which is three times the main frequency, linked to the second harmonic. This result suggest the periodic nature of this solution. As in the first two modes, comparing from the third to the sixth modes with $EVP_1$ case, the high intensity structures are smaller, constrained to the centerline and the maximum of the amplitude appears nearer the cylinder in the streamwise direction. 

Increasing the value of the power law index to a shear-independent fluid $n = 1$ ($EVP_4$), the energy of the first two modes decreases to $7.5\%$ of the total, with a cumulative energy of $14.9\%$, as seen in Fig. \ref{fig:POD_EVPc1}. The energy gap virtually disappears, as the third and fourth mode contributes $5.3\%$ of the kinetic energy. The $EVP_4$ case does not have a periodic behaviour as in the diluted case $EVP_3$. Nevertheless, the first four modes present similar shape, antisymmetric structures along the streamwise direction, as shown in Fig. \ref{fig:POD_c2_2}. Both pairs of time coefficients form annular shape, but with scattered data, suggesting a more complex nature of the flow dynamics than for cases with smaller values of $n$. As depicted in \ref{fig:POD_c3_2}, the FFT shows that the first two modes present the same frequency $\omega = 0.66$ and the third and fourth, presents a frequency value slightly larger, $\omega = 0.74$. This is linked to the increasingly complex dynamics of the flow. Accurately representing the evolution of the flow field requires a larger number of POD modes with similar frequencies and low energy, corresponding to smaller flow scales. These modes present the structures far away from the cylinder, as in the case $EVP_5$.

%%%%%%%%%%%%%%%%%%%%%%%%%%%%%%%%%%%%%%%%%%%%%%%%%%%%%%%%%%%%%%%%%%%
%%%%%%%%%%%%%%%%%%%%%%%%%%%%%%%%%%%%%%%%%%%%%%%%%%%%%%%%%%%%%%%%%%%
 
Increasing the power law index the flow becomes more and more complex. The shear-thickening fluid $n=2.0$ ($EVP_7$) is the one in which the first two modes accumulate less energy, with just $4.6\%$ of the total energy and a sum of $9.1\%$ of the cumulative energy. An energy gap is not observed and the contribution of the third and fourth mode is $4\%$. The shear-thickening $n = 2.0$ case ($EVP_7$) is the most complex case, as shown in Fig. \ref{fig:POD_EVPc1}. The first two modes are similar, with high intensity structures appearing after the cylinder, in the wake. The third mode is formed by large structures appearing further away from the cylinder in the streamwise direction. Finally modes 4, 5 and 6 are concentrated in the part of the domain located between the cylinder and the recirculation bubble. The pairs of time coefficient do not form a clear shape and the FFT show several frequencies for every mode, suggesting the complex quasi-periodic nature of the flow, which tends towards chaotic dynamics as $n$ increases. The analysis carried out in the next section using HODMD, will shed light on the underlying mechanisms of these highly complex dynamic cases. This last case exhibits similar behavior to the previously discussed case $EVP_6$, although this one is more complex. They are characterized by broad-band frequency content and the dominance of a very low-frequency component. This spectral distribution indicates a loss of coherence in the dynamics, with energy spread across multiple frequencies rather than concentrated at discrete peaks. Such behavior suggests the presence of uncorrelated events and a high degree of flow complexity, which are hallmarks of transitional or turbulence-like regimes.

\section{Temporal analysis of flow dynamics using HODMD \label{sec:DMD}}

In this section, an examination of the outcomes obtained from using HODMD is performed. The interaction of viscoelasticity and plasticity introduces a multitude of instabilities with varying spatio-temporal scales and frequencies, resulting in a complex detection of flow patterns. In order to capture all the different dynamics presented in these complex flows, HODMD is applied to the different databases. In each database, HODMD is calibrated varying the values of the tolerances $\varepsilon_1$ and $\varepsilon_2$ and the parameter $d$, following the advice on Ref. \cite{VegaLeClaincheBook20}. 

\begin{figure}%[h] %DMD_2
\begin{subfigure}{1.0\textwidth}
\centering
\includegraphics[width=0.49\linewidth,height=2.5cm]{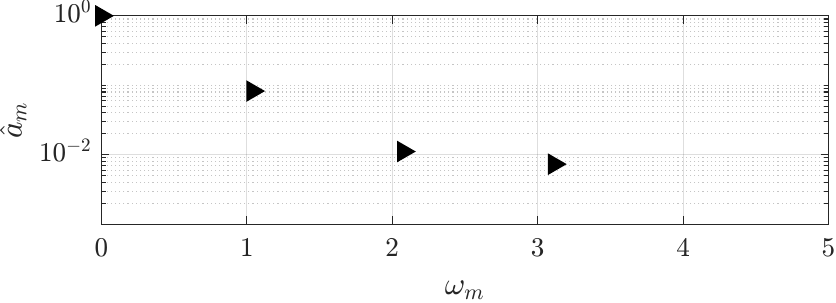}
\caption{Newtonian Fluid}
\label{fig:DMD_2_1}
\end{subfigure}
\begin{subfigure}{0.49\textwidth}
\centering
\includegraphics[width=1.0\linewidth, height=2.5cm]{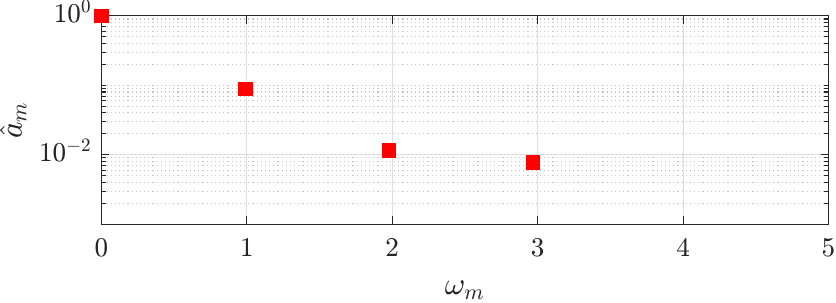}
\caption{$VE_1$ : $\beta_{s}=0.9,~n=1.0,~Bn=0.0$ }
\label{fig:DMD_2_2}
\end{subfigure}
\begin{subfigure}{0.49\textwidth}
\centering
\includegraphics[width=1.0\linewidth, height=2.5cm]{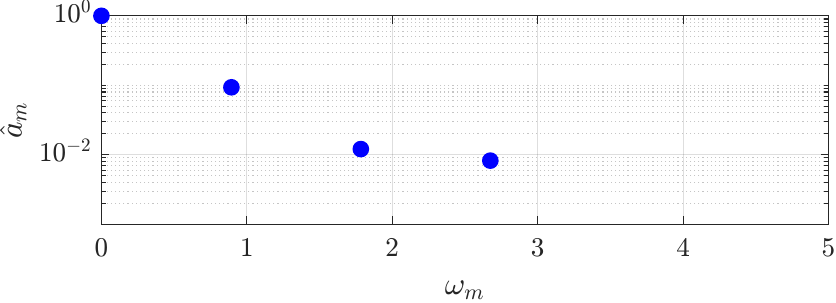}
\caption{$VE_2$ : $\beta_{s}=0.5,~n=1.0,~Bn=0.0$ }
\label{fig:DMD_2_3}
\end{subfigure}
\caption{DMD-d Spectrum. Amplitudes normalized with the maximum value ($\hat{a}_m=a_m/a_0$) vs. frequencies $\omega_m$ obtained calibrating the algorithm with different tolerances and order $d$ for Newtonian (\Figref{fig:DMD_2_1}), $VE_{1}$ (\Figref{fig:DMD_2_2}), and $VE_{2}$ (\Figref{fig:DMD_2_3}).} 
\label{fig:DMD_2}
\end{figure}

\begin{figure}%[h!] %DMD_3
\begin{subfigure}{1.0\textwidth}
\centering
\includegraphics[width=0.7\linewidth]{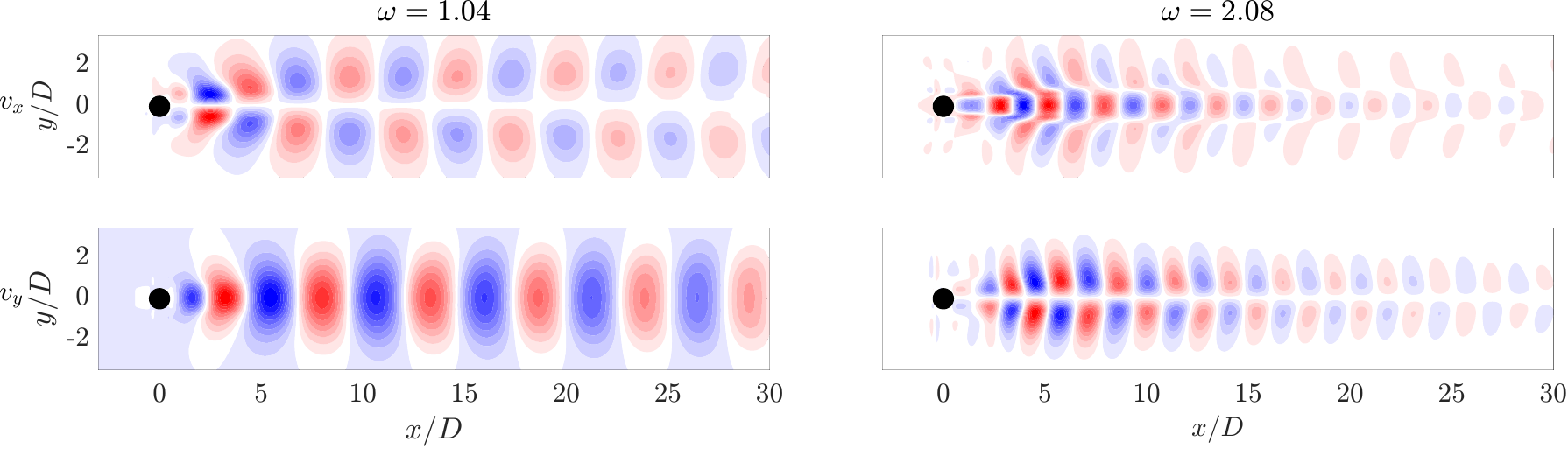}
\caption{Newtonian Fluid}
\label{fig:DMD_3_1}
\end{subfigure}

\begin{subfigure}{1.0\textwidth}
\centering
\includegraphics[width=0.7\linewidth]{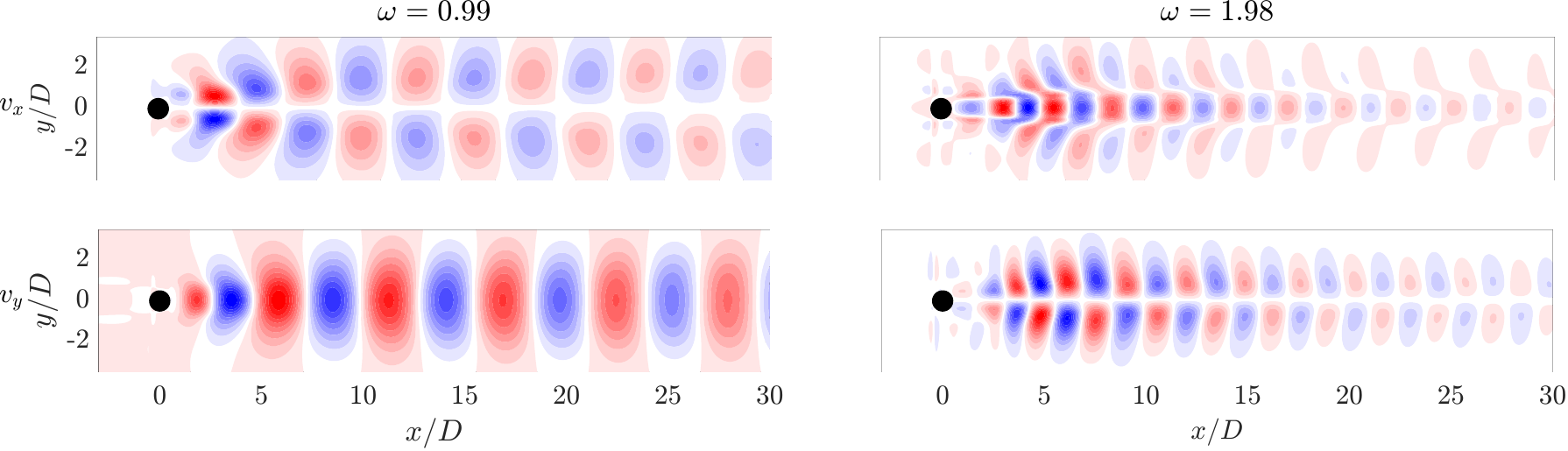}
\caption{$VE_1$ : $\beta_{s}=0.9,~n=1.0,~Bn=0.0$ }
\label{fig:DMD_3_2}
\end{subfigure}

\begin{subfigure}{1.0\textwidth}
\centering
\includegraphics[width=0.7\linewidth]{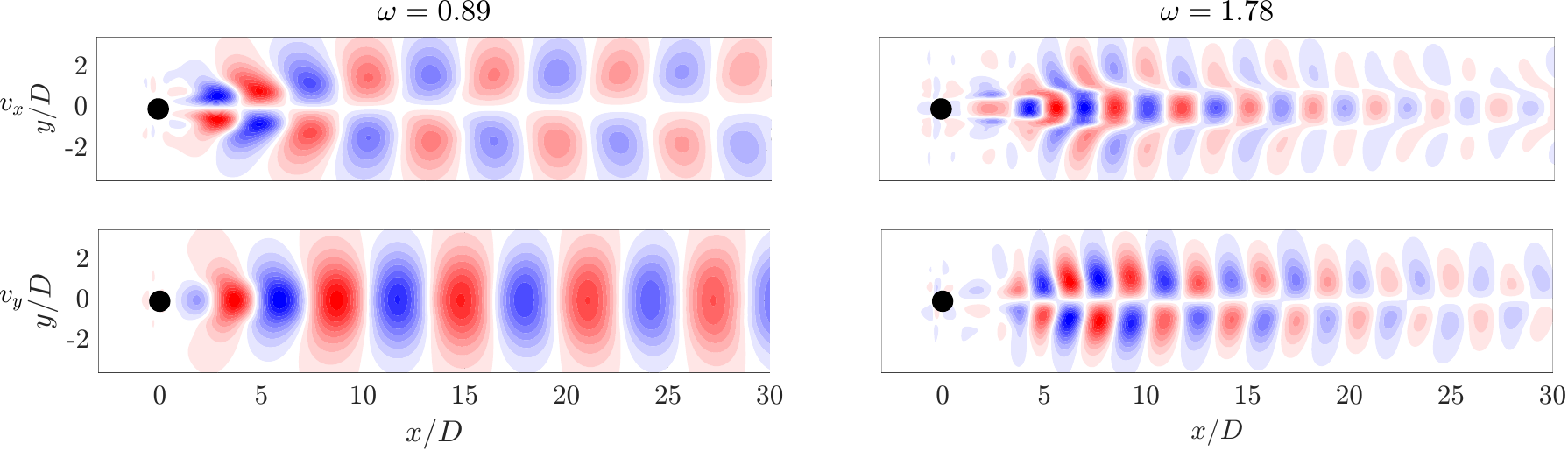}
\caption{$VE_2$ : $\beta_{s}=0.5,~n=1.0,~Bn=0.0$ }
\label{fig:DMD_3_3}
\end{subfigure}

\caption{HODMD modes of the streamwise $v_x$ and normal $v_y$ velocity fields. The main frequency and the first harmonic are represented for Newtonian (Fig. \ref{fig:DMD_3_1}), $VE_1$ (Fig. \ref{fig:DMD_3_2}), and $VE_2$ (Fig. \ref{fig:DMD_3_3}).} 
\label{fig:DMD_3}
\end{figure}

\subsection{HODMD of Newtonian and viscoelastic fluids \label{sec:HODMD_Ve}}

Figure \ref{fig:DMD_2} presents the spectrum of the Newtonian and viscoelastic cases. As seen, the three cases show clear periodic behaviour, only changing the value of the main frequency $\omega_{1}$, while the rest of the modes are harmonic of the first one ($\omega_{n} = n \omega_{1}, n \in \mathbb{N}$). As previously calculated in the POD analysis, the main frequency is $\omega = 1.04$ for the $N$ case, $\omega = 0.99$ for $VE_1$ and $\omega = 0.89$ for $VE_2$. Adding viscoelastic effects to the fluid slows the main frequency, as well as decreasing solvent viscosity ratio $\beta_s$ (which corresponds to increasing the concentration of the polymer). 

The two dominant DMD modes are displayed in \Figref{fig:DMD_3}. The selected modes represent the main frequency and the second harmonic for each case. The two velocity components are shown. For the Newtonian flow case, the mode with frequency $\omega_{1} =1.04$ in \Figref{fig:DMD_3} (left-hand column) is related to the von Kármán vortex street of the two-dimensional flow, which is the first flow bifurcation predicted by the linear theory. The spatial structure of the first dominant mode is approximately circular and well-organized for Newtonian ($N$) and high-$\beta_s$ viscoelastic fluid $VE_{1}$, with frequency $\omega_{1} =0.99$, which gradually shrinks in the streamwise direction. However, for low-$\beta_s$ viscoelastic fluid $VE_{2}$, the mode structure deteriorates slightly close to the cylinder, appearing as a complex shape mode that gradually retains its circular shape and regular pattern by increasing the distance from the cylinder, with a frequency of $\omega_{1}=0.89$. The addition of viscoelasticity, as well as the increment of the concentration of the polymer also extends the intensity of the mode further downstream the domain. The shape of these modes are closely related to the first two modes obtained in the POD analysis, as expected, however, the frequency obtained in both analysis is slightly different, as HODMD needs less snapshots to provide accurate results of the frequency \cite{VegaLeClaincheBook20}.

As shown in \Figref{fig:DMD_3} (right-hand column), the spatial structure of the second dominant mode (the first harmonic) is symmetrical and organized in all three cases: Newtonian ($N$), high-$\beta_s$ viscoelastic fluid $VE_{1}$, and low-$\beta_s$ viscoelastic fluid $VE_{2}$. Unlike the first mode, the second mode is more concentrated at the center-line of the domain. The shape of these modes are closely related to the third and fourth modes obtained in the POD analysis, as expected, since they were associated to a single frequency each

\begin{figure}
\centering
\includegraphics[width=0.45\linewidth]{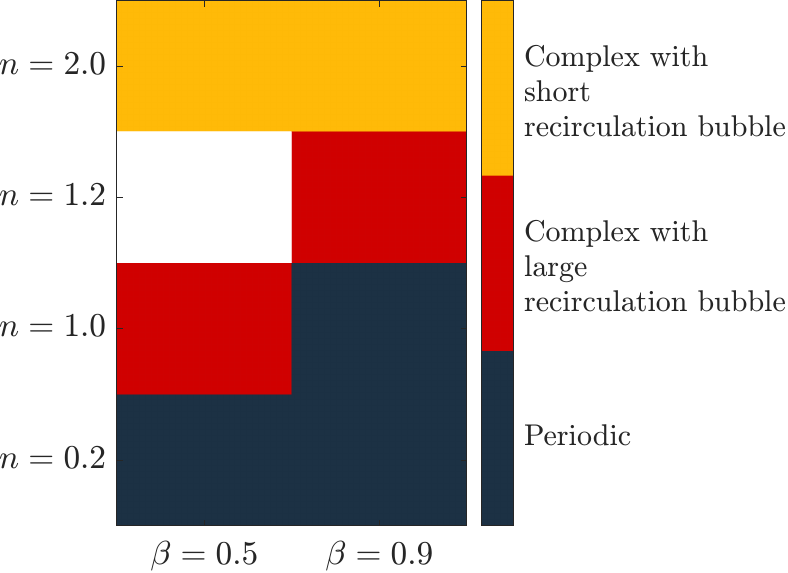}
\caption{Summary of the complexity in the dynamics of the different EVP solutions.}
\label{fig:POD_summary}
\end{figure}

\begin{figure} %[h] DMD_4
\begin{subfigure}{0.49\textwidth}
\centering
\includegraphics[width=1.0\linewidth, height=2.5cm]{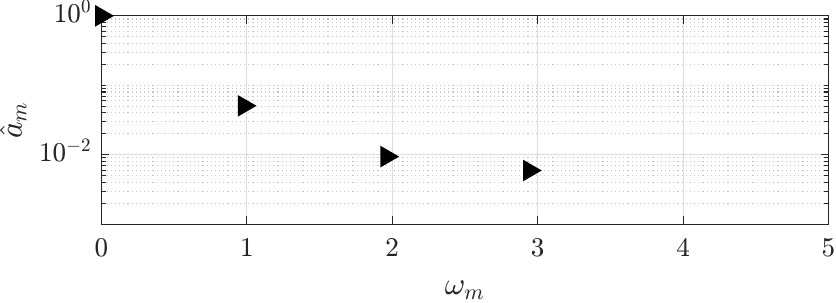}
\caption{$EVP_1$: $\beta_{s}=0.9,~n=0.2,~Bn=2.0$  }
\label{fig:DMD_4_1}
\end{subfigure}
\begin{subfigure}{0.49\textwidth}
\centering
\includegraphics[width=1.0\linewidth, height=2.5cm]{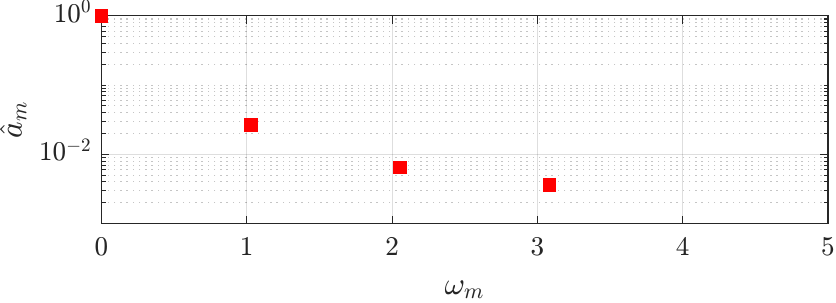}
\caption{$EVP_2$: $\beta_{s}=0.5,~n=0.2,~Bn=2.0$  }
\label{fig:DMD_4_2}
\end{subfigure}
\begin{subfigure}{1.0\textwidth}
\centering
\includegraphics[width=0.49\linewidth, height=2.5cm]{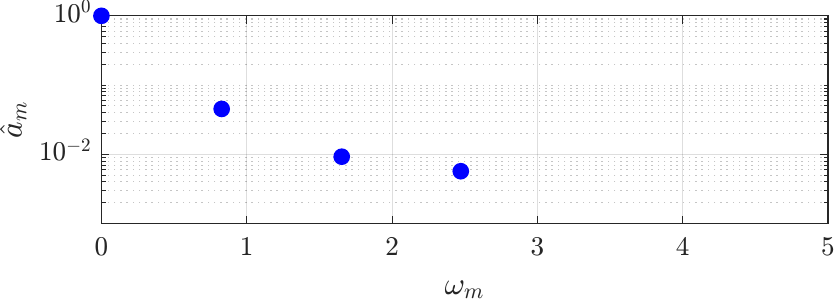}
\caption{$EVP_3$: $\beta_{s}=0.9,~n=1.0,~Bn=2.0$  }
\label{fig:DMD_4_3}
\end{subfigure}
\caption{DMD-d Spectrum. Amplitudes normalized with the maximum value ($\hat{a}_m=a_m/a_0$) vs. frequencies $\omega_m$ obtained with different tolerances and order $d$ for $EVP_{1}$ (\Figref{fig:DMD_4_1}), $EVP_2$ (\Figref{fig:DMD_4_2}), and $EVP_{3}$ (\Figref{fig:DMD_4_3}).} 
\label{fig:DMD_4}
\end{figure}

\begin{figure} %[h] DMD_5
\begin{subfigure}{1.0\textwidth}
\centering
\includegraphics[width=0.7\linewidth]{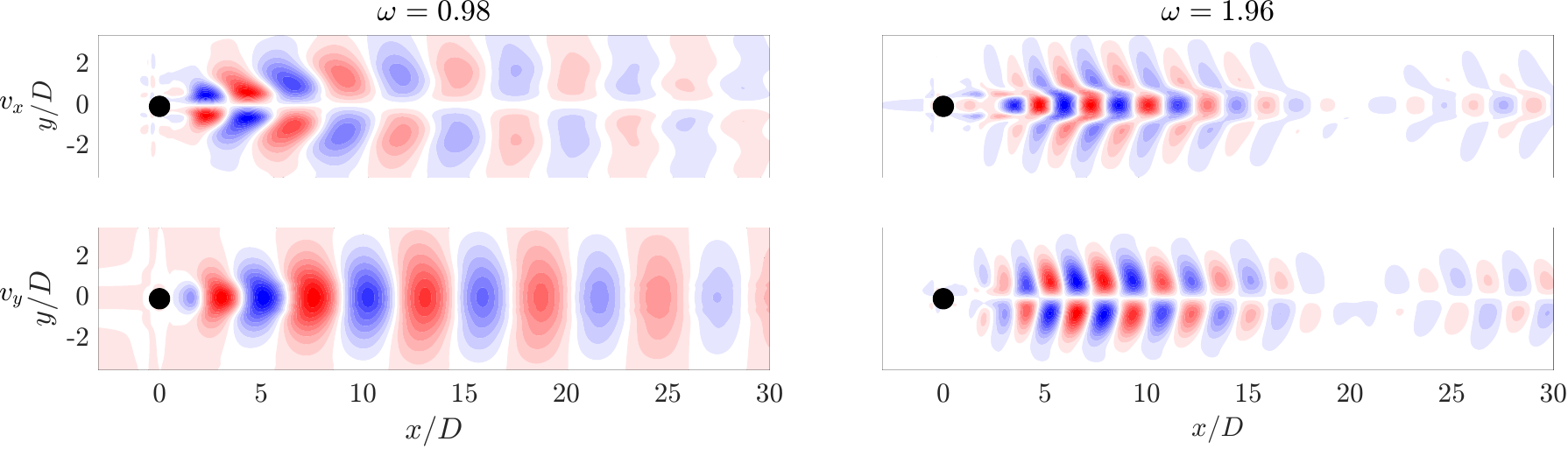}
\caption{$EVP_1$: $\beta_{s}=0.9,~n=0.2,~Bn=2.0$  }
\label{fig:DMD_5_1}
\end{subfigure}
\begin{subfigure}{1.0\textwidth}
\centering
\includegraphics[width=0.7\linewidth]{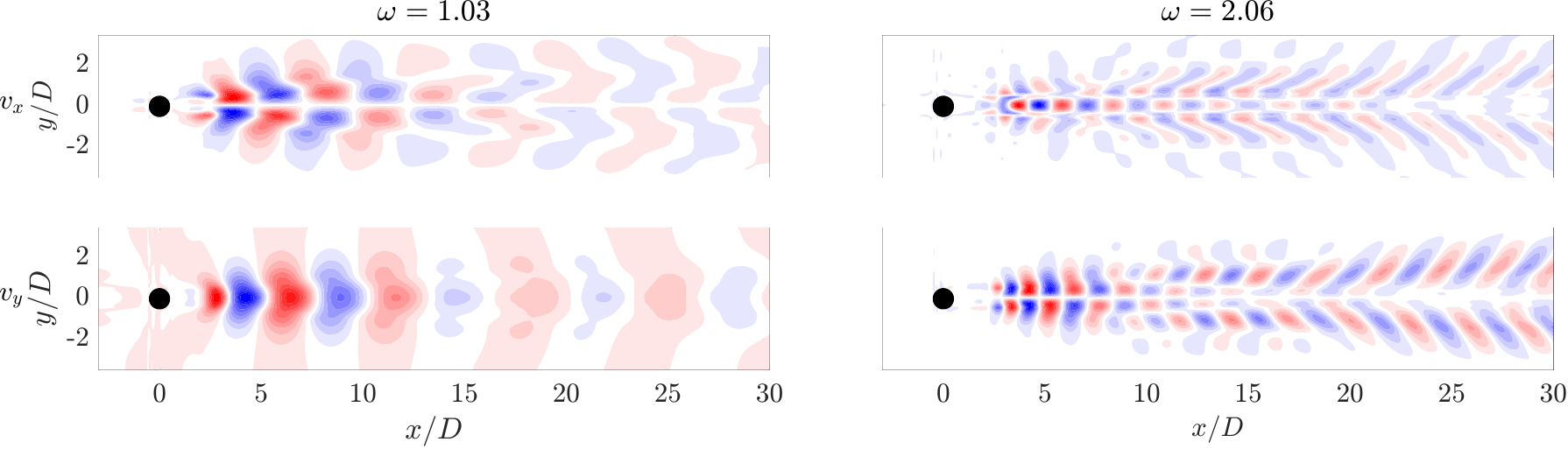}
\caption{$EVP_2$: $\beta_{s}=0.5,~n=0.2,~Bn=2.0$  }
\label{fig:DMD_5_2}
\end{subfigure}
\begin{subfigure}{1.0\textwidth}
\centering
\includegraphics[width=0.7\linewidth]{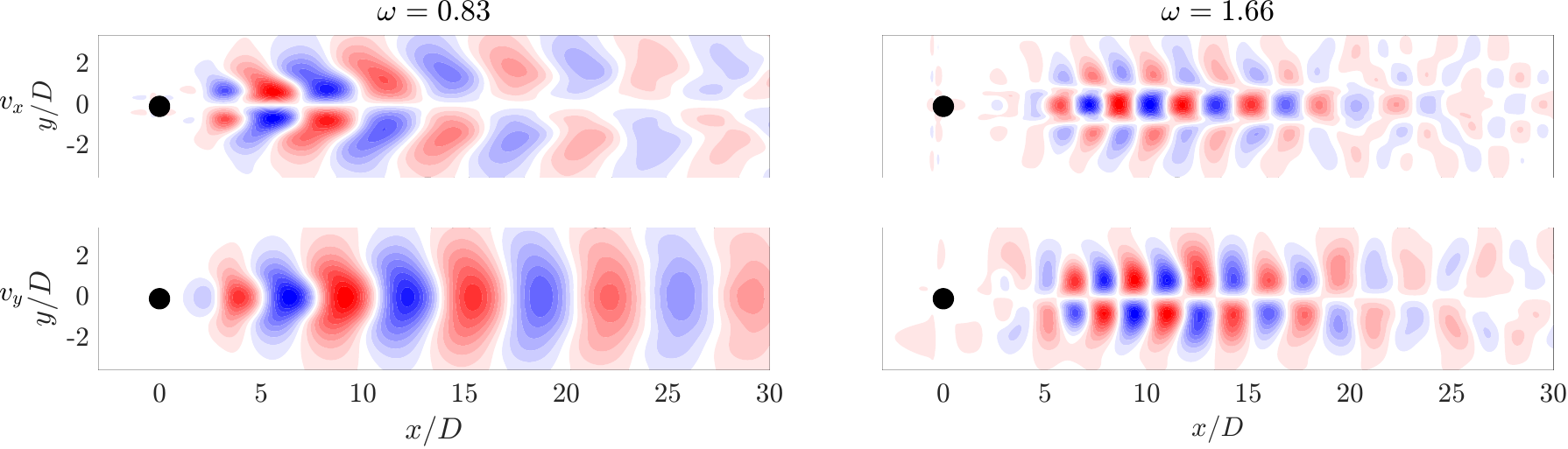}
\caption{$EVP_3$: $\beta_{s}=0.9,~n=1.0,~Bn=2.0$  }
\label{fig:DMD_5_3}
\end{subfigure}
\caption{HODMD modes of the streamwise $v_x$ and normal $v_y$ velocity fields. The main frequency and the first harmonic are represented for $EVP_1$ (Fig. \ref{fig:DMD_5_1}), $EVP_2$ (Fig.\ref{fig:DMD_5_2}), and $EVP_3$ (Fig. \ref{fig:DMD_5_3}).} 
\label{fig:DMD_5}
\end{figure}

\subsection{HODMD of elastoviscoplastic fluids \label{sec:HODMD_EVP}}

In this section, the impact of the interaction between elastic and plastic forces on flow structure is investigated. 
A summary of the EVP cases is shown in Fig. \ref{fig:POD_summary}. As observed, increasing the power-law index $n$ leads to a rise in the complexity of the cases for values of $\beta_s$. As $n$ increases, the recirculation bubble becomes progressively longer, as shown in Fig. \ref{fig:image300}. This continues until the recirculation bubble becomes too long and eventually breaks down. EVP cases with $\beta_s = 0.9$ show that the recirculation bubble persists longer as the power-law index \( n \) increases. Since the solvent and total viscosities are more similar in this case, the influence of \( n \) is attenuated.
Even though, it is evident that complex dynamics emerge before this breakdown. Therefore, cases with long and short recirculation bubbles will be analyzed in separate sections. HODMD will provide insight into these complex cases.

%\\ A. Periodic regime \\
%\paragraph{A. Periodic regime} 
\subsubsection{A. Periodic regime}

The main dynamics of this regime are the same as in the Newtonian and viscoelastic cases, i.e. the bidimensional von Kármán street. The periodic cases of the EVP fluid are two shear-thinning simulations at both values of $\beta_s$, and the shear-independent case at $\beta_s=0.9$; $EVP_1$,$EVP_2$ and $EVP_3$ respectively. In  \Figref{fig:DMD_4}, the spectrum of these cases are shown. As explained earlier, the spectrums show a clearly dominant frequency and its harmonics. The main frequency is $\omega = 0.98$ for $EVP_1$, $\omega = 1.02$ for $EVP_2$ and $\omega = 0.83$ for $EVP_3$ case.

The introduction of plasticity in the viscoelastic case at  $\beta_s=0.9$ ($EVP_3$ compared to $VE_1$) leads to a slower system response, resulting in a reduced dominant frequency (from $\omega = 0.99$ in the $VE_1$ case to $\omega = 0.83$ in the $EVP_3$ case). Lowering the power law index $n$ ($EVP_1$) implies an increase in the main frequency. Lastly, decrease in solvent viscosity contribution (to $\beta_s=0.5$) gives an increase in the main frequency, a behaviour opposed to the one seen in the viscoelastic cases, where lowering $\beta_s$ (concentrated polymer solution) meant a decrease in the frequency (Fig. \ref{fig:DMD_2}).

The shape of the main mode and the first harmonic is  presented in Fig. \ref{fig:DMD_5}. Both velocity components are displayed, showing that all cases exhibit modes resembling the Newtonian case associated with the von Kármán vortex street, though slightly altered.

As shown in the left-hand column of figure \ref{fig:DMD_5}, we observe that the first dominant mode still exhibits a regular, anti-symmetric pattern with a complex shape throughout the entire domain for shear-independent EVP fluid with $n=1$ (denoted by $EVP_{3}$) with a frequency of $\omega_{1}=0.83$. Compared with the viscoelastic case $VE_1$, the addition of plasticity stretches the mode in a region nearer the center line. As the recirculation bubble is larger than in the Newtonian case (\Figref{fig:image300}), the region of high intensity of the mode is further from the cylinder.

When the fluid is more shear-thinning with $n=0.2$, the mode pattern becomes slightly more regular compared to the $n=1$ case. Additionally, the size of the high intensity areas in the modes is smaller for $EVP_{1,2}$ compared to $EVP_{3}$ with $n=1$. The mode related to higher solvent viscosity, $EVP_1$, with a frequency of $\omega_{1}=0.98$, still have bigger structures than the lower solvent viscosity $EVP_2$, with a frequency of $\omega = 1.03$. This last case is concentrated in a region nearer the cylinder an it is stretched in the streamwise direction. In all these three cases, the main mode can be related to the first two POD modes obtained in the previous analysis. 

The right-hand column of \Figref{fig:DMD_5} shows that in high-$\beta_s$ shear-independent EVP fluid case with $n=1$ ($EVP_{3}$), the second robust mode has a regular, symmetric pattern with a circular shape, similar to the previous case, with frequency $\omega_{2} =1.66$. For the shear-thinning fluid case, with $n=0.2$, the pattern of the mode is also regular and symmetric, with a frequency of $\omega_{2} =1.96, 2.06$, for high-$\beta_s$ and low-$\beta_s$ EVP solutions $EVP_{1,2}$ cases, respectively. The differences in the shape of the mode are the same as the differences in the main mode. This mode can be related to the third and fourth mode in the POD analysis. The presence of plasticity increases the flow complexity, which is manifested in the emergence of higher-order harmonics. This observation suggests that the transition to more complex regimes, further detailed in the following sections, is associated with non-linear interactions that become significant at higher frequencies.

%\\ B. Quasiperiodic \\
\subsubsection{B. Complex dynamics with elongated recirculation bubble}

Increasing the power-law index $n$ or lowering the solvent viscosity ratio ($\beta_s$), the complexity of the flow increases, as the elastic instability is triggered. This complex flow regime can be divided in two different subregimes. If $n$ is not big enough, the recirculation bubble is elongated but not broken, and the wake behind this recirculation bubble is complex, with more than one frequency leading the dynamics. The cases which involve this subregime are the low-$\beta_s$ shear-independent EVP fluid $EVP_4$ and the high-$\beta_s$ EVP solution with weak shear-thickening $EVP_5$. 

\begin{figure} %[h] DMD_6
\begin{subfigure}{0.49\textwidth}
\centering
\includegraphics[width=1.0\linewidth, height=2.5cm]{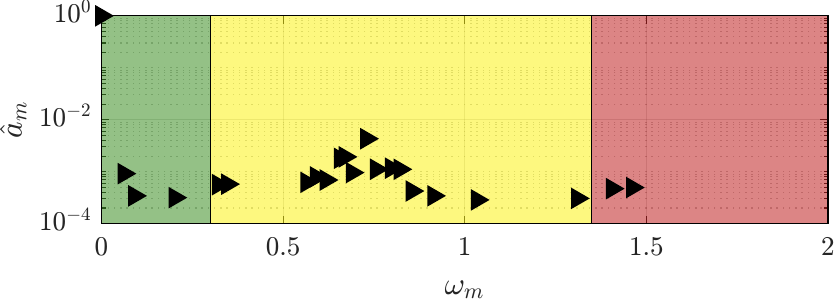}
\caption{$EVP_4$: $\beta_{s}=0.5,~n=1.0,~Bn=2.0$  }
\label{fig:DMD_6_1}
\end{subfigure}
\begin{subfigure}{0.49\textwidth}
\centering
\includegraphics[width=1.0\linewidth, height=2.5cm]{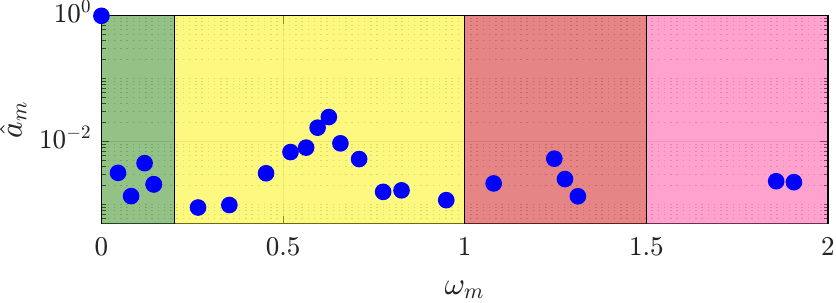}
\caption{$EVP_5$: $\beta_{s}=0.9,~n=1.2,~Bn=2.0$  }
\label{fig:DMD_6_2}
\end{subfigure}
\caption{DMD-d Spectrum. Amplitudes normalized with the maximum value ($\hat{a}_m=a_m/a_0$) vs. frequencies $\omega_m$ obtained with different tolerances and order $d$ for $EVP_{4}$ (\Figref{fig:DMD_6_1}) and $EVP_{5}$ (\Figref{fig:DMD_6_2}). In green, Region I; in yellow, Region II; in red, Region III; and in pink, Region IV.} 
\label{fig:DMD_6}
\end{figure}

\begin{figure} %[h] DMD_7
\begin{subfigure}{1.0\textwidth}
\centering
\includegraphics[width=0.7\linewidth]{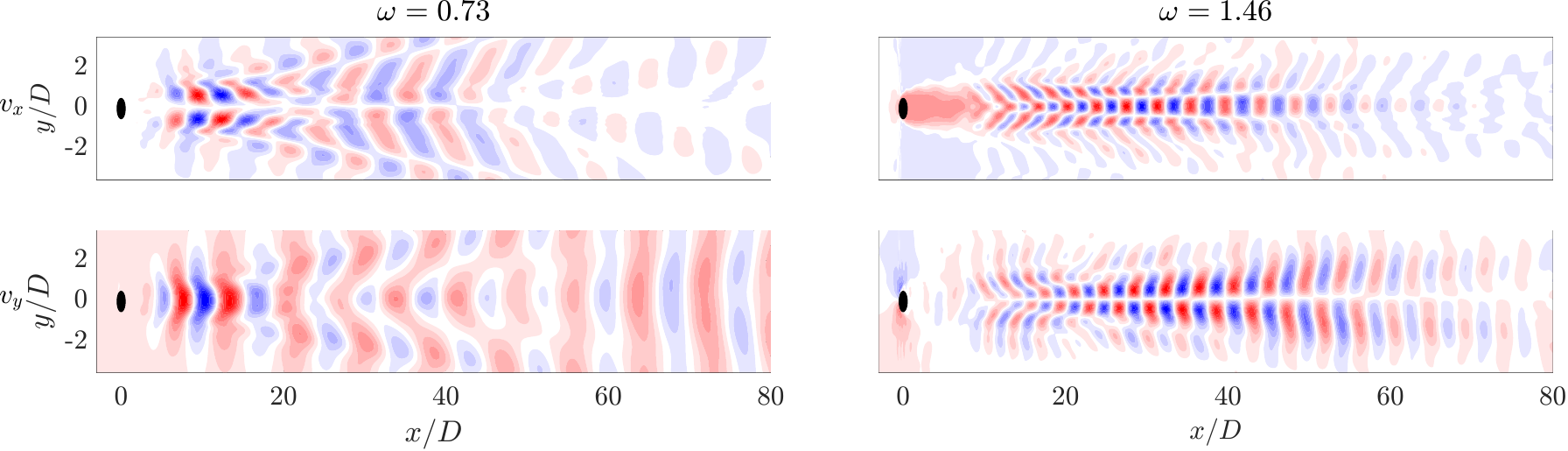}
\caption{$EVP_4$: $\beta_{s}=0.5,~n=1.0,~Bn=2.0$ }
\label{fig:DMD_7_1}
\end{subfigure}
\begin{subfigure}{1.0\textwidth}
\centering
\includegraphics[width=0.7\linewidth]{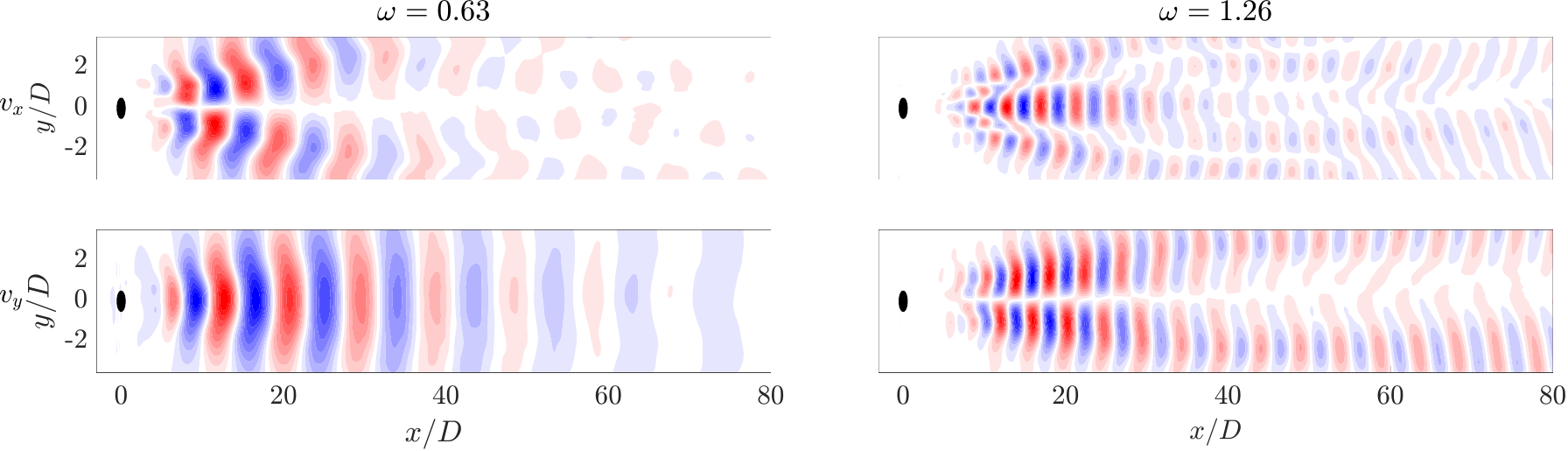}
\caption{$EVP_5$: $\beta_{s}=0.9,~n=1.2,~Bn=2.0$ }
\label{fig:DMD_7_2}
\end{subfigure}
\caption{HODMD modes of the streamwise $v_x$ and normal $v_y$ velocity fields. The main frequency and the mode with doubled frequency are represented for $EVP_4$ (Fig. \ref{fig:DMD_7_1}) and $EVP_5$ (Fig.\ref{fig:DMD_7_2}).} 
\label{fig:DMD_7}
\end{figure}

In Fig. \ref{fig:DMD_6}, the amplitude-frequencies, $a-\omega$, diagram for the $EVP_{4,5}$ cases are displayed. In this case, the spectra are more complex than in the previous cases. The highest amplitude mode in the  shear-independent concentrated case $EVP_4$ (Fig. \ref{fig:DMD_6_1}) has a frequency of $\omega = 0.73$. The shape of this mode, presented in Fig. \ref{fig:DMD_7}, is similar to the ones showed in previous cases, with a regular antisymmetric pattern. This mode can be related to the von Kármán street, however in this case, there is a range of frequencies where the mode related is similar to the one presented, but with lower amplitude. This region comprehends between $0.35\leq \omega \le 1.35$, i.e., Region II (yellow region) in Fig. \ref{fig:DMD_6_1}. Higher frequencies present modes with shape similar to the first harmonic in periodic cases (Region III in Fig. \ref{fig:DMD_6_1}). These modes have a symmetric pattern in the streamwise velocity. 

Comparing these two modes with the shear-thinning concentrated $EVP_2$ case, the mode extends longer in the streamwise direction, as the maximum intensity is presented at $x/D = 35$, while in the $EVP_2$ it is at $x/D = 5$. This can be understood as the recirculation bubble is elongated in this direction, therefore, the wake starts further away from the cylinder.

\begin{figure} %[h] DMD_7
\begin{subfigure}{0.49\textwidth}
\centering
\includegraphics[width=0.8\linewidth]{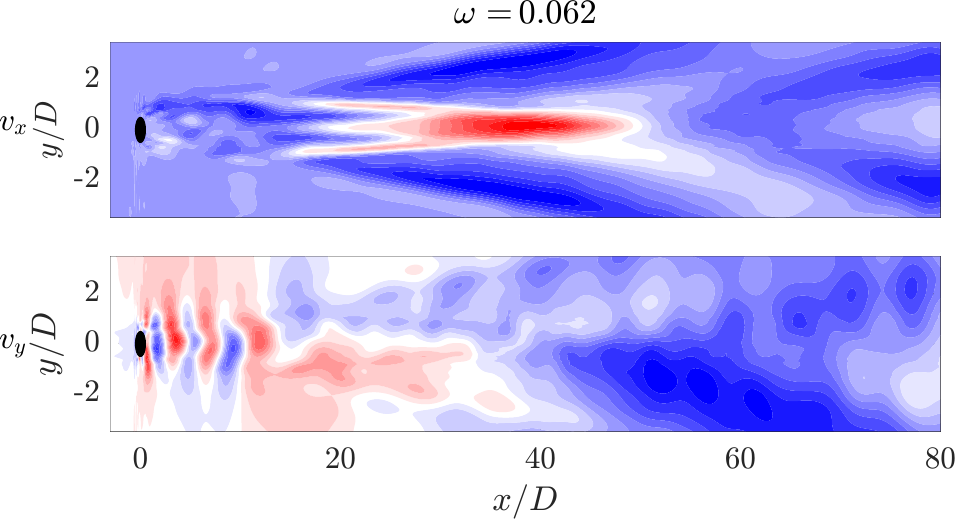}
\caption{$EVP_4$: $\beta_{s}=0.5,~n=1.0,~Bn=2.0$ }
\label{fig:DMD_8_1}
\end{subfigure}
\begin{subfigure}{0.49\textwidth}
\centering
\includegraphics[width=0.8\linewidth]{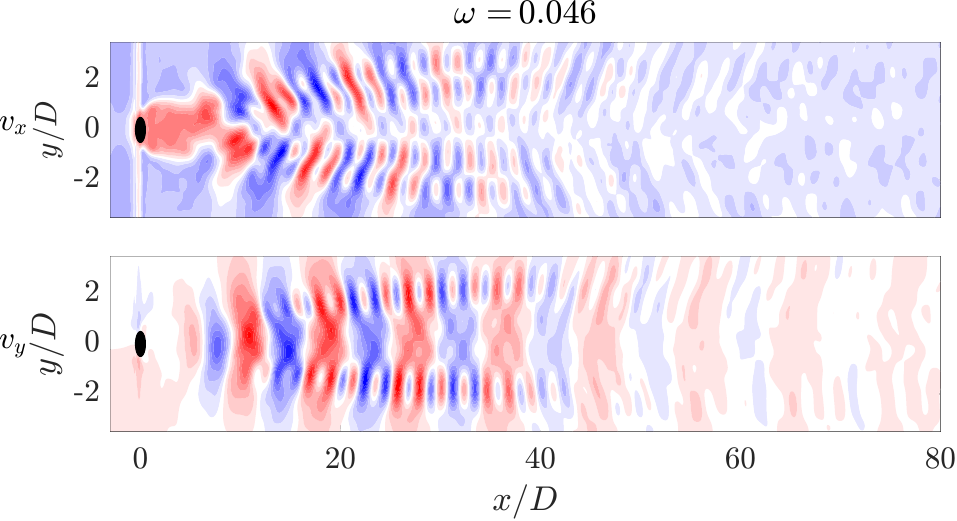}
\caption{$EVP_5$: $\beta_{s}=0.9,~n=1.2,~Bn=2.0$ }
\label{fig:DMD_8_2}
\end{subfigure}
\caption{HODMD modes of the streamwise $v_x$ and normal $v_y$ velocity fields. The lowest frequency is represented for $EVP_4$ (Fig. \ref{fig:DMD_8_1}) and $EVP_5$ (Fig.\ref{fig:DMD_8}).} 
\label{fig:DMD_8}
\end{figure}

Apart from the Regions II and III; in this case, a new region appears with modes with low frequencies (Region I in Fig. \ref{fig:DMD_6_1}). The frequency with the largest amplitude in this region is $\omega = 0.062$. The shape of this mode, shown in Fig. \ref{fig:DMD_8}, presents a high intensity area formed by a large size structures in the far field of the cylinder wake for the streamwise component, while for the normal component, non-symmetric flow structures focus on the near field.

The shear-thickening $n=1.2$ high-$\beta_s$ case $EVP_5$ presents a similar behavior. The spectrum (Fig. \ref{fig:DMD_6_2}) identifies in this case four different regions. The highest amplitude mode is $\omega = 0.63$ (found in Region II), way lower than the shear-independent case $EVP_3$. The shape of this mode, shown in Fig. \ref{fig:DMD_7_2}, is similar to the leading mode in the periodic EVP cases, but the size of the high intensity structures is larger than the $EVP_3$ case. The shape of the rest of the modes in Region II, which comprehends between $0.2 \leq \omega \le 1$, is similar to this mode. Higher frequencies present modes with shape similar to the first harmonic in periodic cases (Region III in Fig. \ref{fig:DMD_6_2}) or to the second harmonic in periodic cases (Region IV in Fig. \ref{fig:DMD_6_2}).

There is a region of low frequency modes (Region I in Fig. \ref{fig:DMD_6_2}), as in the $EVP_4$ case. The modes have lower frequencies in this case, as well as different shape. These modes demonstrate the underlying complexity of the problem, where the elastoviscoplasticity of the fluid induces the transition to turbulence. This is reflected on the modes, which are non-symmetric and present small scales.

%\\ C. Chaotic \\
\subsubsection{C. Complex dynamics with shear-thickening}

Continuing with the increase of the power law index $n$, the recirculation bubble breaks down and the complexity of the problem sharply increases. This new regime encloses the high power law index EVP cases $EVP_{6}$ and $EVP_{7}$ with both values of $\beta_s$.

\begin{figure} %[h] DMD_9
\begin{subfigure}{0.49\textwidth}
\centering
\includegraphics[width=1.0\linewidth, height=2.5cm]{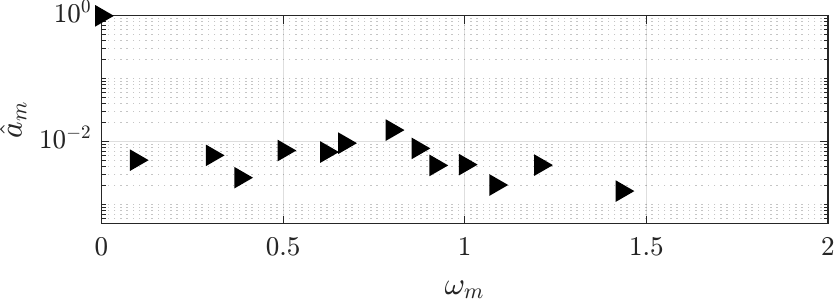}
\caption{$EVP_6$: $\beta_{s}=0.9,~n=2.0,~Bn=2.0$ }
\label{fig:DMD_9_1}
\end{subfigure}
\begin{subfigure}{0.49\textwidth}
\centering
\includegraphics[width=1.0\linewidth, height=2.5cm]{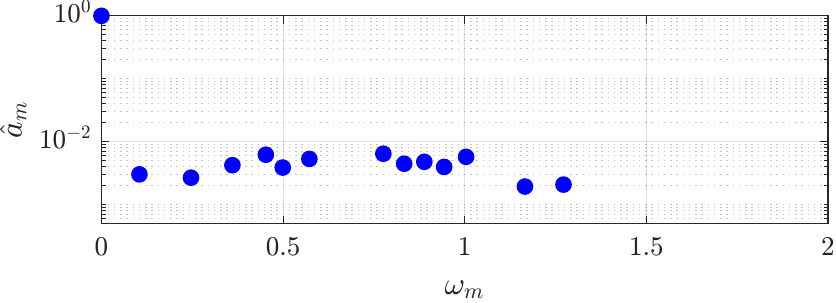}
\caption{$EVP_7$: $\beta_{s}=0.5,~n=2.0,~Bn=2.0$ }
\label{fig:DMD_9_2}
\end{subfigure}
\caption{DMD-d Spectrum. Amplitudes normalized with the maximum value ($\hat{a}_m=a_m/a_0$) vs. frequencies $\omega_m$ obtained with different tolerances and order $d$ for $EVP_{6}$ (\Figref{fig:DMD_9_1}) and $EVP_{7}$ (\Figref{fig:DMD_6_2}).} 
\label{fig:DMD_9}
\end{figure}

The spectrum for both cases is presented in Fig. \ref{fig:DMD_9}. As seen, the spectra are more complex than in previous cases.  The frequencies present similar amplitudes and there is no a clear hierarchy on the modes, suggesting once more that highly complex nature of the flow, which is driven by several frequencies modelling small size flow structures. However, one frequency has a slightly higher amplitude than the others in both cases, i.e., $\omega = 0.8$ in the $EVP_6$ case and $\omega = 0.77$ in the $EVP_7$ case. Additionally, another significant mode is the one with the lowest frequency ($\omega = 0.1$ in $EVP_6$ and $\omega = 0.11$ in $EVP_7$), as it is the first to appear in the spectrum and its frequency governs the periodicity of the main dynamic. In the $EVP_6$ case, this frequency is $1/8$ of the main frequency and in the $EVP_7$ case is $1/7$ of the main one.

\begin{figure} %[h] DMD_7
\begin{subfigure}{1.0\textwidth}
\centering
\includegraphics[width=0.7\linewidth]{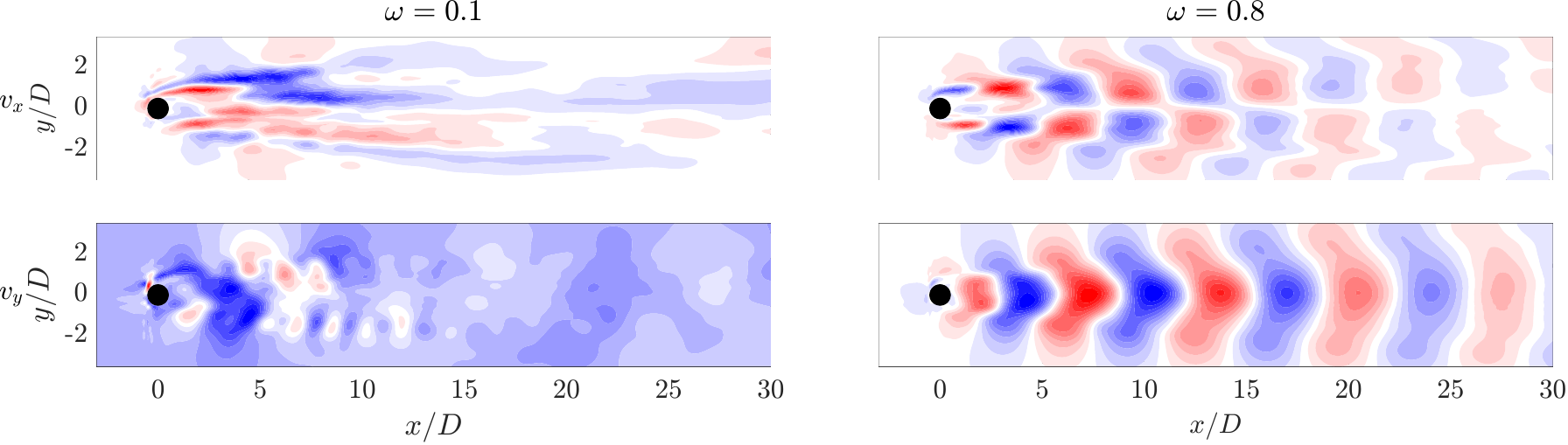}
\caption{$EVP_6$: $\beta_{s}=0.9,~n=2.0,~Bn=2.0$ }
\label{fig:DMD_10_1}
\end{subfigure}
\begin{subfigure}{1.0\textwidth}
\centering
\includegraphics[width=0.7\linewidth]{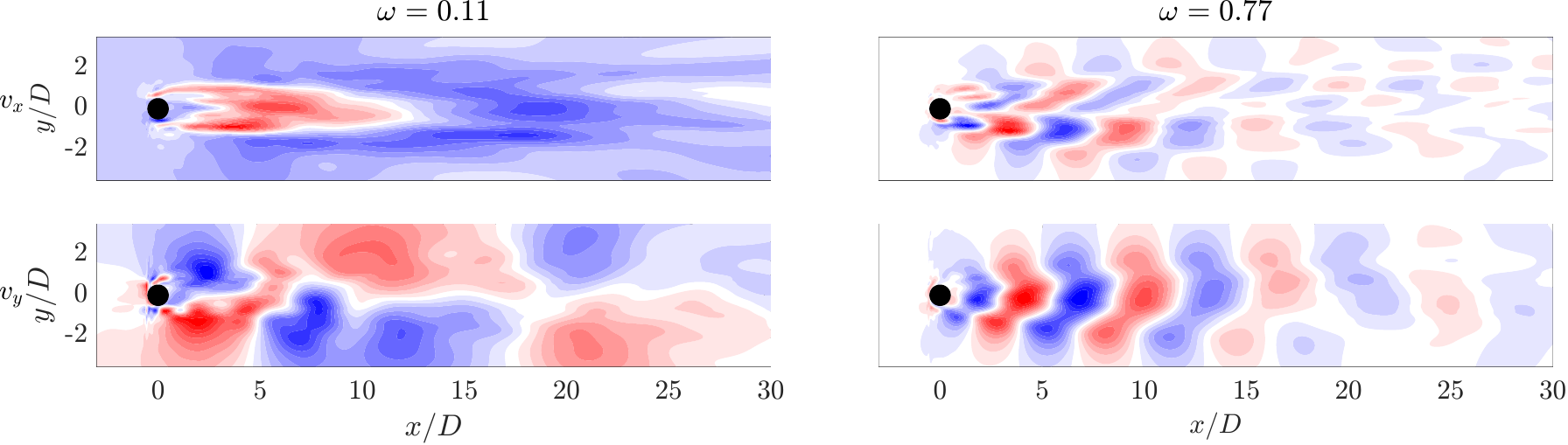}
\caption{$EVP_7$: $\beta_{s}=0.5,~n=2.0,~Bn=2.0$ }
\label{fig:DMD_10_2}
\end{subfigure}
\caption{HODMD modes of the streamwise $v_x$ and normal $v_y$ velocity fields. The lowest frequency mode (left) and the highest amplitude mode (right) are represented for $EVP_6$ (Fig. \ref{fig:DMD_10_1}) and $EVP_7$ (Fig.\ref{fig:DMD_10_2}).} 
\label{fig:DMD_10}
\end{figure}

The modes in which frequency is mentioned above are presented in Fig. \ref{fig:DMD_10}. The mode related to the main frequency is similar in both cases. These modes present an antisymmetric pattern in the streamwise direction, similar to the pattern related to the von Kármán street but in this case the structures start from the cylinder. The size of the structures are smaller than the previous cases, while the main frequency is slightly greater than the cases presented in the previous section.
On the other hand, the low frequency modes present large structures along the streamwise direction. In this region, the complexity of the system is high, and once, ther recirculation bubble is broken, new dynamics appear on the flow with the same importance as the mode related to the von Kármán street.

\begin{figure} %[h] DMD_11
\centering
\includegraphics[width=0.45\linewidth]{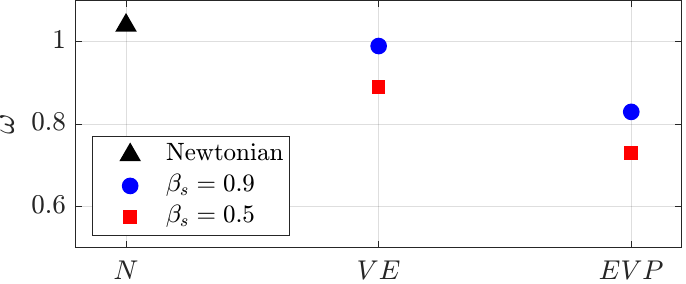}
\caption{Main frequency from the DMD-d analysis of five cases, N, $VE_{1,2}$, and $EVP_{3,4}$ fluids, all shear-independent fluids ($n = 1$).}
\label{fig:DMD_11}
\end{figure}

To sum up, the main frequency of all the cases analysed, i.e., Newtonian, viscoelastic and elastoviscoplastic, is analysed. Figure \ref{fig:DMD_11} shows the main frequencies for Newtonian and viscoelastic flows with $Bn=0$ and shear-independent EVP fluid with $Bn=2.0$ for both values of $\beta_s$ ($EVP_3$ and $EVP_4$, respectively). The results show that the frequency decreases from Newtonian to viscoelastic and further decreases in the EVP case with $n=1$. In all cases, the low-$\beta_s$ solution has a lower frequency compared to the high-$\beta_s$ solution.

\begin{figure} %[h] DMD_12
\centering
\includegraphics[width=0.45\linewidth]{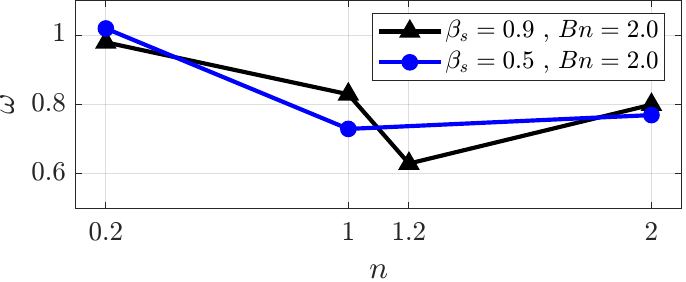}
\caption{Main frequency from the DMD-d analysis of EVP fluids cases $EVP_{1-7}$ fluids, and show the Shear-thinning $n=0.2$ and Shear-thickening $n=2.0$ effects}
\label{fig:DMD_12}
\end{figure}

Figure \ref{fig:DMD_12} shows the evolution of the main frequency in the EVP cases. Independently of $\beta_s$, strong shear-thinning with $n=0.2$ increases the frequency of the dominant mode. The lowest frequency is presented in the cases where the recirculation bubble is the largest and the dynamics are not periodic. The shear-thickening effect with $n=2.0$ breaks the recirculation bubble, which increases the frequency, and frequencies of high-$\beta_s$ and low-$\beta_s$ cases become similar. With strong shear-thinning, the fluid with low-$\beta_s$ presents a higher frequency than the fluid with high-$\beta_s$, which is an opposite behaviour to all other cases.

\section{Conclusions\label{sec:conclusions}}

In the present paper, the effects of the elasticity and plasticity on the two-dimensional flow past a circular cylinder have been thoroughly studied. Starting from the well-known Newtonian flow past a circular cylinder at Reynolds number $Re = 100$, different viscoelastic (VE) and elastoviscoplastic (EVP) fluids have been studied. Different numerical simulations have been conducted using the Saramito Herschel-Bulkley constitutive equations to model the behavior of the EVP fluids. The velocity field was extracted from the simulations conducted and it was analysed by means of Proper Orthogonal Decomposition (POD) and Higher Order Dynamic Mode Decomposition (HODMD). These two machine learning techniques are complementary, as POD extract coherent patterns in fluid mechanics which are orthogonal in space, but they may move with different frequencies; and the coherent patterns extracted with HODMD have a single frequency associated.

The influence of the elasticity on the von Kármán street is studied with the numerical simulations conducted in VE fluids. Two different simulations were carried out at Weissenberg number $Wi = 1$, varying the ratio of solvent viscosity to total viscosity ($\beta_s = 0.5, 0.5$). The results in the simulations showed that the length of the recirculation bubble significantly increases, and the recirculation centers is shifted in the streamwise direction, when compared to the Newtonian case. Analysing the database extracted from both simulations, both cases appear to be periodic although the complexity of the flow has increased. Some differences on the main frequency and the main mode associated appear when compared to the Newtonian case. The diluted solution ($\beta_s = 0.9$) presents higher frequency and the mode is shifted downstream, as the recirculation bubble has grown in that direction. This behaviour of the mode is accentuated in the case of the concentrated solution ($\beta_s = 0.5$), even though the frequency associated is lower.

The complexity of the fluid dynamics highly increases when combining fluid elasticity and plasticity. Seven different EVP fluids were simulated at Bingham number $Bn = 2$, varying the parameter $\beta_s$ and the power law index $n$. The power law index roughly indicates if the fluid exhibits shear-thinning behavior ($n < 1$), shear-thickening ($n > 1$) or shear-independent ($n = 1$). For shear-independent cases, the addition of plasticity enlarges the recirculation bubble compared to the purely viscoelastic flow, as well as the center of this bubble is shifted downstream. Shear-thinning fluids present a smaller and thinner recirculation bubble than their shear-independent counterparts. On the other hand, increasing $n$ in EVP fluids results in a growing complexity of the flow, with more complex dynamics than just the von Kármán street, which is the main feature in all other wakes studied here at $Re=100$. The analysis of the recirculation bubble and the simulations indicate that as $n$ increases, the recirculation bubble grows until it reaches a limit, at which point it breaks, leading to the complex dynamics in the two-dimensional flow past a circular cylinder.

It is possible to differentiate three different regimes when applying the two machine learning algorithms to the different simulations of EVP fluids. The first periodic regime appears for the shear-thinning cases and the shear-independent case with $\beta_s = 0.9$. In this regime, the dynamics are similar to the Newtonian case, with the von Kármán street and its harmonics, although the shape and frequency of the mode varies. The second regime is associated to elongated recirculation bubbles and complex dynamics. This regime comprehends the shear-independent case with $\beta_s = 0.5$ and the weak shear-thickening ($n=1.2$) case with $\beta_s = 0.9$. In these cases, in addition to the mode related to the von Kármán street and its harmonics, a low-frequency mode appears, as well as the non-linear interactions between this mode and the others. The third and last regime appears once the recirculation bubble is broken and breaks down, resulting in complex and more chaotic dynamics. This regime comprehends the strong shear-thickening ($n=2$) cases. Here, there are some DMD modes with a shape similar to the von Kármán mode, but in this regime, those modes appear directly downstream of from the cylinder, while in the periodic cases, the mode starts from the recirculation bubble. Overall, increasing the power-law index $n$ leads to higher shear stress and reduced relaxation capability of the fluid. This can be connected with the emergence of elastic instabilities. As a result of these instabilities, the recirculation bubble progressively elongates until a critical point at which it breaks. Following this rupture, the flow complexity initially decreases slightly before increasing again. This pattern of increasing complexity bears resemblance to the transition observed in Newtonian fluids as the Reynolds number increases, where the onset of turbulence is associated with a rise in flow complexity, preceded by a drop once the recirculation bubble collapses.

To sum up, the present paper shows a thorough examination of the two-dimensional flow past a circular cylinder in the inertial vortex shedding regime, at $Re=100$, in viscoelastic and elastoviscoplastic fluids. We identify three different dynamical regimes when varying different governing parameters. The understanding of the flow features appearing in these non-Newtonian flows sheds light on the interplay of inertia, yield stress and elasticity on flows behind obstacles.

\section*{Acknowledgements}
S.L.C. acknowledges the ENCODING project that has received funding from the European Union’s Horizon Europe research and innovation programme under the Marie Sklodowska-Curie grant agreement No. 101072779. S.L.C. acknowledges the MODELAIR project that has received funding from the European Union’s Horizon Europe research and innovation programme under the Marie Sklodowska-Curie grant agreement No. 101072559. The results
of this publication reflect only the author’s view and do not necessarily reflect those of the European Union. The European Union can not be held responsible for them. A.C. and S.L.C. acknowledge the grant PLEC2022-009235 funded by MCIN/AEI/ 10.13039/501100011033 and by the European Union “NextGenerationEU”/PRTR and the grant PID2023-147790OB-I00 funded by MCIU/AEI/10.13039 /501100011033 /FEDER, UE. T. I. and S. P. acknowledge funding from European Research Council through the grant StG2019-952529, and computational resources provided by the National Academic Infrastructure for Supercomputing in Sweden (NAISS). A.C. and S.L.C. gratefully acknowledge the Universidad Politécnica de Madrid (www.upm.es) for providing computing resources on the Magerit Supercomputer.

\bibliographystyle{unsrt}  
\bibliography{mybibfile}  %%% Remove comment to use the external .bib file (using bibtex).

\end{document}